\documentclass{aa}  

\usepackage{graphicx}
\usepackage[flushleft]{threeparttable}
\usepackage{txfonts}
\usepackage{hyperref}
\usepackage[mathcal]{euscript}
\usepackage{varwidth}
\usepackage{booktabs}
\usepackage{multicol}
\usepackage{natbib}
\usepackage[caption=false]{subfig}
\usepackage{multirow}
\graphicspath{{./immagini/}}
\begin{document}

   \title{A joint \textit{XMM}-\textit{NuSTAR} observation of the galaxy cluster Abell 523: constraints on Inverse Compton emission}

   \author{F. Cova\thanks{filippo.cova@inaf.it}
          \inst{1}
          \and
          F. Gastaldello\thanks{fabio.gastaldello@inaf.it}\inst{1}
          \and
		  D. R. Wik\inst{2}  
		  \and
		  W. Boschin\inst{5,6,7}
		  \and
		  A. Botteon\inst{3,4}
		  \and
		  G. Brunetti\inst{4}
		  \and
		  D. A. Buote\inst{8}
		  \and
		  S. De Grandi\inst{9}
		  \and
		  D. Eckert\inst{16}
		  \and
		  S. Ettori\inst{10,11}
		  \and
		  L. Feretti\inst{4}
		  \and
		  M. Gaspari\inst{12}\thanks{\textit{Lyman Spitzer Jr.} Fellow}
		   \and
		  S. Ghizzardi\inst{1}
		  \and
		  G. Giovannini\inst{3,4}
		  \and
		  M. Girardi\inst{13,14}
		  \and
		  F. Govoni\inst{15}
		  \and
		  S. Molendi\inst{1}
		  \and
		  M. Murgia\inst{15}
		  \and
		  M. Rossetti\inst{1}
		  \and
		  V. Vacca\inst{15}
          }

   \institute{IASF - Milano, INAF, Via Corti 12, I-20133 Milan, Italy
              \and
              Department of Physics \& Astronomy, University of Utah, Salt Lake City, UT 84112-0830, USA
              \and
              Dipartimento di Fisica e Astronomia, Università di Bologna, via P. Gobetti 93/2, 40129 Bologna, Italy
			  \and
INAF - Istituto di Radioastronomia, via P. Gobetti 101, 40129 Bologna, Italy
			  \and
			 Fundación Galileo Galilei - INAF, Rambla José Ana Fernandez Pérez 7, E-38712 Breña Baja, TF, Spain
			  \and
			 Instituto de Astrofisica de Canarias, C/Via Lactea s/n, E-38205 La Laguna, TF, Spain
			  \and
			  Dep. de Astrofisica, Univ. de La Laguna, Av. del Astrofisico Francisco Sanchez s/n, E-38205 La Laguna, TF, Spain
			  \and
			  Department of Physics and Astronomy, University of California at Irvine, 4129 Frederick Reines Hall, Irvine, CA 92697-4575		
			  \and
INAF - Osservatorio Astronomico di Brera, via E.Bianchi 46, 23807 Merate, Italy
			  \and
INAF - Osservatorio di Astrofisica e Scienza dello Spazio, via Pietro Gobetti 93/3, 40129 Bologna, Italy 
			  \and
INFN, Sezione di Bologna, viale Berti Pichat 6/2, I-40127 Bologna, Italy
			  \and
			   Department of Astrophysical Sciences, Princeton University, 4 Ivy Lane, Princeton, NJ 08544-1001 USA
			   \and
			   Dipartimento di Fisica dell’Università degli Studi di Trieste - Sezione di Astronomia, via Tiepolo 11, I-34143 Trieste, Italy
			   \and
INAF - Osservatorio Astronomico di Trieste, via Tiepolo 11, I-34143 Trieste, Italy
			   \and
INAF - Osservatorio Astronomico di Cagliari, Via della Scienza 5, I-09047 Selargius (CA), Italy
			   \and
			   Max-Planck Institut für Extraterrestrische Physik, Giessenbachstrasse 1, 85748 Garching, Germany
   }

\date{}

 
  \abstract
   {}
  {We present the results of a joint \textit{XMM-Newton} and \textit{NuSTAR} observation (200 ks) of the galaxy cluster Abell 523 at $z=0.104$. The peculiar morphology of the cluster radio halo and its outlier position in the radio power P(1.4 GHz) -  X-ray luminosity plane make it an ideal candidate for the study of radio - X-ray correlations and for the search of inverse Compton (IC) emission.}
  {We constructed bi-dimensional maps for the main thermodynamic quantities (i.e., temperature, pressure and entropy) derived from the \textit{XMM} observations to describe in detail the physical and dynamical state of the cluster ICM. We performed a point-to-point comparison in terms of surface brightness between the X-ray and radio emissions, to quantify their morphological discrepancies. Making use of \textit{NuSTAR}'s unprecedented hard X-ray focusing capability, we looked for IC emission both globally and locally, after properly modeling the purely thermal component with a multi-temperature description.}
   {The thermodynamic maps obtained from the \textit{XMM} observation suggest the presence of a secondary merging process that could be responsible for the peculiar radio halo morphology. This hypothesis is supported by the comparison between the X-ray and radio surface brightnesses, which shows a broad intrinsic scatter and a series of outliers from the best-fit relation, corresponding to those regions that could be influenced by a secondary merger. The global \textit{NuSTAR} spectrum can be explained by purely thermal gas emission, and there is no convincing evidence that an IC component is needed. The $3\sigma$ upper limit on the IC flux in the 20-80 keV band is in the range $\left[2.2 - 4.0\right] \times 10^{-13} \, \mathrm{erg} \, \mathrm{s}^{-1} \, \mathrm{cm}^{-2}$, implying a lower limit on the magnetic field strength in the range $B > [0.23 - 0.31] \, \mu G$. Locally, we looked for IC emission in the central region of the cluster radio halo finding a $3\sigma$ upper limit on the 20-80 keV non-thermal flux of $3.17 \times 10^{-14} \, \mathrm{erg} \, \mathrm{s}^{-1} \, \mathrm{cm}^{-2}$, corresponding to a lower limit on the magnetic field strength of $B \gtrsim 0.81 \, \mu G$.}
   {}

   \keywords{galaxies:cluster:general - X-rays:galaxies:clusters
               }
\titlerunning{A joint \textit{XMM-NuSTAR} observation of the galaxy cluster A523}
\maketitle
 
%

\section{Introduction}
The presence of relativistic particles and magnetic fields in the intra-cluster medium (ICM) of galaxy clusters has been established by a number of observations at radio frequencies (see \citealp{Feretti12} and \citealp{vanWeeren19} for reviews) and poses important physical questions (see \citealp{Brunetti14} for a review). The extended ($\sim$Mpc), diffuse, low surface brightness and steep-spectrum sychrotron sources known as radio halos are produced by relativistic electrons spiraling around $\sim \mu$G magnetic fields. These structures have so far been detected in $\sim$50 clusters at $z<0.4$ and in few other clusters above that redshift \citep{Yuan15}. All these clusters are characterized by high mass, high X-ray luminosity and temperature, and they show indications of a merger process as probed by X-ray morphology (\citealp{Buote01}, \citealp{Cassano10}), X-ray temperature maps (e.g., \citealp{Govoni04}) and presence of optical substructures (e.g., \citealp{Girardi11}). A causal connection between the hot and relativistic plasma properties is then suggested by the similarities of the radio morphology of the halos with the X-ray structure of the ICM \citep{Govoni01a}, and the total radio halo power $P_{1.4GHz}$ at 1.4 GHz correlates with the cluster total X-ray luminosity $L_{X}$ (e.g. \citealp{Liang00,Giovannini09,Kale13}). The radio power also correlates with the total cluster mass, which is the main physical parameter involved, as X-ray luminosity is only a proxy for the total mass and the one with the greatest scatter (with respect, for example, to X-ray temperature, gas mass or Sunyaev Zel'dovich (SZ) signal, see \citealp{Basu12} and \citealp{Cassano13}).

The synchrotron emission is a combined product of both the particle and the magnetic field density, therefore the latter cannot be globally constrained by radio observations alone. However, the same population of relativistic electrons that produce the radio emission will also produce hard X-ray emission by inverse Compton (IC) scattering photons from the ubiquitous Cosmic Microwave Background (CMB). For a power-law energy distribution, the ratio of IC to synchrotron flux gives a direct and unbiased measurement of the average magnetic field strength in the ICM. The search for non-thermal IC emission in clusters began with the first X-ray sensitive satellites, although the extended $\sim$keV photons from clusters were soon recognized to be of thermal origin (e.g. \citealp{Solinger72}, \citealp{Mitchell76}). However, clusters showing a diffuse and extended radio emission must also present IC emission at some level. Unfortunately, non-thermal emission is hard to detect, due to thermal photons which are simply too numerous below 10 keV, while IC emission should dominate and produce excess emission at higher energies, where the bremsstrahlung continuum falls off exponentially. In addition, the presence of multi-temperature structures, naturally occurring in merging galaxy clusters, could cause a false IC detection and needs to be modeled accurately; this problem is particularly relevant when using non-imaging and high background instruments, like the ones commonly used for the search of non-thermal emission in the hard X-ray band. The first IC searches with \textit{HEAO-1} resulted only in upper limits, and thus lower limits on the average magnetic field of $B \gtrsim 0.1 \, \mu$G (\citealp{Rephaeli87}; \citealp{Rephaeli88}). The next generation of hard X-ray satellites instead - \textit{RXTE} and \textit{Beppo-SAX} - several claimed detections ($10^{-11} - 10^{-12}\, \mathrm{erg}\, \mathrm{s^{-1}}\, \mathrm{cm}^{-2}$), although mostly of marginal significance and controversial (see \citealp{Rephaeli08} and \citealp{Brunetti14} for reviews). More recent observatories however (\textit{Suzaku} and \textit{Swift}) did not confirm IC emission at similar levels (\citealp{Ajello09}; \citealp{Ajello10}; \citealp{Wik12}; \citealp{Ota14}). The only exception, the Bullet cluster, has been falsified by a sensitive observation made with \textit{NuSTAR} \citep{Wik14}, the first satellite with imaging capabilities in the hard X-ray band (3-79 keV).

The galaxy cluster Abell 523 (A523) at $z=0.104$ is a massive system ($M_{200} \sim (7-9) \, \times \, 10^{14} \, M_{\odot}$; \citealp{Girardi16}; hereafter G16) that hosts  extended and diffuse radio emission with a total radio power of $P_{1.4 \, \mathrm{GHz}} = (2.0 \pm 0.1) \, \times \, 10^{24} \, \mathrm{W} \, \mathrm{Hz}^{-1}$ (G16), mainly elongated along the ESE-WNW direction (\citealp{Giovannini11}; hereafter G11) and a minor SSW-NNE elongation. G11 and G16 classified this radio emission as a radio halo because 1) the radio source permeates both merging subclusters and 2) the radio structure does not show any feature typical of radio relics such as high brightness filamentary structure or a transverse flux asymmetry as found in relic sources due to the propagation of a shock wave (see e.g. \citealp{VanWeeren11}).
\cite{VanWeeren11} suggested a possible relic interpretation for this feature, mainly because of its patchy morphology and perpendicular orientation with respect to the ICM and  galaxy distribution. However this interpretation has weaknesses due to its position within the merging subclusters, difficult to reconcile for a bimodal merger scenario between the two main subclusters launching a shock wave generating the relic, as recognized by \cite{Golovich18}. The features of the radio halo in A523, however, are clearly not typical: (i) the radio halo is mainly elongated in the direction perpendicular to the likely merging axis, with only a minor elongation aligned with the main optical/X-ray cluster elongation (see Fig. 1 in G16), while, generally, the opposite phenomenology is observed, with the radio halo being elongated in the same direction as the merger \citep{Feretti12}; (ii) A523 is peculiar in the $P_{1.4\mathrm{GHz}} - L_{X}$ plane, having a higher radio power (or a lower X-ray luminosity) than expected (see Fig. 20 in G16). Moreover, the radio emission is clearly offset from the X-ray emission and the radio/X-ray peaks offset is $\sim 0.3 \, h_{70}^{-1}$ Mpc. These latter peculiarities make A523 one of the best candidates for the search of IC emission, as the purely thermal cluster emission is low and can be separated from the non-thermal emission. G16 also detected a modest polarization (FPOL $\sim 15-20\%$). A polarized signal is unusual, since it has been detected so far only in a couple of other peculiar radio halos: MACS J0717.5+3745 \citep{Bonafede09}, and A2255 \citep{Govoni05}. In the latter case, however, the polarized filaments might be connected to relic emission in the outskirts of the cluster just projected at the center, see \cite{Pizzo11}. Both the observed radio/X-ray offset and polarization might be the result of having most of its magnetic field energy on large spatial scales; however, it is still not obvious how this can explain the peculiar morphology of the radio halo of A523. We also mention that A523 is part of a filamentary structure recently observed with the Sardinia Radio Telescope by \cite{Vacca18}, who found a possible evidence diffuse radio emission which might be connected to a large-scale filament of the cosmic web. Despite its interesting properties, A523 was still lacking deep and good quality X-ray data, the only modern observation being the shallow \textit{Chandra} observation analyzed in G16. For this reason, we requested a joint deep observation of the cluster with \textit{XMM-Newton} and \textit{NuSTAR}. On one hand, the \textit{XMM-Newton} data will allow us to perform a more complete and detailed characterization of the ICM and its connection with the cluster radio halo. On the other hand, making use of \textit{NuSTAR}'s imaging capabilities, we will investigate the presence of IC emission in this ideal system, where the relatively low X-ray luminosity should in principle minimize the thermal contribution. 

We describe the \textit{XMM-Newton} and \textit{NuSTAR} observations and their processing in Section \ref{sec:obs_processing}. In Section \ref{sec:bgd_modeling} we give a brief description of the background modeling for both telescopes. We show images in different energy bands in Section \ref{sec:image_anal}. In Section \ref{sec:thermal_em} we show the results obtained from the \textit{XMM-Newton} observations, accounting for the cluster thermal emission, while the results of the search for a non-thermal emission with \textit{NuSTAR} are presented in Section \ref{sec:nonthermal_em}. Finally, we discuss the implications of our results in Section \ref{sec:discussion} and summarize them in Section \ref{sec:summary}. Throughout this paper we use $H_{0} = 70 \, \mathrm{km} \, \mathrm{s}^{-1} \, \mathrm{Mpc}^{-1}$ and $h_{70} = H_{0}/(70 \, \mathrm{km} \, \mathrm{s}^{-1} \, \mathrm{Mpc}^{-1})$. We assume a flat cosmology, with $\Omega_{m} = 0.27$. In the adopted cosmology, 1' corresponds to $\sim 115 \, h_{70}^{-1}$ kpc at the cluster redshift. Unless otherwise stated, uncertainties are indicated at the 68\% confidence level.

\section{Observation and data processing}
\label{sec:obs_processing}
\subsection{XMM-Newton}
A523 was observed by \textit{XMM-Newton} on February 18th, 2016 (ObsID 0761070101 and ObsID 0761070201) for a total unfiltered exposure time of $\approx 220$ ks. The observations data files (ODF) were processed to produce calibrated event files using the \textit{XMM-Newton} Science Analysis System (XMM-SAS v14.0) and the corresponding calibration files, following the Extended Source Analysis Software scheme (ESAS; \citealp{Snowden08}). The presence of anomalous CCDs was also taken into account, removing them from the analysis, and soft proton flares were filtered out using ESAS tasks \textit{mos-filter} and \textit{pn-filter}. The \textit{cheese} procedure was then used to individuate and mask point sources in the field of view. The resulting clean exposure times for both observations and for each instrument are listed in Table \ref{tab:observations_xmm}. We also performed an estimation of the amount of residual soft proton flares contamination, comparing the measured count rate in a hard band in the exposed and unexposed part of the field of view (Fin/Fout, \citealp{LeccMol08}). We report on the results for the detector MOS as it is clear that soft proton contamination in the pn occurs also in the unexposed area of the detector due to a different camera geometry with respect to the MOS, and MOS1 requires special handling after the loss of two CCDs. The Fin/Fout procedure gives a ratio of 1.048 for the observation 0761070101 and a ratio of 1.051 for the observation 0761070201;
those values are in the range 1.0-1.15 considered to have negligible contamination as in the sample of 48 observations analized in \citealp{LeccMol08} and in the 495 observations, most of which from the XXL survey \citep{Pierre16} comprising more than 5 Ms of data analyzed in appendix A of \citealp{Ghirardini18}.
We therefore conclude that our observations are not effectively contaminated by residual soft protons and we will not include that component in our background modelling.

\begin{table}[t]
\caption{\label{tab:observations_xmm}\textit{XMM-Newton} observations of A523.}
\medskip
\centering
\begin{tabular}{l c c c } 
\toprule
ObsID & EPIC detector & Clean $t_{exp}$\\
 & & (ks) & \\
\midrule
 & MOS1 & 77.3\\
0761070101 & MOS2 & 79.4  \\
 & pn & 63.0  \\
 \\
 & MOS1 & 75.6 \\
0761070201 & MOS2 & 87.3 \\
 & pn & 62.9  \\
\bottomrule
\end{tabular}
\end{table}

Finally, spectra, effective areas and response files (ARF and RMF) for selected regions were extracted using the ESAS tasks \textit{mos-spectra} and \textit{pn-spectra}. The same tasks were used for extracting FOV images in the 0.5-2.5 keV band.

\subsection{NuSTAR}
A523 was observed by \textit{NuSTAR} on 2016 April 02-04 (ObsID 7012001002) and 06-08 (ObsID 7012001004), for a total of unfiltered exposure time of 96 ks for ObsID 7012001002 and 103 ks for ObsID 7012001004. We filtered the events from both observations and both modules A and B with a standard pipeline processing (HEAsoft v6.22.1 and NuSTARDAS v1.8.0) and the 20171002 version of the $NuSTAR$ Calibration Database. In the filtering we adopted strict criteria regarding passages through the South Atlantic Anomaly (SAA) and a “tentacle”-like region of higher activity near part of the SAA; in the call to the general processing routine that creates Level 2 data products, \texttt{nupipeline}, the following flags are included: \texttt{SAAMODE=STRICT} and \texttt{TENTACLE=yes}. The resulting clean exposure times for both observations are listed in Table \ref{tab:observations_nu}.

\begin{table}[th]
\centering
\caption{\textit{NuSTAR} observations of A523.}
\label{tab:observations_nu}
\medskip
\begin{tabular}{l c c c c} 
\toprule
ObsID & \multicolumn{2}{c}{Optical Axis Location} & \multicolumn{2}{c}{Exposure Time} \\
 & $\alpha$ (J2000) & $\beta$ (J2000) & Raw & Clean \\
 & (deg) & (deg) & (ks) & (ks) \\
\midrule
7012001002 & 74.761 & 8.7892 & 96 & 95 \\
7012001004 & 74.762 & 8.7947 & 103 & 91 \\
\bottomrule
\end{tabular}
\end{table}

From the cleaned event files, we directly extracted images using \texttt{xselect}, created exposure maps using \texttt{nuexpomap}, and extracted spectra and associated response matrix (RMF) and auxiliary response (ARF) files using \texttt{nuproducts}. The call to \texttt{nuproducts} includes
\texttt{extended=yes}, most appropriate for extended sources, which weights the RMF and ARF based on the distribution of events within the extraction region, assuming that to be equivalent to the true extent of the source. The effective smoothing of the source due to the point-spread function (PSF) is not folded in with the weighting; however, the relatively narrow FWHM of $\approx 18''$ lessens the impact of this omission. The response across a given detector is uniform, so the RMFs of the four detectors are simply averaged by the weighted fraction each detector contributes to a region. In addition to the mirror response, the RMF includes low energy absorption in the detectors (due to a CdZnTe dead layer and platinum electrodes). The procedures adopted for estimating and extracting the background spectra, necessary for the spectral analysis, are described in Section \ref{sec:bgd_modeling}.

\section{Background modeling}
\label{sec:bgd_modeling}
\subsection{XMM-Newton}
\label{sec:bgd_xmm}
\begin{figure} [th]
\centering
\includegraphics[angle=270,width=\linewidth]{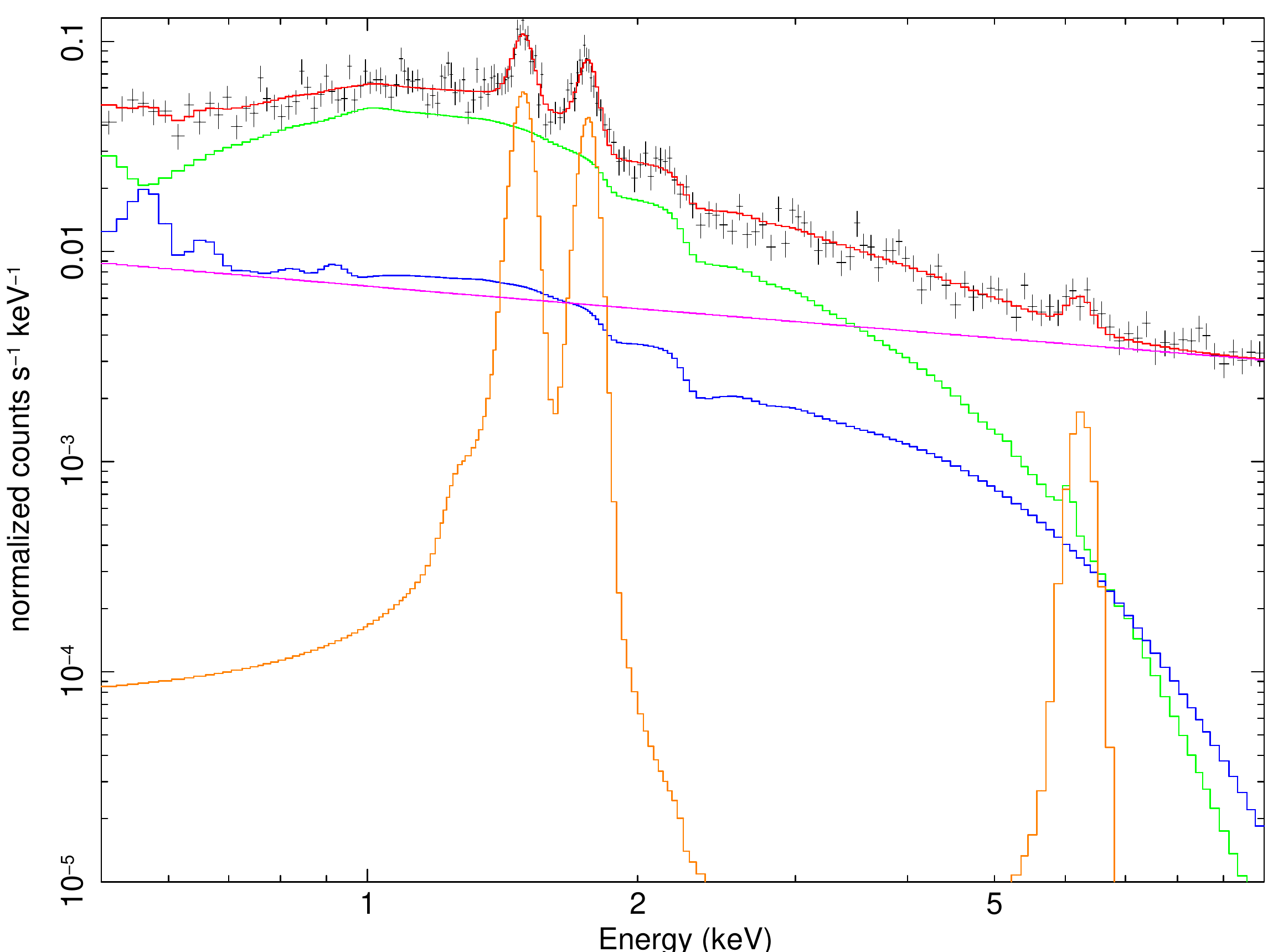}
\caption{Spectrum extracted from a cluster annular region from an \textit{XMM-Newton} observation (MOS1), with all the background components shown separately, derived as described in Section \ref{sec:bgd_xmm}. Red:total resulting model; green:cluster thermal model; blue:X-ray sky background; orange:florescence instrumental lines; magenta: instrumental non X-ray background.}
\label{fig:bgd_comp}
\end{figure}

The \textit{XMM-Newton} background is both of cosmic and instrumental origin. The cosmic background components are well known, but have to be explicitly modeled, as they represent a significant portion of the spectra on a broad range of energies and they can sensibly vary both in intensity and spectral shape in the sky. The model used here consist of three spectral components, as shown in Figure \ref{fig:bgd_comp}: a thermal unabsorbed component at $E\sim 0.1$ keV representing the \textit{Local Hot Bubble} (LHB) emission; a thermal absorbed component at $E\sim 0.25 $ keV accounting for the galactic halo emission; an absorbed power-law (with a photon index fixed at $\alpha=1.46$), representing the Cosmic X-ray Background (CXB). The normalizations for these components were fixed in the cluster spectral analysis to the best-fit parameters found, for each instrument, from an annular region centered in the X-ray surface brightness peak, with inner and outer radii being 12' and 14' respectively, which was considered to be a region with very low cluster emission (it corresponds to 0.9-1.1 $r_{200}$ given the mass estimate based on the mass-temperature relation, see section \ref{subsec:discussion_mass}). Including the uncertainty on these background parameters in the cluster spectral analysis results in a systematic uncertainty on thermal parameters which is on average 5 - 7\% lower than the statistical error in the central cluster region and in all the regions considered for our thermodynamic maps (see section \ref{sec:xmm_therm_maps}), thus we did not considered this uncertainty further. For the absorbed components, we used photoelectric absorption (\texttt{phabs} in XSPEC), fixing the galactic hydrogen column density to $1.06 \times 10^{21} \mathrm{cm}^{-2}$, as tabulated in LAB HI Galactic survey in \citet{Kalberla05}. We also tried to fix this parameter to the value of \cite{Willingale13}, which is $1.57 \times 10^{21} \mathrm{cm}^{-2}$; however, if we use this value the fit to our spectrum gets worse and when we let the parameter free to fit we find $0.97 \pm 0.01 \times 10^{21} \mathrm{cm}^{-2}$, which is closer to the value of \citet{Kalberla05}.

The instrumental background model is then extracted from filter-wheel-closed observations, accounting for its spatial variations; for details on the model components and the procedure used see Appendix \ref{sec:appendixA}.

\begin{figure} [th]
\centering
\includegraphics[angle=270,width=\hsize]{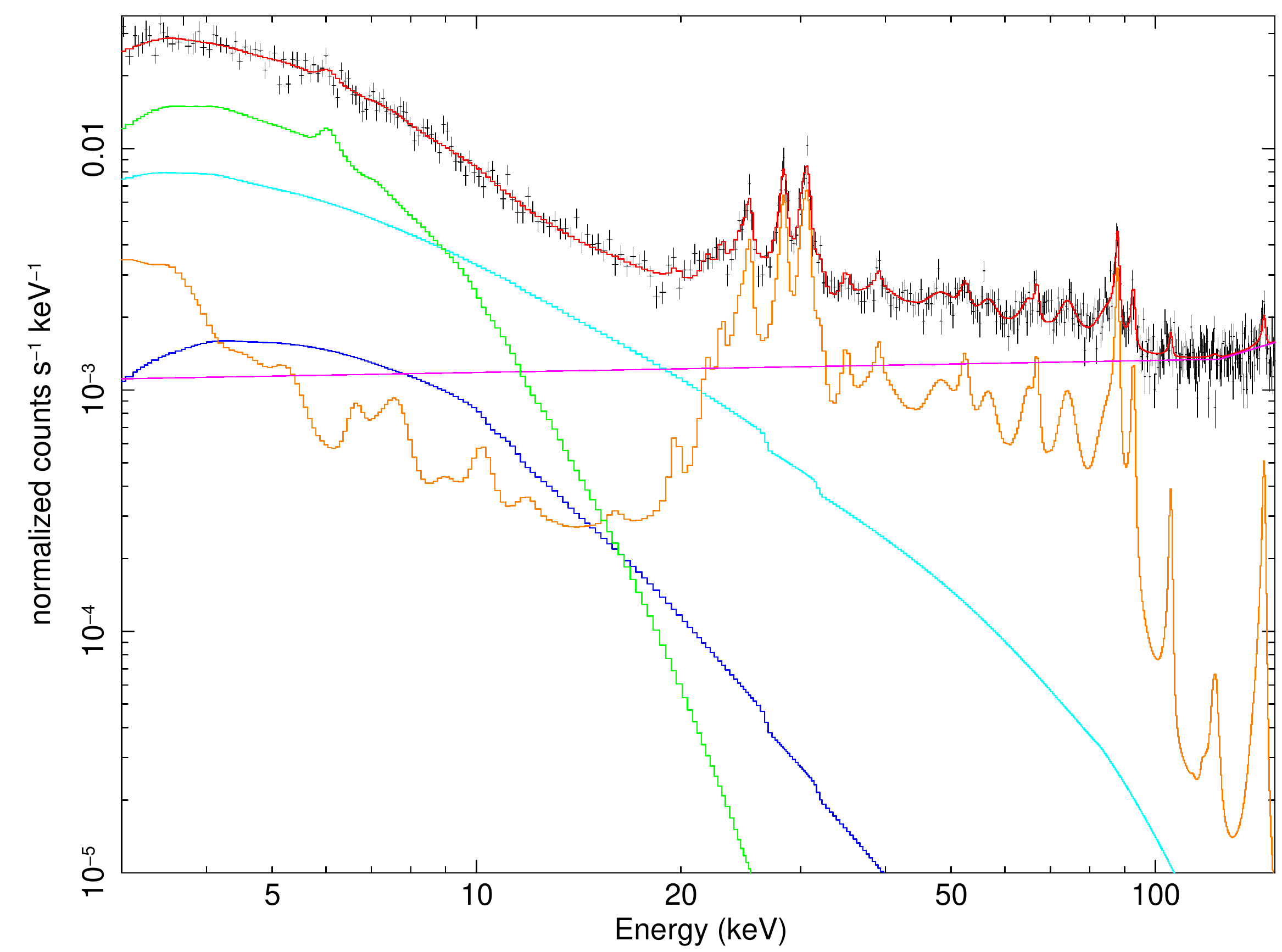}
\caption{Spectrum extracted from a circular region of radius 5' centered on the cluster X-ray brightness peak from a \textit{NuSTAR} observation (detector A), with all background components modeled shown separately, derived as described in Section \ref{sec:bgd_xmm}. Red:total resulting model; green:cluster thermal model; cyan: aperture background; blue: fCXB background; magenta: instrumental particle continuum; orange: instrumental lines and solar reflected component.}
\label{fig:bgd_components}
\end{figure}

\subsection{NuSTAR}
\label{sec:bgd_nu}
The \textit{NuSTAR} background can be described by a number of spectral components, of both cosmic and instrumental origin, all of which can vary to some extent both spectrally, spatially and somewhat temporally. However, the spatial variations of these components are quite well understood, allowing us to extrapolate their parameters to the source regions of interest. The background model, described in detail in \citet{Wik14}, consists of four main components, as shown in Figure \ref{fig:bgd_components}: (1) unfocused cosmic X-ray background from the sky, leaking past the aperture stops (Aperture) dominating the background spectrum below $\approx 20$ keV; (2) focused and ghost-ray cosmic X-ray background (fCXB), also contributing at low energies; (3) several instrumental activation and emission lines, mainly contributing at energies above $\approx 20$ keV; (4) instrument continuum emission, also contributing at higher energies, produced primarily (but probably not entirely) by high energy gamma rays, which either pass through the anti-coincidence shield and Compton scatter in the detector or scatter untriggered in the shield itself.

The standard background treatment suggested in \citet{Wik14} (i.e. \texttt{nuskybgd}), would require the presence of regions free of cluster emission, in order to characterize the background and produce scaled background spectra for the desired region and/or images for the required energy bands. However, A523 covers almost entirely \textit{NuSTAR}'s field of view (see Figure \ref{fig:nuA523}, top panel), with only small regions in the corners presumably source-free. For this reason, we adopted empirical nominal models extracted from blank fields observations for the Aperture and the fCXB components (as done in \citealp{Gasta15}). Details about the background modeling can be found in Appendix \ref{sec:appendixA}, together with a comparison between the model obtained with the procedure adopted by \texttt{nuskybgd} (using the small source-free regions left in the FOV) and the one produced by our procedure. 

\section{Image analysis}
\label{sec:image_anal}
\subsection{XMM-Newton}
\begin{figure} [t]
\centering
\includegraphics[width=\linewidth]{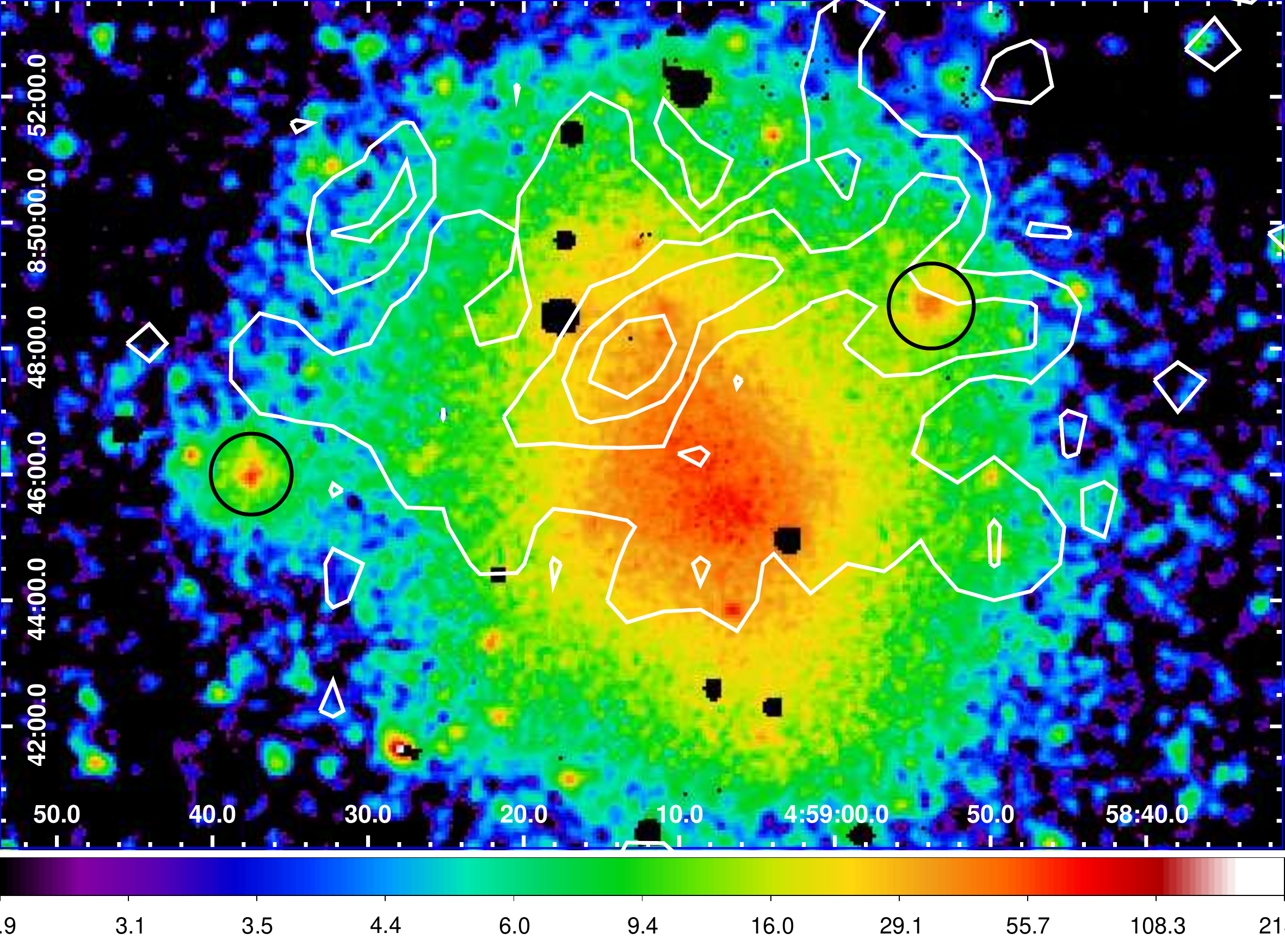}
\caption{\textit{XMM-Newton} background subtracted and exposure corrected combined image for the three EPIC instruments and both observations in the soft 0.5-2.5 keV band, with point sources removed. White contours refer to the low-resolution 1.4 GHz VLA image (with discrete sources subtracted) and highlight the radio halo. The two black circles indicate the positions of the background structures discussed in the text.}
\label{fig:A523}
\end{figure}

In Figure \ref{fig:A523} we show the exposure corrected, background subtracted and point sources removed image for Abell 523, combining all the EPIC (\textit{European Photon Imaging Camera}) instruments and both observations in the soft band 0.5-2.5 keV. The image was created with ESAS task \texttt{adapt-merge}, which adaptively smooths mosaicked images. The estimated surface brightness peak lies at $\mathrm{R.A.} = 04^{h}59^{m}08^{s}.0, \, \mathrm{Dec} = +08^{\circ}46'00'' \,(\mathrm{J}2000.0)$, which is consistent with what was found in G16. The cluster is quite disturbed (as confirmed quantitatively by the morphological indicators reported in G16), with an elliptical morphology; two regions of extended emission are also evident, coincident with the background optical substructures described in G16 (labeled \textit{BACKstruct}, see Section \ref{sec:backstruct}), that were not characterized before due to the low signal-to-noise of their shallower \textit{Chandra} observation.

\begin{figure}[t]
\centering
\includegraphics[width=0.72\linewidth]{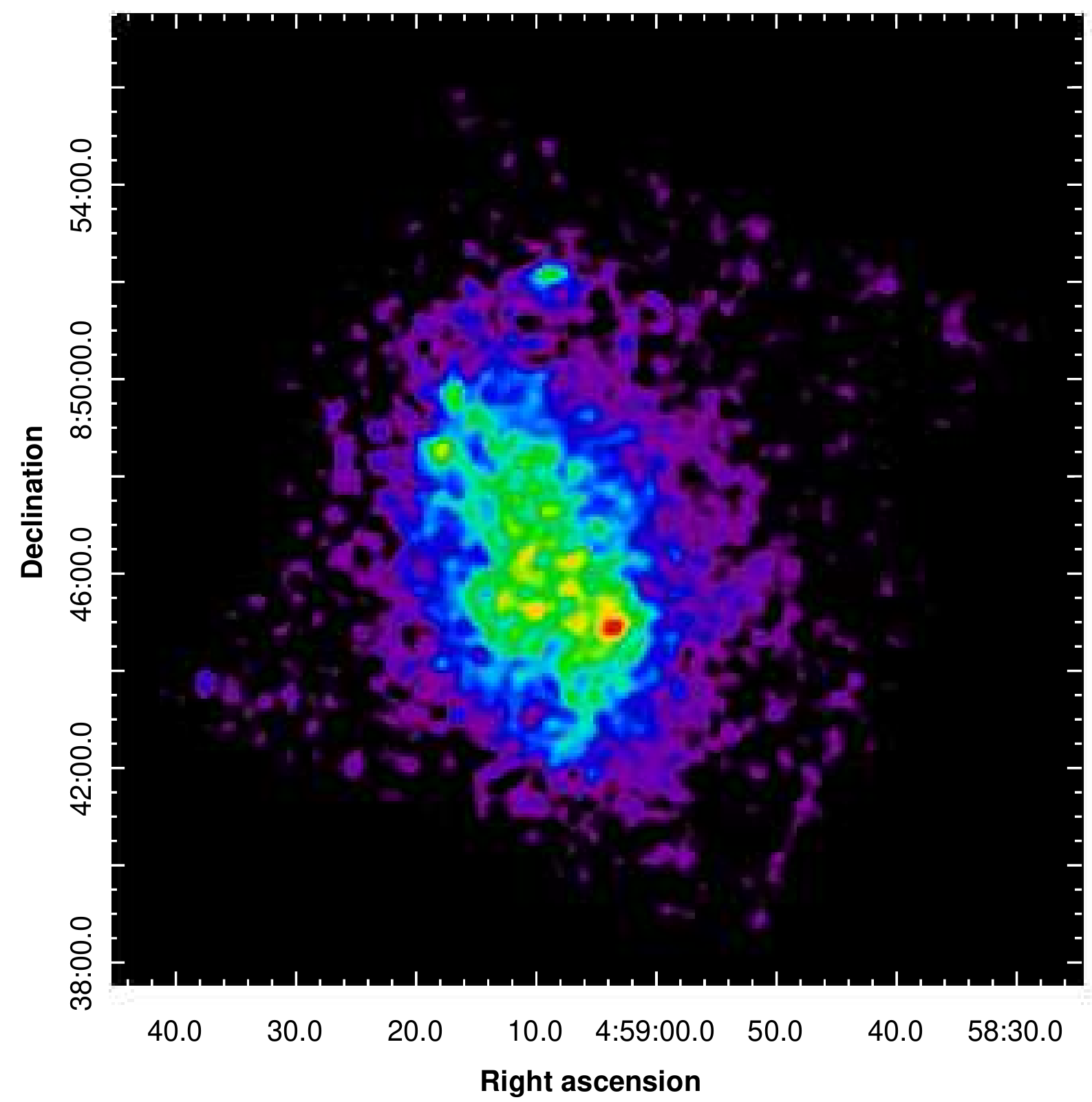}
\includegraphics[width=0.72\linewidth]{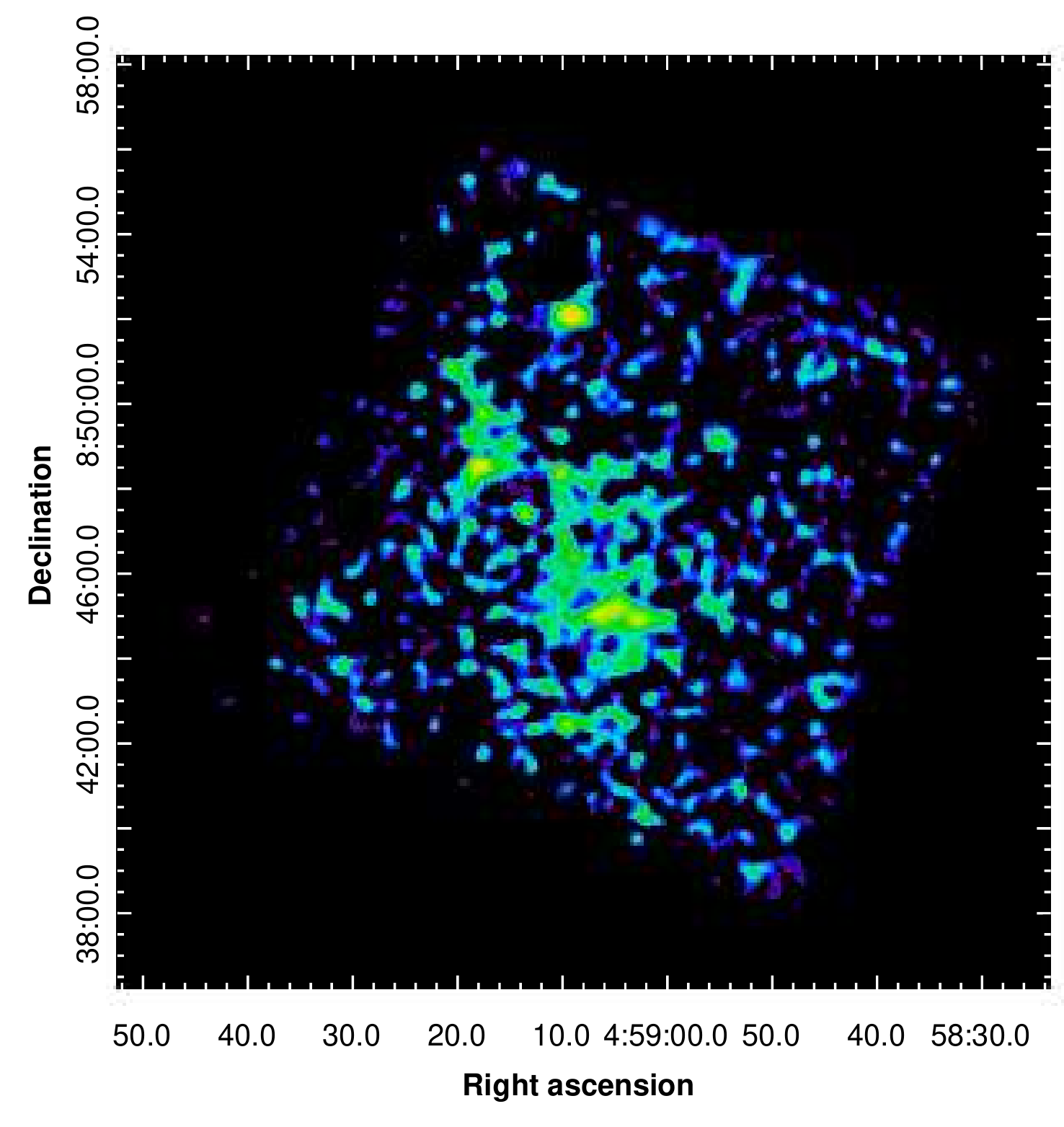}
\includegraphics[width=0.72\linewidth]{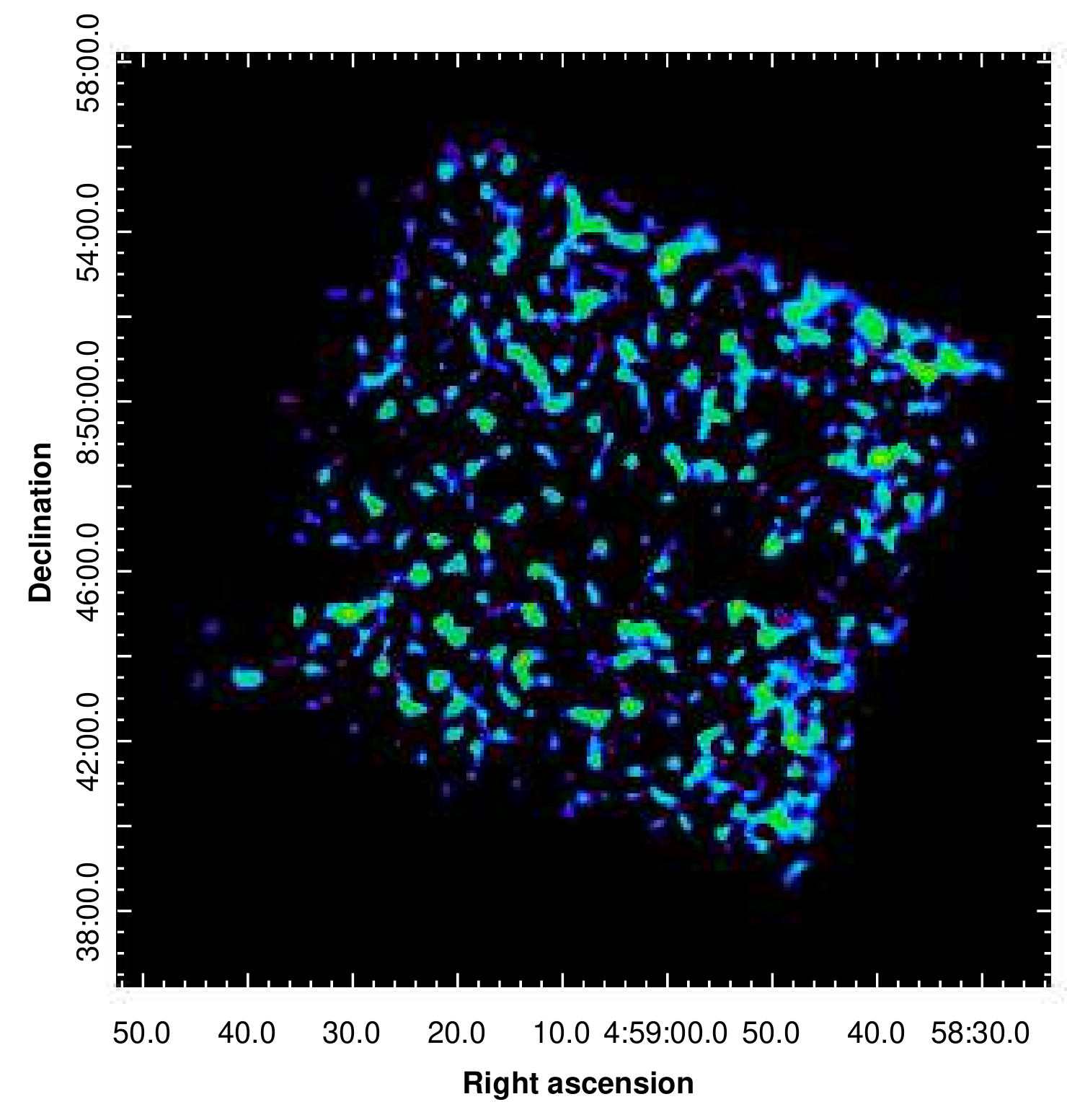}
\caption{\textit{NuSTAR} background-subtracted and exposure-corrected images combined from both observations and telescopes. Images are presented on a logarithmic scale and have been smoothed with a Gaussian kernel of 17''.2 (7 pixels). The energy band for each image is, from top to bottom: 3–10 keV, 10–20 keV, and 20–50 keV.}
\label{fig:nuA523}
\end{figure}

\subsection{NuSTAR}
We extracted images with \texttt{nuproducts}. The exposure maps were obtained using the task \texttt{nuexpomap}, and were created at single energies for each band, roughly corresponding to the mean energy of the
band. Background images were produced using \texttt{nuskybgd} \citep{Wik14}, as described in Section 2.

Background-subtracted and exposure-corrected images in three energy bands (top: 3-10 keV; middle: 10-20 keV; bottom: 20-50 keV) combined from both observations and telescopes are presented in Figure \ref{fig:nuA523}; the images have been Gaussian smoothed by 17''.2 (7 pixels). The 3-10 keV image show the morphology of the hot gas in A523 and resembles the \textit{Chandra} and \textit{XMM-Newton} images, blurred by the larger \textit{NuSTAR} PSF. In the 10-20 keV image some cluster emission is still faintly visible, while the 20-50 keV image is completely noise dominated and no cluster emission is visible. 

\section{Thermal emission}
\label{sec:thermal_em}
We use the deep \textit{XMM-Newton} observation to study in detail A523's thermal emission, as it dominates at energies $\lesssim 10$ keV.
\subsection{Global analysis}
We extracted a spectrum in an annular region corresponding to $0.05 \, R_{180} < r < 0.2 \, R_{180}$, with $R_{180} = 1780 \,(kT/5\mathrm{keV})^{1/2} \, h(z)^{-1}\,\mathrm{kpc}$ and $h(z)=(\Omega_{\mathrm{M}}(1+z)^{3}+\Omega_{\mathrm{\Lambda}})^{1/2}$ (based upon the scaling relation of \citealp{Arnaud05}), using an iterative procedure (OUT region; \citealp{Leccardi10}). The spectral analysis resulted in an average temperature estimation of $4.29 \pm 0.04$ keV. Based on the classification scheme of \cite{Leccardi10}, the system can be classified as a High Entropy Core (HEC) system, with a calculated pseudo-entropy ratio of $\sigma = 0.66 \pm 0.03$, which is consistent with the hypothesis of a merger. This quantity represents the ratio between the pseudo-entropy calculated in the OUT region to the pseudo-entropy calculated in the IN region (corresponding to $r<0.05 \, R_{180}$). We extracted the surface brightness profile from the mosaicked and exposure corrected image in the 0.5-2.5 keV band. The best-fitting $\beta$-model has a core radius of $r_{c} = (219 \pm 12) \, h_{70}^{-1}$ kpc (i.e., $114'' \pm 6''$) and $\beta = 0.47 \pm 0.02$. Assuming that the cluster emission profile follows this model, the estimated luminosity in the 0.1-2.4 keV rest frame band within $R_{500}$ is $L_{X,500} = 1.39 \pm 0.04 \times 10^{44} \, h_{70}^{-2} \, \mathrm{erg} \, \mathrm{s}^{-1}$, which is consistent with the previous Chandra estimate of G16 .

\subsection{Radial analysis}
\begin{figure} [t]
\centering
\includegraphics[width=0.49\textwidth]{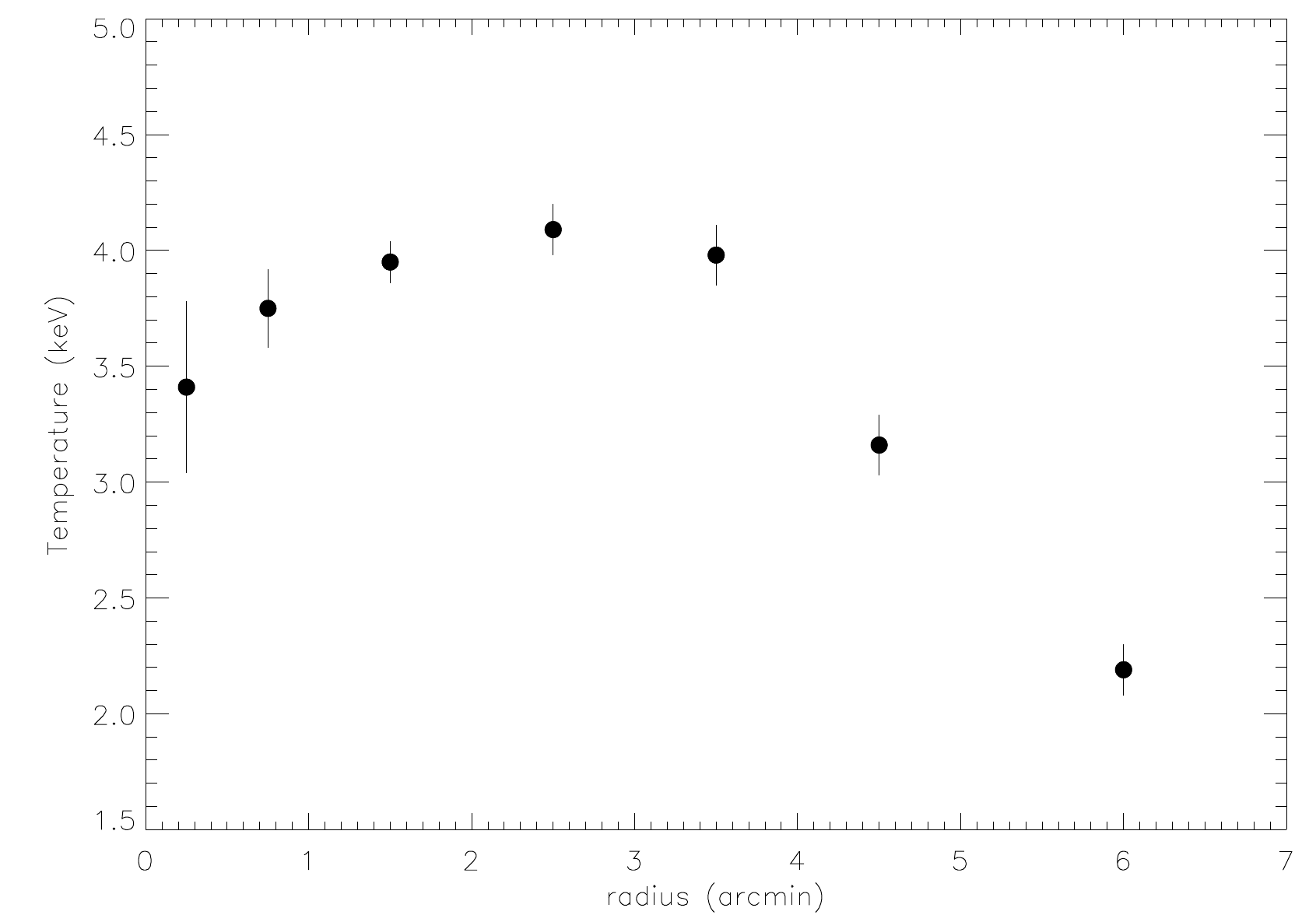}
\includegraphics[width=0.49\textwidth]{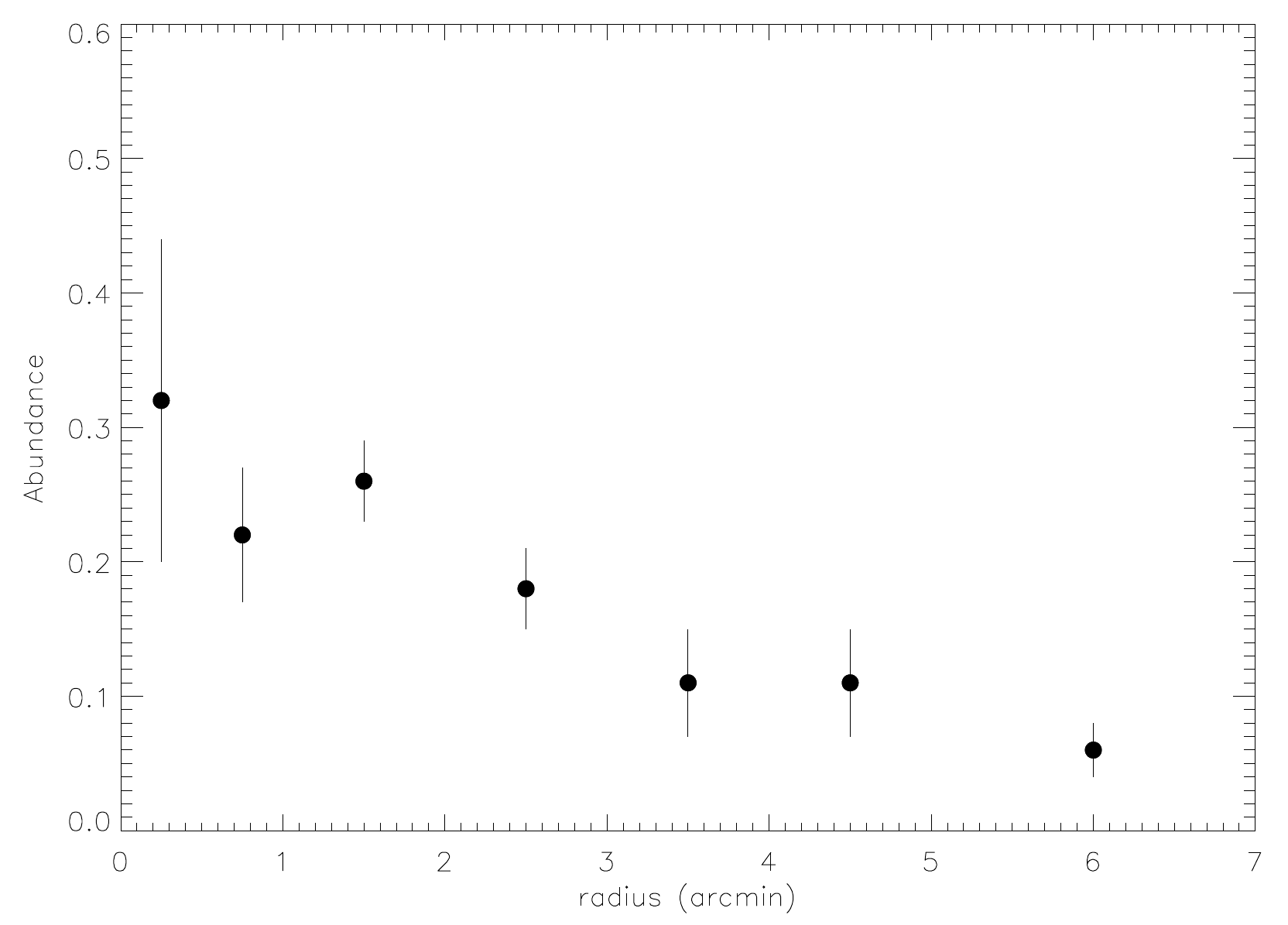}
\caption{Mean temperature (top panel) and abundance (bottom panel) radial profiles. The best-fit parameters are plot in function of the mean radius for each annulus, expressed in arcmin.}
\label{fig:radial_profiles_a523}
\end{figure}

We performed a radial spectral analysis of A523 extracting spectra from ten concentric annuli, centered on the peak of the cluster X-ray brightness. 
The radii considered are the same used by \cite{Snowden08}. In Figure \ref{fig:radial_profiles_a523} we plot the resulting radial profiles for the best-fit temperature and abundance (using tables from \citealp{Anders89}) parameters for the three detectors combined. Data are shown for the first seven annuli, for which a significant cluster emission is detectable. The remaining annuli were used to constraint the cosmic background, as described in Section \ref{sec:bgd_modeling}.

The radial profiles essentially confirm the merger scenario: (1) despite not being completely flat, the temperature profile shows only a mild central gradient, which is still consistent with a disturbed system, compared to the more steep gradients observed in cool-core clusters \citep{LeccMol08}; (2) the abundance profile shows a modest central gradient, consistent with what found in \cite{Leccardi10} for Non Cool Core (NCC) systems. The abundance values in the three outer annuli are lower than the typical cluster values of 0.2-0.3 solar found in cluster outer regions. This could just be an indication that the abundance parameter in this observation is prone to systematic errors, as the cluster emission is comparable to the background level. This is indeed confirmed when we fix the abundance parameter to a nominal value of 0.2 $Z_{\odot}$, as in this case we find that the temperature and normalization values show an average variation of $\approx$ 5\%. Moreover, the fit statistic does not show any appreciable improvement when we leave the parameter free to fit.

\subsection{Thermodynamic maps}
\label{sec:xmm_therm_maps}
\begin{figure} [t]
\centering
\includegraphics[width=\linewidth]{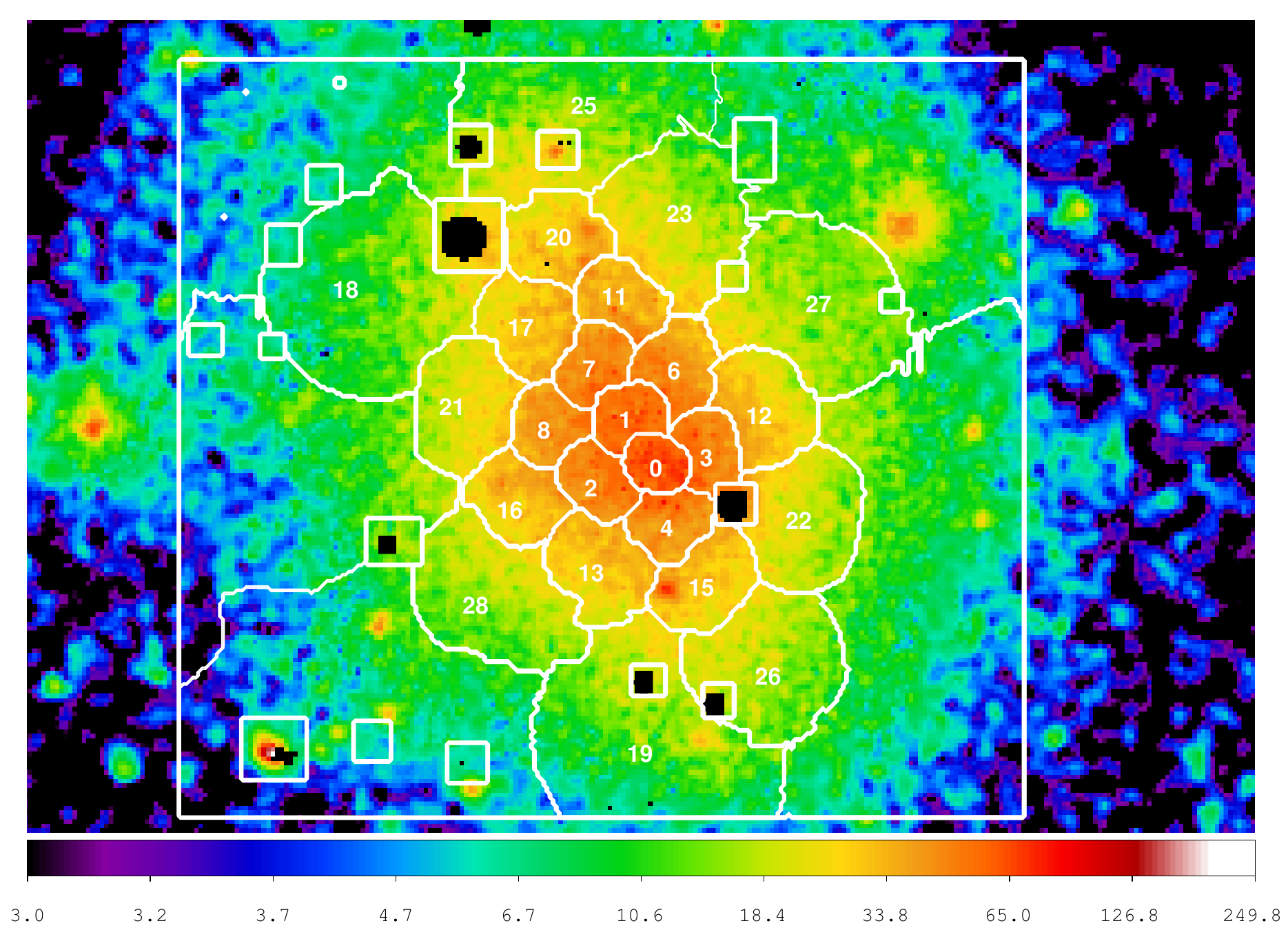}
\caption{Regions individuated from \texttt{contbin} overlaid to the \textit{XMM-Newton} surface brightness image, numbered in order of creation. For each region we performed a spectral analysis to build the thermodynamic maps presented in Section 5.}
\label{fig:map_over_xmm}
\end{figure}

\begin{figure*} [th]
 \subfloat[Temperature map]{
\includegraphics[width=0.5\linewidth]{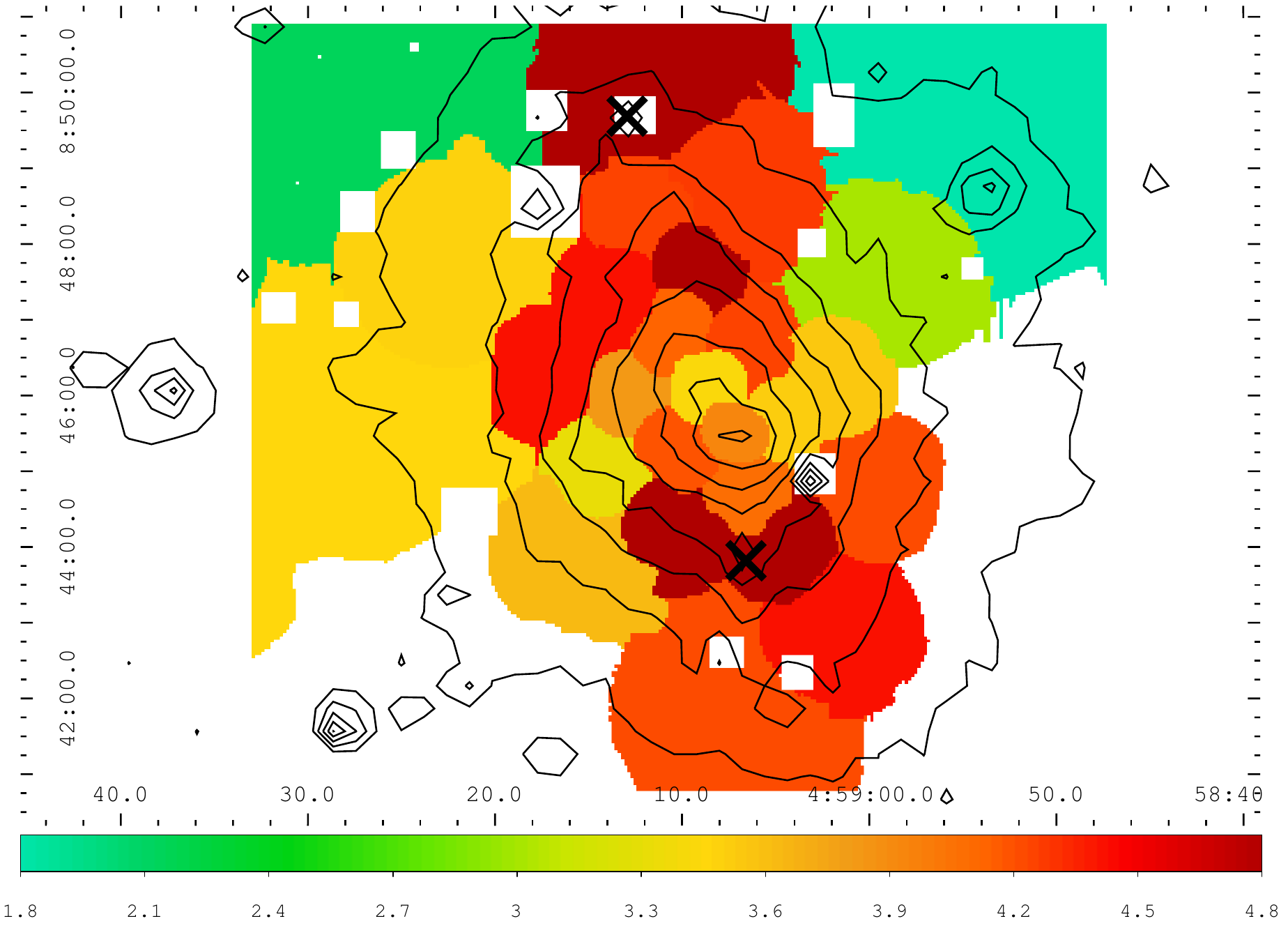}
}
\subfloat[Abundance map]{
\includegraphics[width=0.5\linewidth]{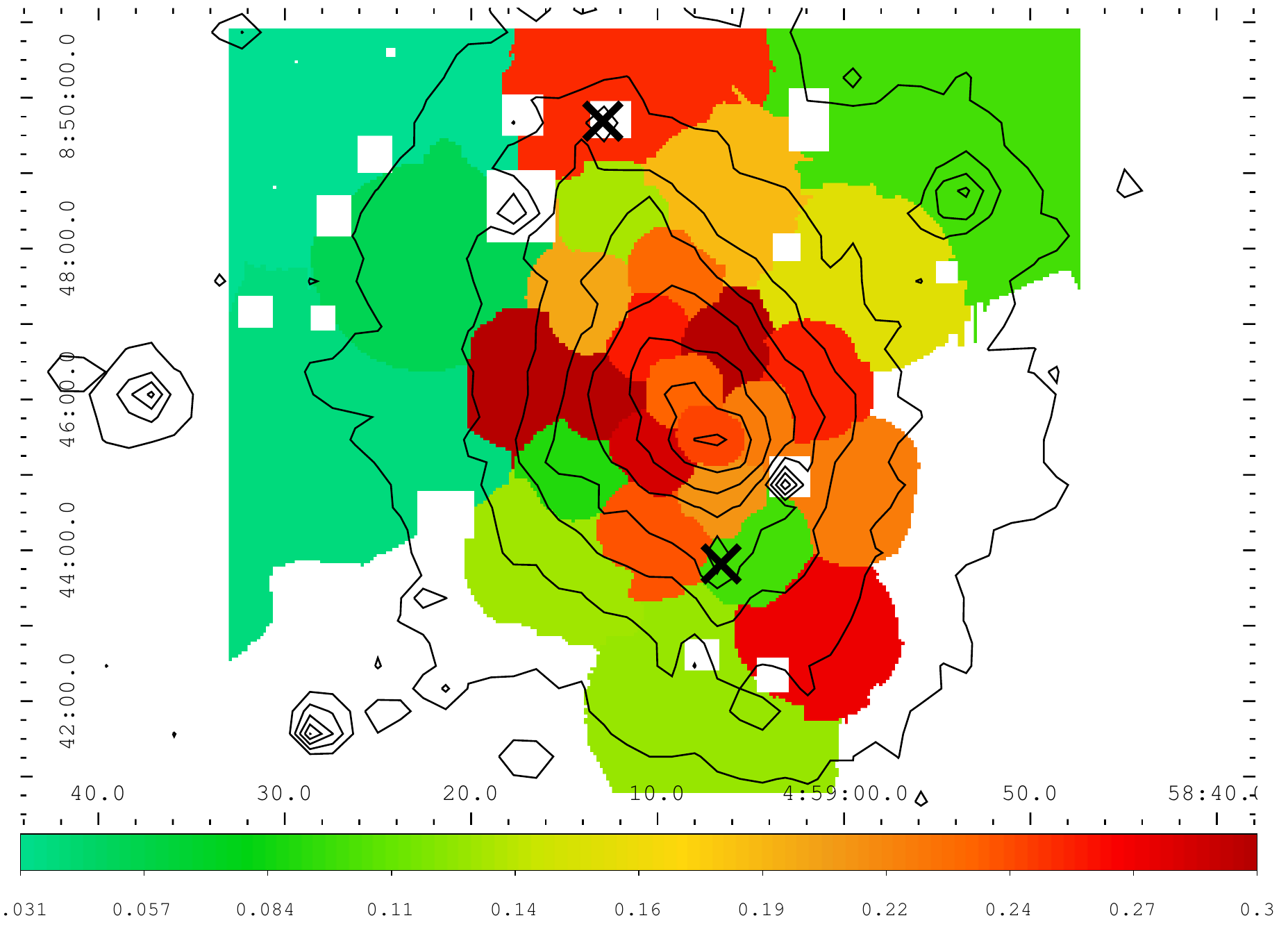}
}\par
\subfloat[Pressure map]{
\includegraphics[width=0.5\linewidth]{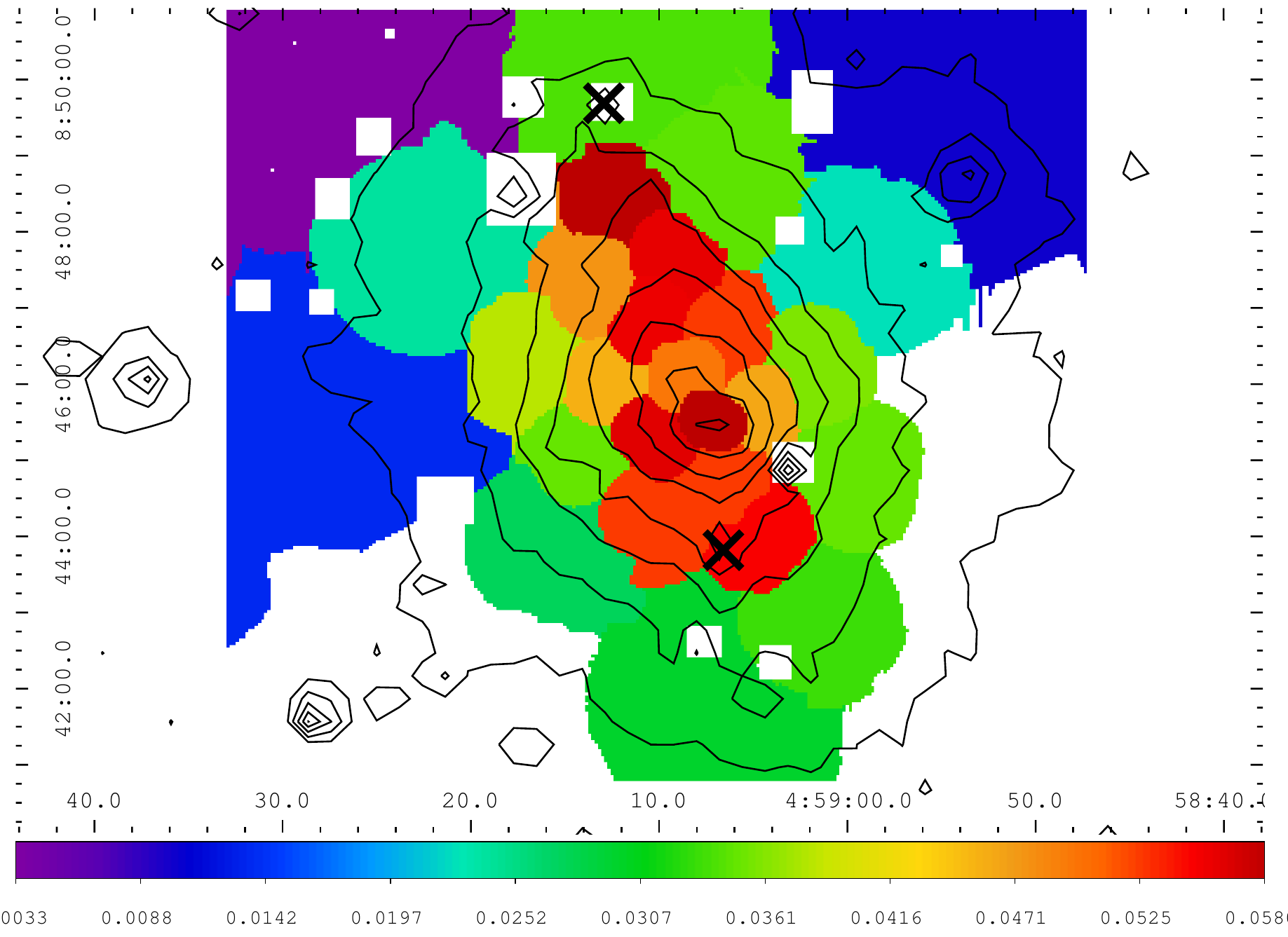}
}
\subfloat[Entropy map]{
\includegraphics[width=0.5\linewidth]{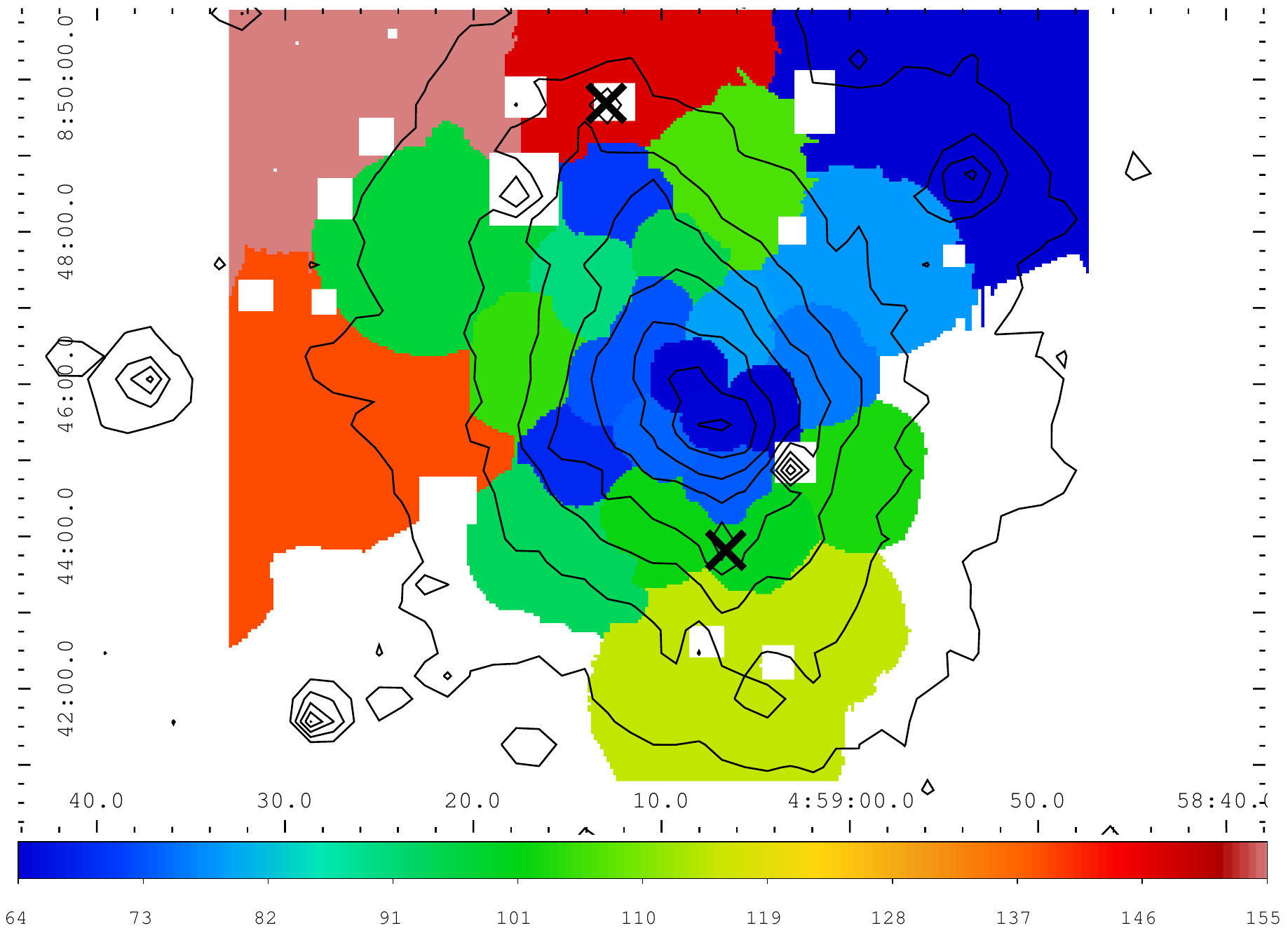}
}
\caption{Thermodynamic maps of A523. (a) temperature map (keV); (b) metal abundance map ($Z_{\odot}$); (c) pseudo-pressure map (arbitrary units); (d) pseudo-entropy map (arbitrary units). Overlaid to each image are the surface brightness contours obtained by the \textit{XMM-Newton} image and the positions of the two brightest cluster galaxies are indicated by two black crosses. }
\label{fig:thermo_maps_xmm}
\end{figure*}

A proper description of a disturbed system like Abell 523 requires a detailed bi-dimensional analysis. We used \texttt{CONTBIN v1.4} \citep{Sanders06} to produce projected maps of temperature, abundance, pressure and entropy, requiring a threshold for signal-to-noise ratio of 100 and masking residual point sources. We also appropriately set a geometric parameter to produce circular regions. This procedure bins the X-ray image using contours from an adaptively smoothed map such that the generated bins closely follow the surface brightness. The procedure selected 24 regions or bins, well covering the X-ray observed surface brightness (see Figure \ref{fig:map_over_xmm}). For each bin we extracted and fitted spectra as described in Section 2, combing all EPIC instruments and both observations. The resulting maps are shown in Figure \ref{fig:thermo_maps_xmm}. 

While temperatures and abundances are obtained as a direct result of the spectral fitting, entropy \textit{s} and pressure \textit{P} were calculated following \cite{Rossetti07}. These quantities are often called pseudo-entropy and pseudo-pressure, the prefix pseudo- referring to the fact the they are projected along the line of sight. They are computed as follows:
\begin{equation}
	P = T \times \mathrm{EM}^{1/2} \quad (\mathrm{keV} \, \mathrm{cm}^{-5/2} \, \mathrm{arcmin}^{-1})
\end{equation}
\begin{equation}
	s = T \times \mathrm{EM}^{-1/3} \quad (\mathrm{keV} \, \mathrm{cm}^{5/3} \, \mathrm{arcmin}^{-2/3})
\end{equation}
where EM is the projected emission measure, defined as:
\begin{equation}
	\mathrm{EM} = \mathcal{N}/A \quad (\mathrm{cm}^{-5} \, \mathrm{arcmin}^{-2})
\end{equation} 
with $A$ the area of each bin expressed in $\mathrm{arcmin}^{2}$ and proportional to the square of the electron density integrated along the line of sight. $\mathcal{N}$ is the normalization of the thermal model, i.e. 
\begin{equation}
\mathcal{N} = \frac{10^{-14}}{4\pi \left[D_{A}(1+z)\right]^{2}} \int n_{e}n_{H} \, dV \quad (\mathrm{cm}^{-5})
\end{equation}
where $D_{A}$ is the angular size distance of the source, $n_{e}$ and $n_{H}$ are the electron and hydrogen density, respectively.

The temperature map has a non-symmetric morphology, with regions at higher temperature localized near the subcluster cores seen in optical band (G16). These hotter regions are probably a consequence of the merging process, as they are mainly located along the NE-SW direction. The range in percent errors for the various temperature regions is 3-7\%.

The abundance map exhibits higher metallicity regions displaced from the subcluster cores. The average abundance value over the whole map is consistent with the typical value found in the outer regions of galaxy clusters $Z = 0.23 \pm 0.01$ \citep{Leccardi10}. Higher abundance regions can however be seen in the proximity of the subcluster cores, probably tracing the past activities of their BCGs. The range in percent errors for the innermost abundance regions is 20-30\%, while the outermost regions have error up to 40-60\%.

The pseudo-pressure map quite well highlights the merging axis connecting the two substructures. This quantity is a good tracer of mass in the system, so the relatively higher pressure regions found near the two BCGS are consistent with this picture. The highest pressure region is found in the cluster center (the peak of the X-ray surface brightness) and is closer to the south subcluster. No high pressure regions are evident in the region where the radio halo is detected. The range in percent errors for the various pseudo-pressure regions is 3-6\%.

The pseudo-entropy map has a quite uniform morphology, with a central and extended lower entropy region mainly tracing the SSW-NNE direction, coincident with the merging axis. Some higher entropy regions are also visible in the outer parts of the cluster, almost in agreement with the temperature map. In particular, in the upper NNE regions, a high temperature-high entropy bin is visible close to the BCG1, while in the SSW regions the entropy is more modestly increased. This behavior looks consistent with the entropy being a tracer of the dynamical and physical processes that happens in the ICM, as heating/cooling processes or merger shocks. The inner low-entropy regions, in particular, could be tracing some stripped gas from the subcluster cores (as in \citealp{RossMol10}) or the initial phase of a subsequent more relaxed state. However, as already noticed, in these regions the metallic abundances are still too low to claim for an actual cool core remnant. Aside from the main SSW-NNE elongation of the lower entropy regions, a second elongation in the NW direction is somewhat visible, coincident with the axis of the radio halo of A523 (see G16). The range in percent errors for the various pseudo-entropy regions is 3-7\%.


\subsection{Analysis of background structures}
\label{sec:backstruct}
From both optical and X-ray observations, two clear structures are evident in the WNW and ESE regions. These groups have been classified as background structures in G16 (and dubbed \textit{BACKstruct}), as their estimated mean redshift is $z\approx 0.14$. The nature of these structures is still unclear, however Girardi et al. proposed that they can be connected to the outer regions of the cluster Abell 525, as this system is part - with A523 - of the super-cluster SCL62 and has a mean photometric redshift of $z_{phot} \approx 0.14$.

To test this hypothesis we analyzed spectra extracted from two circular regions of radius $\sim50$ kpc centered in the two background groups, combining both observations. The spectra were analyzed separately with a thermal model (in case of thermal emission produced from the hot intracluster gas) and with a power-law model (testing the hypothesis that these were actually point sources, e.g. AGN). For the WNW-group, the analysis was performed on the MOS instruments only, as for the pn this structure is located in a gap between two CCDs. The resulting C-statistic over degrees of freedom ratio for both groups shows that both structures are much better fitted by thermal models, confirming the thermal nature of their emission. In particular, for the WNW-group spectrum we find a ratio of 22/18 for the thermal model, and a ratio of 53/19 when using a single power-law model. Similarly, for the ESE-group we find a ratio of 22/15 for the thermal model and a ratio of 40/16 for the power-law model. From the joint spectral analysis we can also extract some main parameters, such as temperature and unabsorbed flux in the 0.1-2.4 keV band. The best-fit values found for these parameters and for both groups are listed in Table \ref{tab:backstruct_par}. 

\begin{table}[th]
\caption{Best-fit parameters for the two background groups discussed in the text.}
\label{tab:backstruct_par}
\medskip
\centering
\renewcommand\arraystretch{1.7}
\begin{tabular}{l c c c}
\toprule
Group & kT & Abund. & Flux [0.1-2.4 keV]\\
	& (keV) & ($Z_{\odot}$)	&	($10^{-14} \,  \mathrm{erg} \, \mathrm{cm}^{-2} \, \mathrm{s}^{-1}$)\\
\midrule
WNW-group &  $1.79^{+0.11}_{-0.16}$ & $0.26^{+0.12}_{-0.06}$  & $5.50 \pm 0.02$\\
ESE-group &  $1.45^{+0.08}_{-0.14}$& $0.12^{+0.05}_{-0.04}$ & $5.31 \pm 0.02$\\
\bottomrule
\end{tabular}
\end{table}


\subsection{Radio and X-ray comparison}
\label{sec:radio_x_comp}
Radio and X-ray thermal emissions in galaxy clusters are often deeply correlated, both morphologically and in terms of their surface brightnesses as seen in \citet{Govoni01a}. Merger-driven turbulent re-acceleration models (e.g., \citealp{Brunetti07}) that nowadays provide the accepted scenario for the origin of radio halos predict a connection between thermal and non-thermal emission simply because a fraction of the kinetic energy of the ICM is dissipated into non-thermal components. Thermal--nonthermal correlations are thus quite useful because they have the potential to constrain ICM microphysics and the degree of coupling between thermal and non-thermal quantities.
The radio-Xray surface brightness correlation is quite useful, as it can be directly extracted from observable quantities, can be predicted from theoretical models and does not depend on stringent hypotheses. To do that, we fit radio and X-ray surface brightness data with a power-law in the form $F_{\mathrm{Radio}} = a (F_{X})^{b}$. The best-fit parameters are then calculated with a least squares method.

\begin{figure} [t]
\centering
\includegraphics[width=\linewidth]{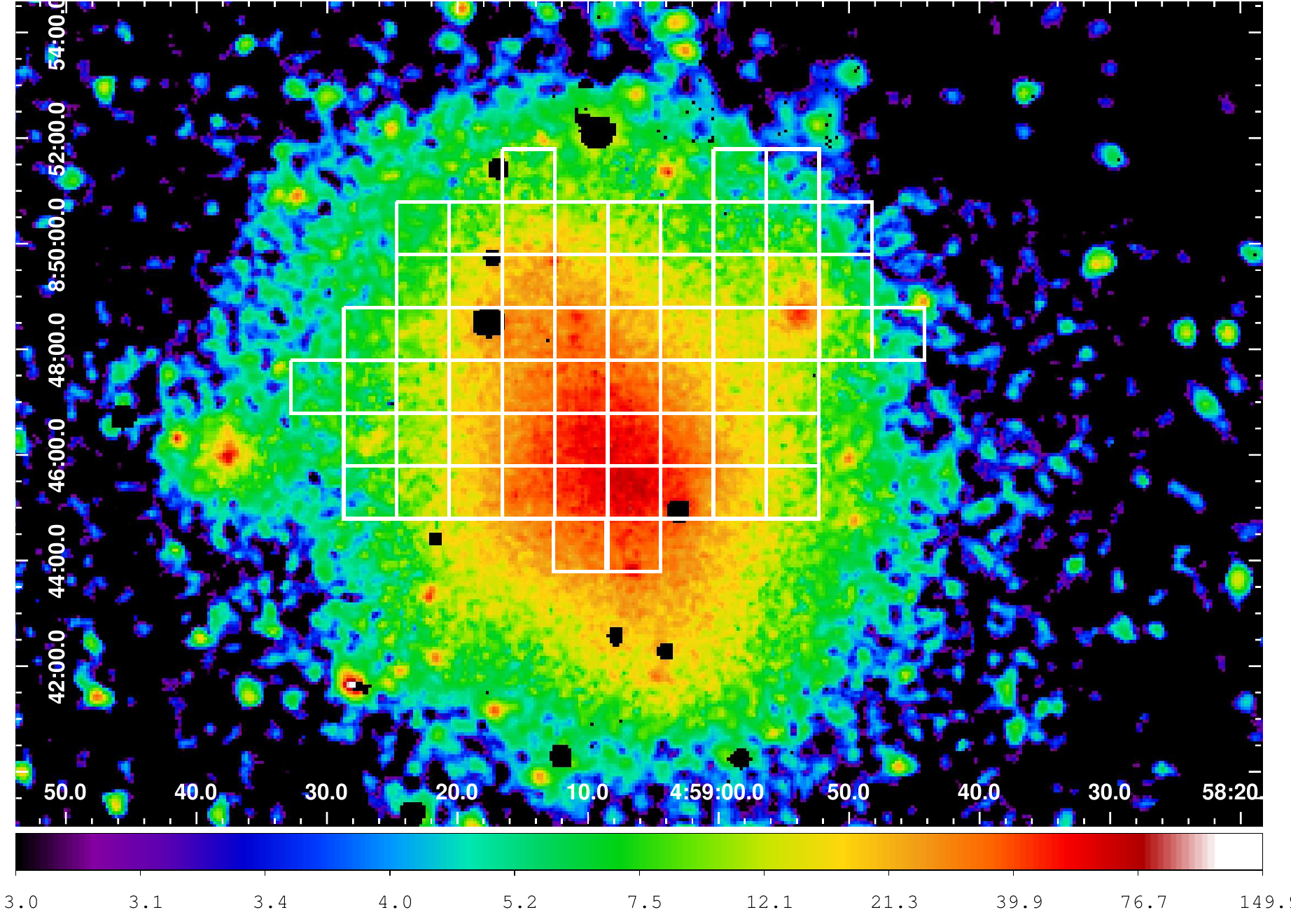}
\caption{Square grid adopted for the radio-X-ray surface brightness comparison, overlaid to the \textit{XMM-Newton} surface brightness image. Each cell has a size of 60''.}
\label{fig:radiogrid_over}
\end{figure}

For A523 we considered a square grid (built with \texttt{Synage++}; \citealp{Murgia01}) which covers the whole radio-halo projected area (Figure \ref{fig:radiogrid_over}). For each grid cell we then extracted mean values and rms for the radio surface brightness at 1.4 GHz from archival VLA data (see G16 for details). The cell size was chosen in order to have a good resolution and at the same time a reasonable signal-to-noise ratio. We then obtained X-ray surface brightness data by extracting the net counts number in each cell from the exposure corrected and background subtracted image, for both observations combined and in the 0.5-2.5 keV energy band. Counts were then converted to surface brightness values using the thermal models obtained from the temperature map. This procedure is faster than extracting and fitting the single spectra for each cell of the grid. In order to estimate the goodness of this method, we compared count rates for the regions considered for the thermodynamic maps, both from the image and the spectra extracted. The two rates show a very good agreement, with the slope of the best-fit regression line being $0.97 \pm 0.06$.

\begin{figure} [t]
\centering
\includegraphics[width=\linewidth]{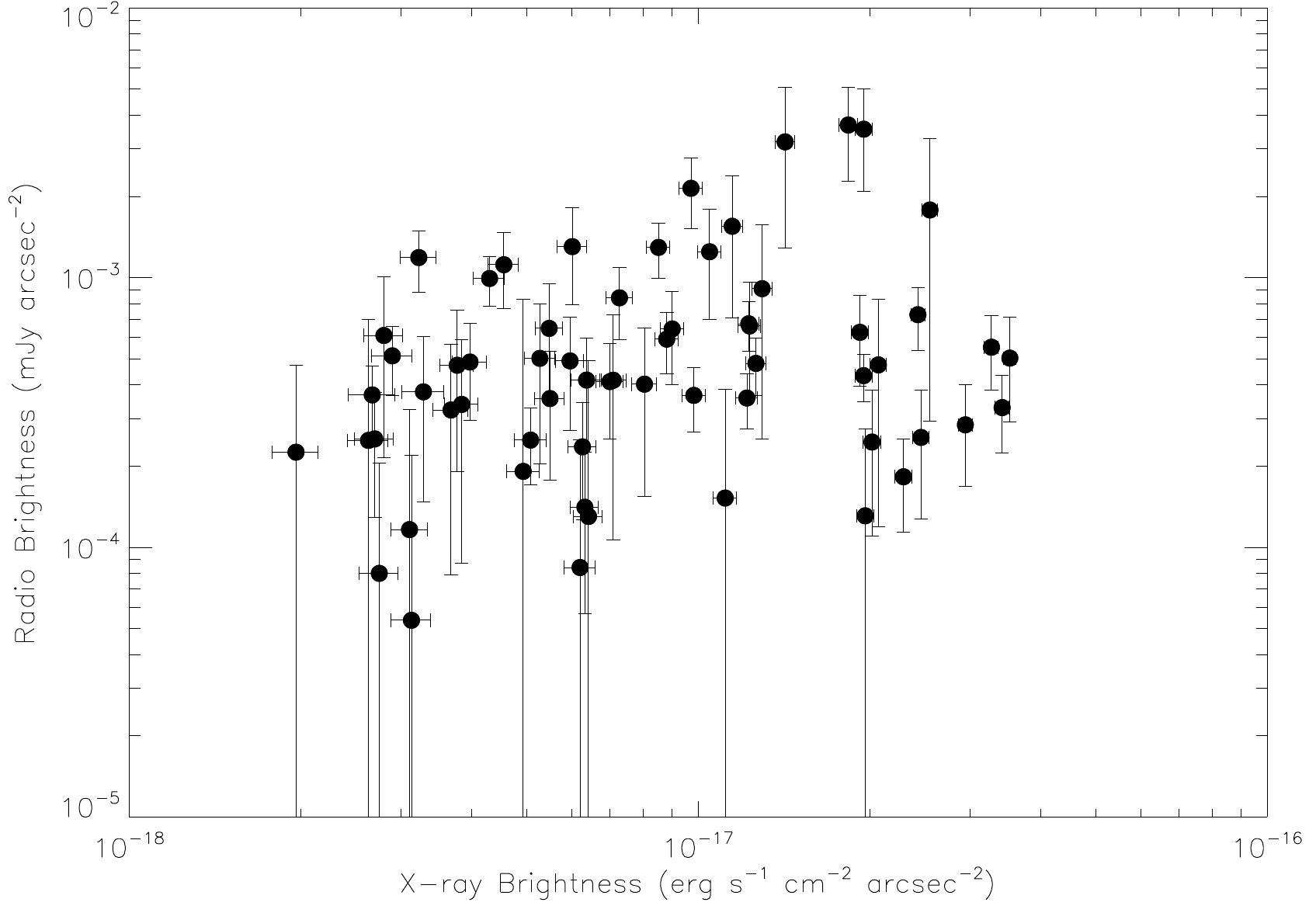}
\caption{Radio and X-ray surface brightnesses comparison for A523. Each point shows the mean of the brightness obtained within each cell of the grid with error bars indicating the rms of the brightness distribution.}
\label{fig:SBcomp_A523}
\end{figure}

\begin{figure} [t]
\centering
\includegraphics[width=\linewidth]{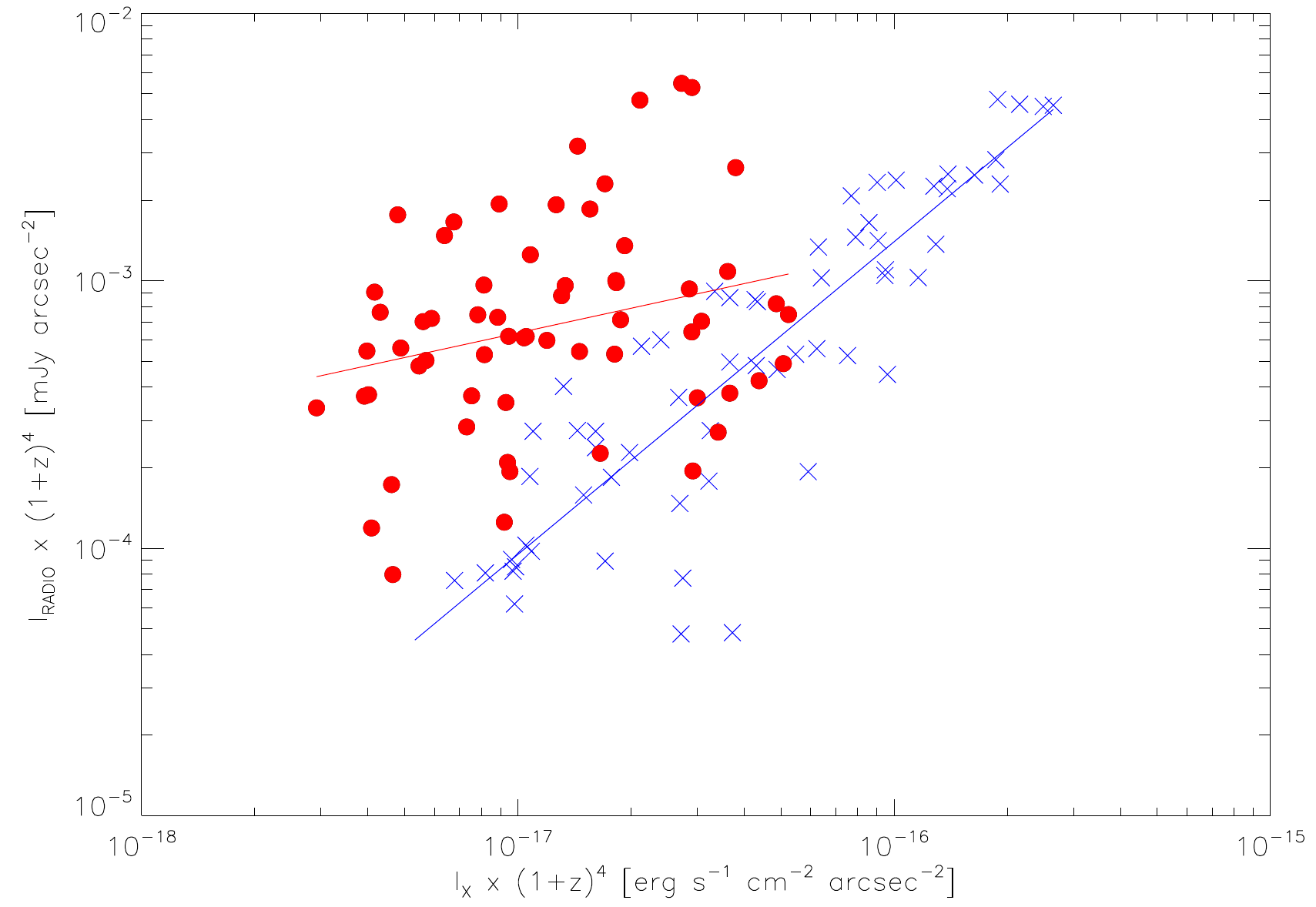}
\caption{Radio and X-ray surface brightness comparison for A523 (red point) and A2744 (blue crosses). For both clusters the best-fit relation estimated with the least square method is indicated by a solid line. Data were corrected for cosmological dimming.}
\label{fig:A523_A2744_comp}
\end{figure}

In Figure \ref{fig:SBcomp_A523} we show the point-to-point comparison between the radio and X-ray surface brightnesses. No clear correlation is evident, and the data have a broad intrinsic scatter, compared to what found in typical radio-halo clusters. This comparison confirms locally the discrepancy already noted for the global X-ray and radio emissions. This peculiar behavior can be further highlighted by the comparison with a typical radio-halo cluster: Abell 2744, studied by \citet{Govoni01a}. For this system radio and X-ray emissions (from \textit{ROSAT} data) appear morphologically very similar and their surface brightnesses correlation can be well described by a linear relation (with $b = 0.99 \pm 0.05$). In Table \ref{tab:A2744_A523} we compare some main properties for the two clusters and the plot of the comparison using least squares fit to the radio-X-ray surface brightness comparisons of the two clusters is shown in Figure \ref{fig:A523_A2744_comp}.  

\begin{table}[th]
\caption{Main properties for the two clusters considered in this paper: A523 and A2744 (data from Govoni et al. 2001 and references therein). The 4th column refers to the total radio flux density at 1.4 GHz.}
\label{tab:A2744_A523}
\medskip
\centering
\begin{tabular}{l c c c c} 
\toprule
Cluster & redshift & $L_{\mathrm{X}}$ [0.1-2.4 keV] & $kT$ & $P_{1.4 \, GHz}$ \\
	     &   & ($10^{44} \, \mathrm{erg}/\mathrm{s}$) & (keV) & $10^{24}$ W $\mathrm{Hz}^{-1}$ \\
\midrule
A523 & 0.104 & 1.39 & 4.29 & 2.0 \\
A2744 & 0.308 & 22.05 & 11.0 & 17.0 \\
\bottomrule
\end{tabular}
\end{table}

\begin{figure*} [t]
\begin{multicols}{2}
\includegraphics[width=\linewidth]{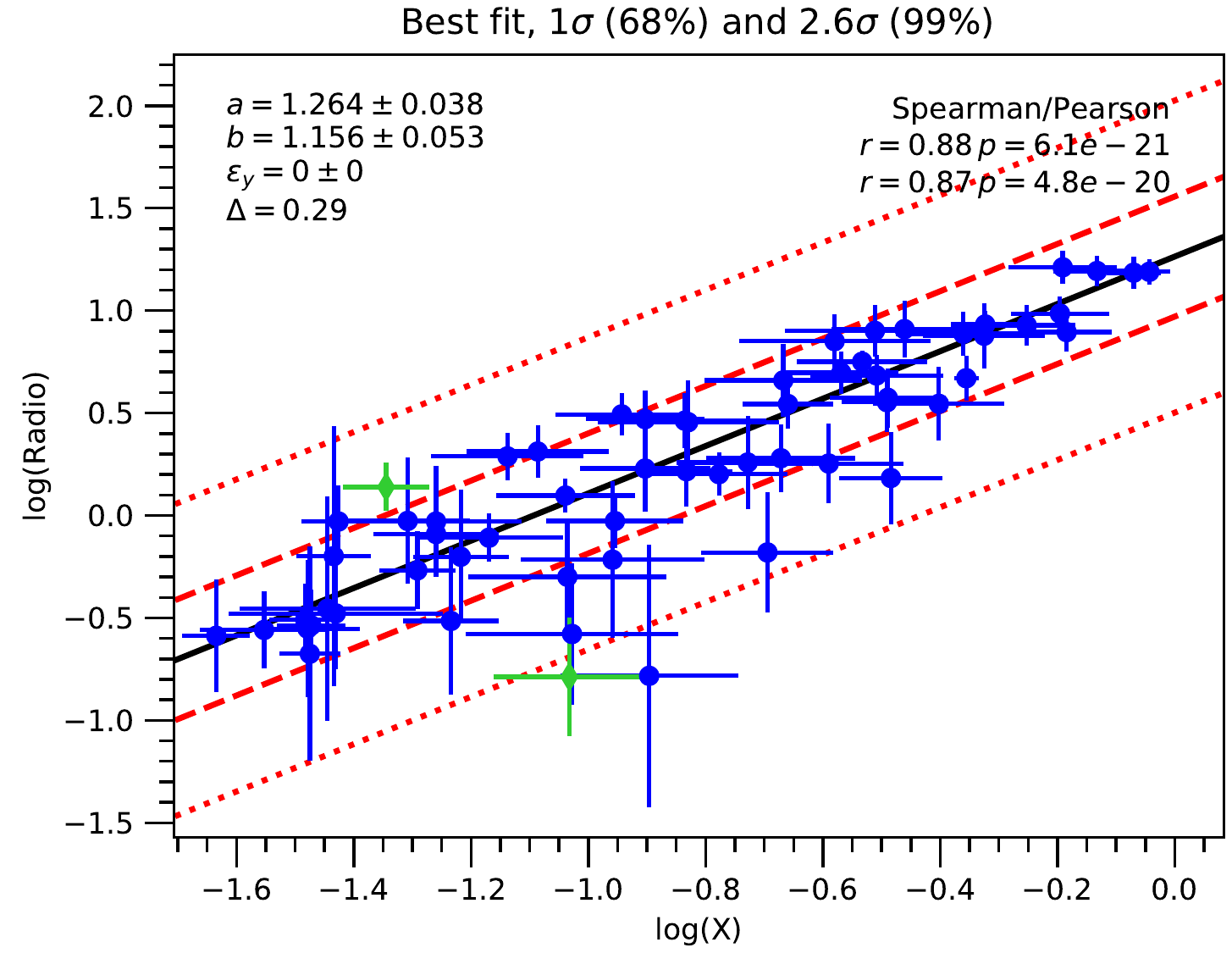}\par
\includegraphics[width=\linewidth]{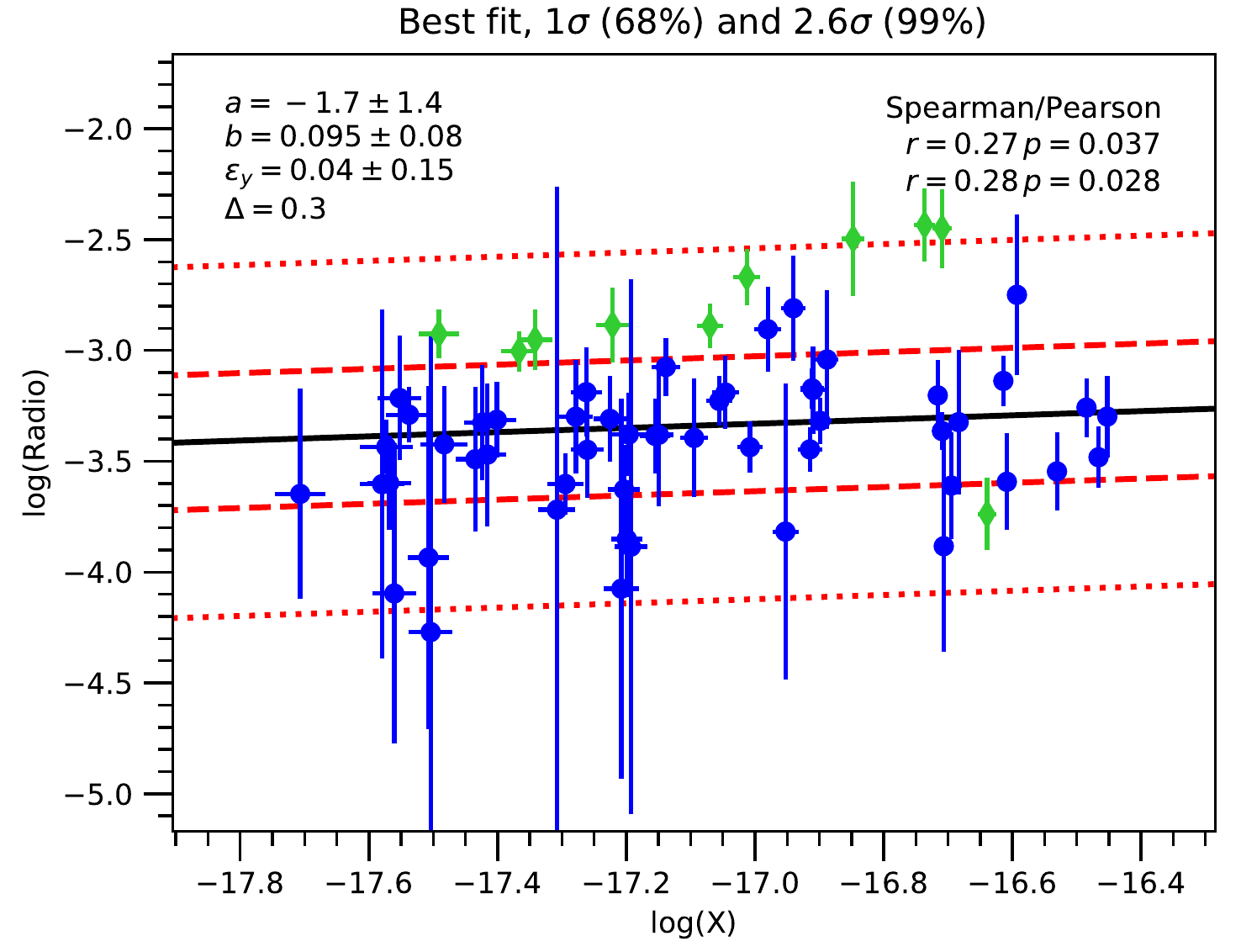}\par
\end{multicols}
\caption{Best-fit for the radio-X-ray surface brightnesses relation with the \textit{lts-linefit} method by \citet{Cappellari13} for the clusters A2744 (left panel) and A523 (right panel). Best-fit parameters and observed intrinsic scatter $\Delta$ are indicated in the upper left corner. Red dashed lines indicate $1\sigma$ confidence levels, while red dotted lines indicate the $2.6\sigma$ confidence levels. Green points indicate the outliers from the best-fit relation (black solid line).}
\label{fig:lts_linefit}
\end{figure*}

However, we note that this method is not particularly suited for this cluster, given the large error bars in the radio data. In addition to that, a least squares linear fit assumes that the scatter present in the data is purely of statistical origin, but as evidenced before A523 shows a significant intrinsic scatter in its data, due to its very own peculiarity. A proper fitting method should then take this intrinsic scatter into account. To do that, we adopted the \texttt{lts-linefit} method by \citet{Cappellari13}, which performs a robust linear fit on data with error bar on both variables, possible presence of large outliers (defined as those data deviating more than $2.6 \sigma$ from the best-fit linear relation) and an unknown intrinsic scatter. We used this technique for both A2744 and A523; the obtained results are shown in Figure \ref{fig:lts_linefit} left and right panel, respectively. For the former cluster, we find a linear relation consistent with the simple least square fit, according to the fact that this system shows a low intrinsic scatter. For the latter cluster we find an even flatter relation than the one previously observed, a significant number of outliers and a wider intrinsic scatter. The individuated outliers correspond to grid cells characterized by higher radio emission than the average, given their X-ray emission and are located in a direction perpendicular to the merging axis. 

If interpreted via turbulent reacceleration this distribution would suggest that in the outskirts of the cluster the specific turbulent energy flux, $\approx  V_A^3 / l_A$ ($V_A$ is the Alfven velocity and $l_A$ is the MHD scale, see \citealp{Brunetti16}), is larger or that the fraction of the turbulent energy flux that goes into particle reacceleration is larger in the outskirts. Alternatively our results might also suggest that the halo is in fact a ghost/revived radio plasma in the cluster periphery that is seen in projection. Future studies, including robust measurements of the spectrum of the halo will clarify its origin.

\section{Non-thermal emission}
\label{sec:nonthermal_em}
\textit{NuSTAR}'s unprecedented hard X-ray focusing capability makes it the ideal instrument for the search for a diffuse non-thermal inverse Compton (IC) emission above $\approx10$ keV. In this section we will present the upper limits obtained for this emission in A523 both globally and locally, obtained taking fully into account the cluster multi-temperature thermal emission. In Appendix \ref{sec:appendixB}, we also present a set of ad-hoc simulations performed in order to test the reliability of our results.
\subsection{Global spectrum}
The global spectral analysis was performed on the spectra extracted from an inner circular region of 5 arcmin, centered in the peak of the X-ray cluster's surface brightness. The spectra were extracted for both telescopes and observations using \texttt{nuproducts}, for a total of four spectra. We jointly fit the spectra with an APEC thermal plasma, with Galactic absorption fixed at $N_{\mathrm{H}} = 1.06 \times 10^{21} \mathrm{cm}^2$; the fit was performed in \textit{Xspec} in the 3-100 keV band, using C-statistics and metallicities relative to the abundances of \citet{Anders89}. The higher part of the spectra ($> 79$ keV), where used to constrain the instrumental background. Although not strictly necessary for a fit using the C-statistic, we re-binned the data to ensure a minimum 30 counts per bin, reducing the time required to perform fits and emphasizing differences between the model and the data. The background was accounted for by including all the spectral components of the best-fit model used to produce the background spectrum as described in Section \ref{sec:bgd_modeling}. We decided to model the background, rather than subtracting its simulated spectrum to the source spectrum as done in \citet{Wik14}, in order to have a better control over its parameters.

\begin{figure} [th]
\centering
\includegraphics[angle=270,width=\linewidth]{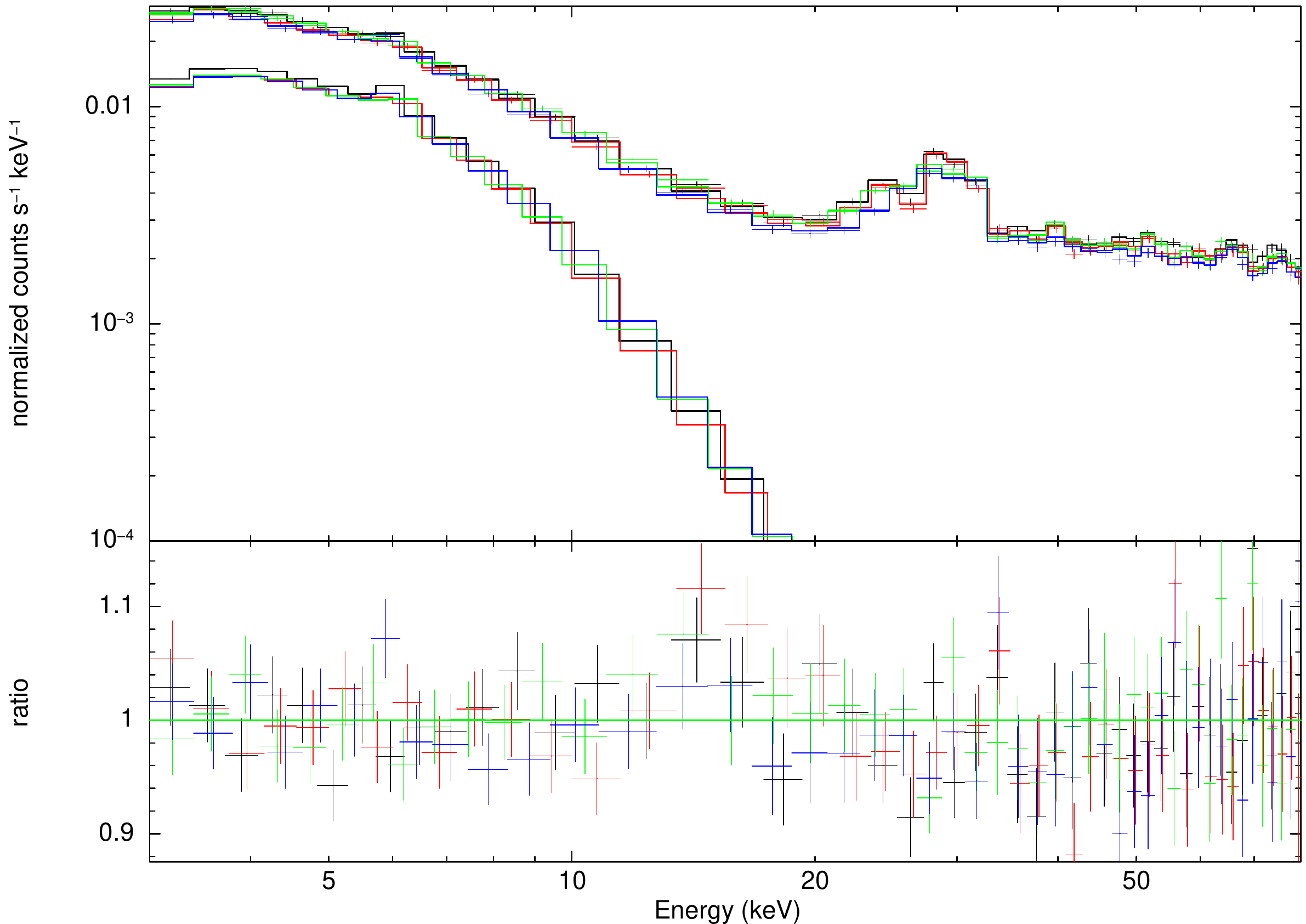}
\includegraphics[angle=270,width=\linewidth]{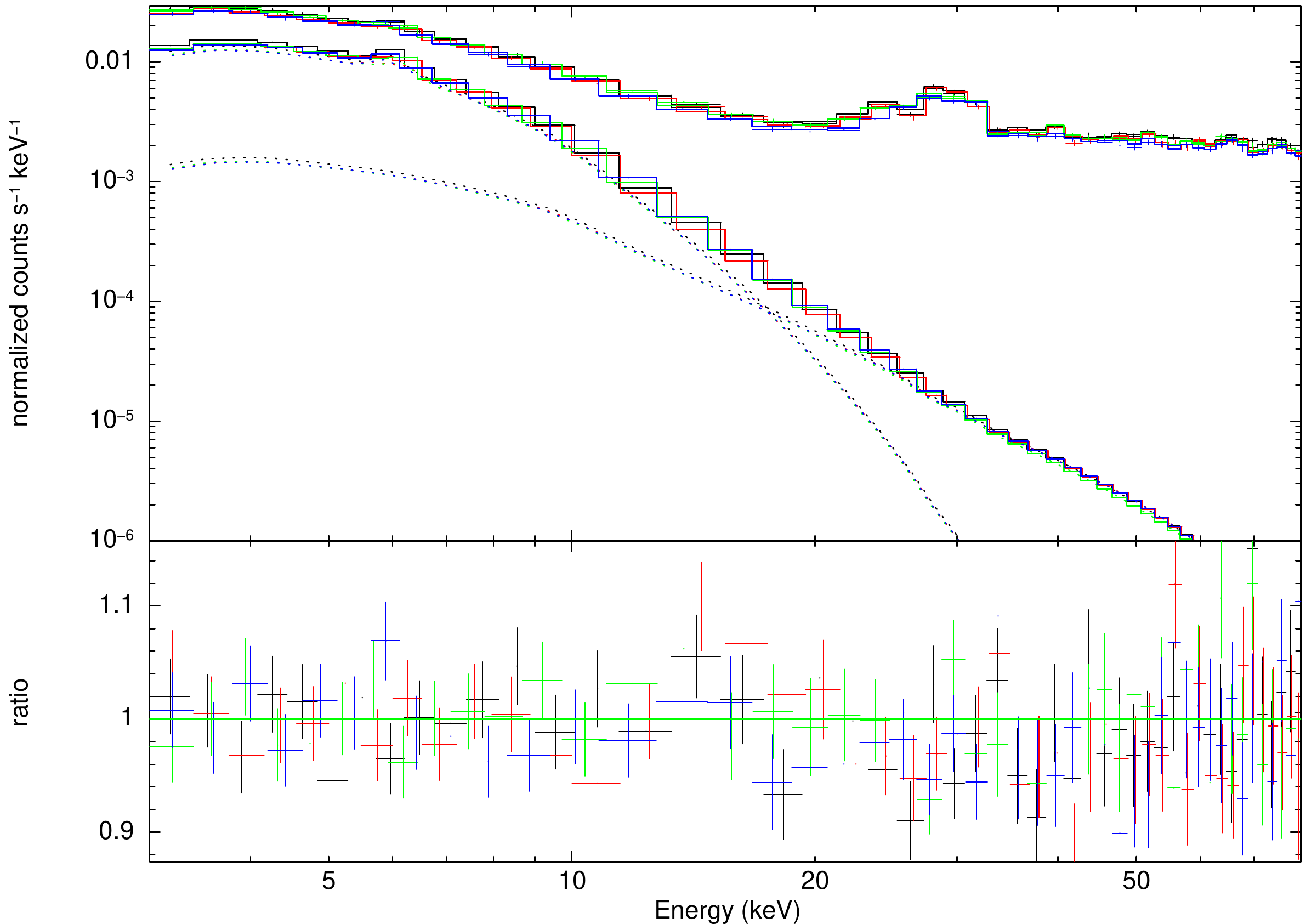}
\caption{Best-fit models and ratio with data for the 5 arcmin spectra extracted from both telescopes and both observations. Top panel: single temperature thermal model. Bottom panel: single temperature thermal model plus power-law with photon spectral index $\Gamma = 2.0$ (fixed); dashed lines refer to the contribution to the total model from the thermal (upper) and the power-law (lower) components.}
\label{fig:5min_fits}
\end{figure}

The spectral fit was performed using different models, and the resulting best-fit parameters are shown in Table \ref{tab:fit_param_5min}. The first is a single temperature model, which is the simplest description of a thermal plasma. This description is likely unrealistic, especially in disturbed and merging clusters, with temperature variations across its volume. This is in fact the case of A523, where the $XMM-Newton$ observation highlighted modest but non-negligible temperature variations, requiring the construction of a multi-temperature map to describe its thermal properties. However, 1T models can sometimes be a good approximations for multi-temperature and featureless spectra (as shown in e.g. \citealp{Mazzotta04}). In Figure \ref{fig:5min_fits} (top panel), we show the best-fit model and its ratio to the data of this single temperature approach. The fit seems to perform quite well, however some excess is somewhat visible in the 10-20 keV band, indicating that the spectrum is not of a truly isothermal plasma.
The second model consists of a single temperature plus power-law, in order to investigate the possible presence of non-thermal IC emission in the $20 - 80$ keV band. The photon spectral index was fixed at $\Gamma = 2$, based on the radio analysis in G16. The best-fit model is shown in Figure \ref{fig:5min_fits} (bottom panel) and in Table \ref{tab:fit_param_5min} we quote the $3\sigma$ upper limit on the 20-80 keV non-thermal flux. Despite not being statistically significant, the addition of a power-law model seems to slightly better reproduce the $>10$ keV part of the spectrum. Nevertheless, a counts excess between 10-20 keV is still visible; this residual excess could either be due to the fact that more thermal components are required to better reproduce the cluster emission or some residual not fully taken into account in the background model, or an instrumental feature not well calibrated and accounted for in the ARF. In support of this latter hypothesis, we note that a similar feature in the same energy range and of the same level is visible in the \textit{NuSTAR} Coma Cluster spectrum of \cite{Gasta15}.

In order to put more solid constraints on this last value and calculate its confidence intervals, we  used Bayesian statistics and a Markov chain Monte Carlo (MCMC) technique. The fits were performed in \texttt{Xspec} using the \cite{Goodman10} algorithm implementation, which evolves a series of \textit{walkers} (i.e. vectors of the fit parameters) using random steps given by the difference between two walkers. For this simulation we used ten walkers and a total of $10^4$ steps, with a burning phase of 5000 steps, to ensure that the chain converged to a steady state. The Bayesian statistic was turned on setting constant priors for the APEC thermal components, while we used Gaussian priors for the background components, centered on the nominal values obtained from the simulated background model from the \textit{fixed} method described in Section 2. The widths of the Gaussian priors were set to the expected systematic errors: 8\% for the aperture component, 3\% for the instrumental continuum and 10\% on the fCXB. The results for the posterior best-fit values using the MCMC technique are also listed in Table \ref{tab:fit_param_5min}; the parameters found are very similar to those previously extracted, meaning that the likelihood is dominating the Bayesian statistic. We then marginalized over all the other parameters to obtain the posterior probability distribution for the IC flux, using the \texttt{margin} command. From this distribution we then obtained the 68\% confidence intervals, which resulted in: [0.366-0.385] $10^{-12} \, \mathrm{erg} \, \mathrm{s}^{-1} \, \mathrm{cm}^{-2}$.

The power-law flux found with this initial spectral analysis seems quite robust to MCMC tests. However, its normalization is not sensibly affected by background fluctuations (which should be tested in the MCMC procedure), like we would expect with a true non-thermal excess. In addition, compared to the Aperture and fCXB components, its statistical significance in the fit is indeed quite low. This may indicate that its behavior more likely mimics additional thermal components missing from the simple single temperature model rather than interpreting a true non-thermal component. To confirm this hypothesis we fitted more complicated models going beyond the single temperature description. We tried: i) a two-temperature (2T) model. this is the simplest multi-temperature model and it may just reduce the residuals but it may not represent a true range in temperature, as shown in \cite{Gasta15}; ii) a detailed multi-temperature map model, representing the range of temperatures in the various regions of the volume of the cluster as measured by the 2D spectral analysis.

The 2T thermal model (two APEC components with tied abundances) should in principle well describe a multi-temperature component spectrum, as thermal continua are quite featureless. However, when a bright non-thermal IC emission is present, the higher temperature component of the 2T model can obtain an unphysically high value. This in fact seem to happen in A523's case, with a best-fit higher temperature component of $\approx8.5$ keV, and both temperatures poorly constrained. Such a high temperature value is not consistent with the extracted \textit{NuSTAR} temperature map (see Section \ref{sec:temp_map_nu}), and is then not very realistic. However, the addition of a power-law component, accounting for a non-thermal emission, to this model does not impact on the best-fit thermal parameters and we find a $3\sigma$ upper limit on the 20-80 keV non-thermal flux much lower than before. This could probably indicate that the higher value for the IC upper limit previously found with the 1T+IC model was actually an artifact due to the fitting procedure accommodating for the residuals in the low energy part of the spectrum, not well reproduced by a single temperature model. A more detailed thermal model could then be sufficient to properly describe the cluster global spectrum, without the need of the inclusion of a non-thermal power-law component.

\subsection{Temperature map}
\label{sec:temp_map_nu}
\begin{figure} [t]
\centering
\includegraphics[width=\linewidth]{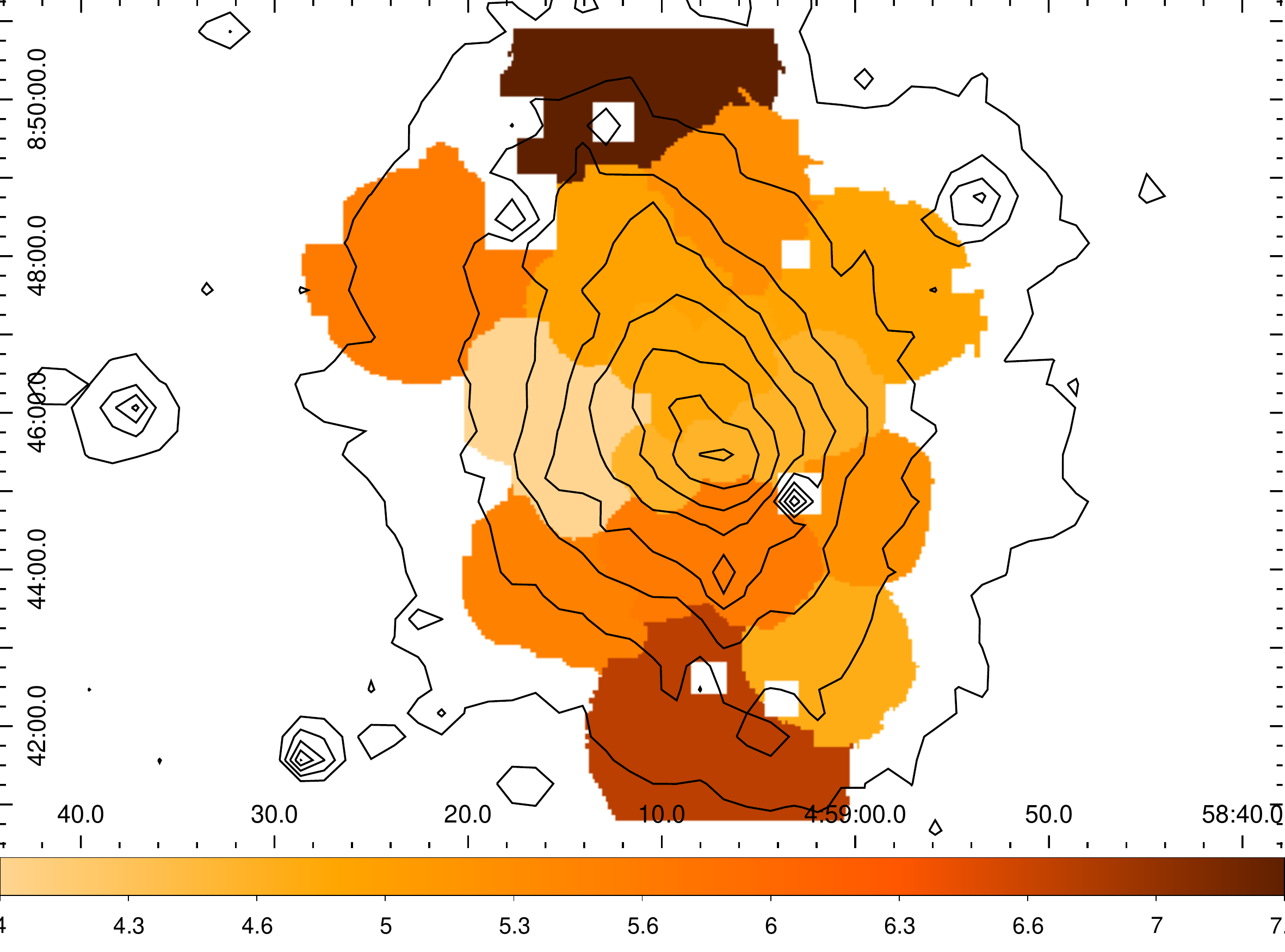}\par
\caption{A523's temperature map extracted from the \textit{NuSTAR} analysis, with the cluster's X-ray surface brightness contour levels overlaid. }
\label{fig:temp_map_nu}
\end{figure}
In order to put more stringent constraints on the cluster multi-temperature thermal emission, we built and included in the spectral fit a temperature map obtained from \textit{NuSTAR}'s observations. To produce the map we initially adopted the very same regions individuated from the \textit{XMM-Newton} analysis (given by with the \texttt{contbin} algorithm by \cite{Sanders06}, following the cluster X-ray surface brightness). However, the smaller inner regions taken singularly were not suited for a \textit{NuSTAR} analysis, given the larger instrument PSF compared to \textit{XMM}'s one, with very low total counts. We then grouped some of the inner regions until reasonable counts for each instrument were reached. The resulting map is shown in Figure \ref{fig:temp_map_nu}. Following \cite{Gasta15}, we summed the 13 APEC models with fixed temperatures, abundances and normalizations, allowing a global normalization constant free to fit to account for possible discrepancies between the map and the global spectrum. In order to better reproduce the actual cluster thermal emission we removed some regions from the 5 arcmin circle that are not covered by the temperature map, and where the cluster X-ray emission is not or only faintly present. With this procedure the overall adjustment constants were reduced to a 8-10\% level. In Table \ref{tab:fit_param_5min} we show the resulting best-fit statistic value. This model represent a reasonable description of the multi-temperature components in A523, and should in principle put a very good constraint on the cluster thermal emission, allowing for a non-thermal flux detection if present. The inclusion of a power-law model, however, does not result in a statistically significant improvement of the fit and the 20-80 keV flux of the non-thermal power-law component has a $3\sigma$ upper limit significantly lower that the one previously found with the simple 1T+IC model.

\begin{table*}[th]
\centering
\caption{Global spectrum fit parameters.}
\label{tab:fit_param_5min} 
\begin{varwidth}[b]{\linewidth}
\renewcommand\arraystretch{1.2}
\begin{threeparttable}
\medskip
\begin{tabular}{l c c c c c c} 
\toprule
Model & $kT$ & Abund. & Norm.\tnote{a} & $\Gamma$ & IC flux\tnote{b} & C-stat/dof \\
 & (keV) & ($\mathrm{Z}_{\odot}$) & ($10^{-3} \mathrm{cm^{-5}}$) &  & ($10^{-12} \mathrm{erg} \, \mathrm{s}^{-1} \, \mathrm{cm}^{-2}$) & \\
\midrule
1T & $5.2 \pm 0.1$ & $0.127 \pm 0.017$ & $5.6 \pm 0.2$ & $\cdots$ & $\cdots$ & 9939/9694 \\
1T+IC  & $4.7 \pm 0.3$ & $0.145 \pm 0.020$ & $5.3 \pm 0.2$ & 2.0 (fixed) & $<$ 0.94 & 9935/9693\\
1T+IC\,\tnote{c} & $4.7 \pm 0.1$ & $0.143^{+0.008}_{-0.007}$ & $5.5 \pm 0.1$ & 2.0 (fixed)& $0.376 \pm 0.009$ & 9753/9681\\
\multirow{2}{*}{2T+IC} & $3.2 \pm 1.3$ & \multirow{2}{*}{$0.159 \pm 0.022$} & $4.6 \pm 2.0$ & \multirow{2}{*}{2.0 (fixed)} & \multirow{2}{*}{< 0.34} & \multirow{2}{*}{9931/9692} \\
 & $8.5 \pm 5.0$ &  & $2.0 \pm 2.5$ &  &  &  \\
$\mathrm{T}_{\mathrm{map}}$+IC\, \tnote{d} & $\cdots$ & $\cdots$ & $\cdots$ & 2.0 (fixed) & < 0.22 & 10227/9691\\
$\mathrm{T}_{\mathrm{map}}$+IC\,\tnote{e} & $\cdots$ & $\cdots$ & $\cdots$ & 2.0 (fixed)& < 0.35 & 10225/9691 \\
$\mathrm{T}_{\mathrm{map}}$+IC\,\tnote{c}\, \tnote{e} & $\cdots$ & $\cdots$ & $\cdots$ & 2.0 (fixed)& < 0.25 & 10225/9679 \\
$\mathrm{T}_{\mathrm{map}}$+IC\,\tnote{e} & $\cdots$ & $\cdots$ & $\cdots$ & $2.1 \pm 0.9$& < 0.40 & 10255/9690 \\
\bottomrule
\end{tabular}
\begin{tablenotes}
\item \textbf{Notes.}
  \item[a] Normalization of the APEC thermal spectrum, which is given by $\left\lbrace 10^{-14}/\left[4\pi (1+z)^{2}D_{A}^{2}\right]\right\rbrace \, \int n_{e} n_{H} dV$, where $z$ is the redshift, $D_{A}$ is the angular diameter distance, $n_{e}$ is the electron density, $n_{H}$ is the ionized hydrogen density, and $V$ is the volume of the cluster.
   \item[b] 20-80 keV.
   \item[c] Obtained with the Bayesian MCMC analysis. Statistic includes a Bayesian contribution of -182, due to Gaussian priors.
   \item[d] For the map models the only parameters left free to fit are the overall adjustment constants and the power-law normalization.
   \item[e] Obtained with the re-measured temperature map, with parameters estimated in the 3.0-8.5 keV band (see text).
\end{tablenotes}
\end{threeparttable}
\end{varwidth}
\end{table*}

We investigated the possibility that a non-thermal bias could arise when fitting the map regions spectra with 1T thermal models only, biasing upward their estimated temperatures. We then re-calculated the best-fit temperatures and normalizations of the map regions, fitting their spectra only in the 3.0-8.5 keV energy band in order to minimize the impact of the possible presence of a non-thermal component. The high end of the energy range was chosen at 8.5 keV, as at that value the aperture background component starts dominating over the cluster thermal emission (see Figure \ref{fig:bgd_components}). With this temperature map model, we have a $3\sigma$ upper limit on the 20-80 keV non-thermal flux of $3.54 \times 10^{-13} \, \mathrm{erg} \, \mathrm{s}^{-1} \, \mathrm{cm}^{-2}$. If we allow the power-law photon spectral index is left free to vary, we find a best-fit value of $\approx 2.1$, consistent with the initial fixed index, and even the estimated upper limit is consistent with the one previously found. Even though this method gives a higher upper limit on the IC flux, we still cannot state that a true non-thermal bias is present, as also in this case we don't find a statistically significant detection of the component. 

\subsection{Maximum halo region}
\label{sec:max_halo}
\begin{figure} [t]
\includegraphics[width=\linewidth]{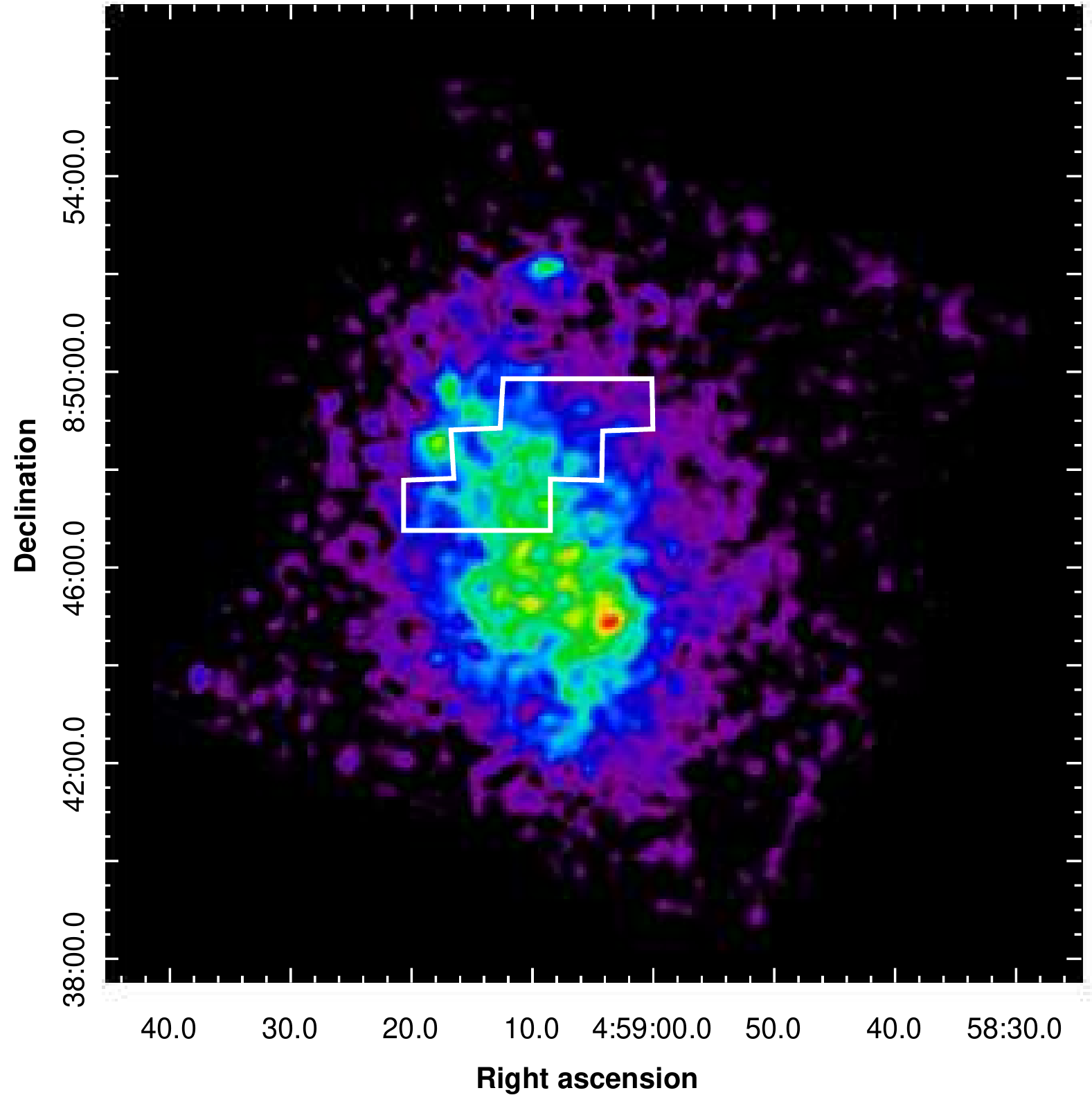}
\caption{Grid cells where the radio-halo emission is maximum (see text), overlaid on the \textit{NuSTAR} 3-10 keV image of A523.}
\label{fig:max_halo}
\end{figure}

Given A523's peculiarity in the radio-X-ray morphology, we can restrict our search for IC emission in those regions where the radio surface brightness is at its maximum (defined as those cells in the grid showed in Figure \ref{fig:radiogrid_over} where the surface brightness is greater or equal than $10^{-3} \, \mathrm{mJy}\, \mathrm{arcsec}^{-2}$). These regions do not coincide with the peak of the X-ray brightness and are located almost at the cluster periphery, but not in a background dominated region (see Figure \ref{fig:max_halo}). This peculiarity should in principle represent the ideal condition for a detection of a non-thermal excess, as the thermal contribution is minimized.

Despite these conditions, the spectrum extracted seems to be well represented by a simple single temperature thermal model. The best-fit thermal model has a temperature of $6.0 \pm 0.3$ keV, which is an intermediate value consistent with the temperature map showed in Figure \ref{fig:temp_map_nu}. Neither the inclusion of an additional thermal model or a power-law model with a fixed photon spectral index at $\Gamma = 2.0$ result in a statistically significant improvement of the fit. The estimated $3\sigma$ upper limit on the 20-80 keV non-thermal flux is $3.17 \times 10^{-14} \, \mathrm{erg} \, \mathrm{s}^{-1} \, \mathrm{cm}^{-2}$.

\section{Discussion}
\label{sec:discussion}

\subsection{Scaling relations and cluster mass estimate}
\label{subsec:discussion_mass}
In the introduction of this paper, we addressed the peculiarities of A523 as a radio halo cluster and in particular its position in the $P_{1.4 \, GHz} - L_{X}$ plane, which is confirmed by our luminosity estimate, which is perfectly consistent with the one found in G16. According to the best-fit relation of \cite{Cassano13}, A523 is under luminous in X-ray by a factor of $\gtrsim 4$ or over luminous in radio by a factor of $\gtrsim 20$.
However the scaling relation that should be taken into account is the scaling with mass such as a $P_{1.4 \, GHz} - M_{500}$ relation since
the energy budget available to non-thermal components is directly connected with cluster mass rather than with the X-ray luminosity \cite{Brunetti14}.

In order to assess the position of A523 in the $P_{1.4 \, GHz} - M_{500}$ plane, we need to estimate its mass from our measurements. From our measure of $kT_{\mathrm{OUT}}$, and using the $M-T$ scaling relation of \cite{Arnaud05}, the derived cluster mass is $M_{500} = 3.1 \pm 0.5 \times 10^{14} \, M_{\odot}$. If, however, we use our revised estimate of $L_{X,500}$ and the $L-M$ scaling relation of \cite{Pratt09}, we find a cluster mass of $M_{500} = 2.3 \pm 0.1 \times 10^{14} \, M_{\odot}$. 
We also measured the low scatter mass proxy $Y_{\rm X}$ by measuring iteratively
the gas mass within $R_{500}$ and the X-ray temperature from a fit to the spectrum extracted in the [0.15-0.75] $R_{500}$ annulus. We then used the scaling relation $M_{500}-Y_{\rm X}$ \citep{Arnaud10}
to derive a cluster mass of $M_{500} = 2.9 \pm 0.1 \times 10^{14} \, M_{\odot}$.

Despite the uncertainty in the cluster mass estimate, all our predicted values are much lower than the $M_{500} \gtrsim 5.0 \times 10^{14} \, M_{\odot}$ value predicted by the best-fit relation of \cite{Cassano13}, confirming that A523 is an outlier even in the $P_{1.4\, GHz} - M_{500}$ plane. 

\subsection{Thermodynamical maps and the merging scenario}
The thermodynamical maps of A523 obtained from the \textit{XMM-Newton} observation give a clear insight on the cluster's dynamical status.

The temperature map suggests a scenario which is compatible with an off-axis merger between two subcluster of similar mass and with a non-zero impact parameter ($b\approx 2 r_{s}$, with $r_{s}$ being the typical scale length associated with the NFW profile; see \citealp{Ricker01}). It is worth noticing that high temperature regions are also present between the two cores, suggesting that the merging process could be in a phase immediately subsequent to the core passage.

The abundance map confirms a disturbed status, being quite uniform and without regions at very high metallicity. This is an evidence that the merging process effectively perturbed the ICM, and has diluted the metal abundances injected from the cluster's galactic population. This hypothesis is further confirmed by the fact that none of the analyzed regions present high abundance levels, with average metallicities always below $0.4 \, Z_{\odot}$ within $1\sigma$, which is often considered a threshold value for a distinction between CC and NCC systems, e.g. \cite{RossMol10}. 

The pseudo-pressure map has a quite asymmetrical morphology, with regions at higher pressure located along the merging axis and tracing the mass of the two subclusters, supporting the disturbed scenario. Regions of higher pressure are located along the merging axis and tracing the mass of the two subclusters, moreover the core of the southern subcluster shows high pressure. This could indicate that this structure could be more massive than the one in the north region, contributing more to the global pressure distribution. The absence of high pressure regions along the radio-halo main axis is an evidence that a secondary process in this direction, if present, involves much less mass than the main event.

\begin{figure} [t]
\includegraphics[width=\linewidth]{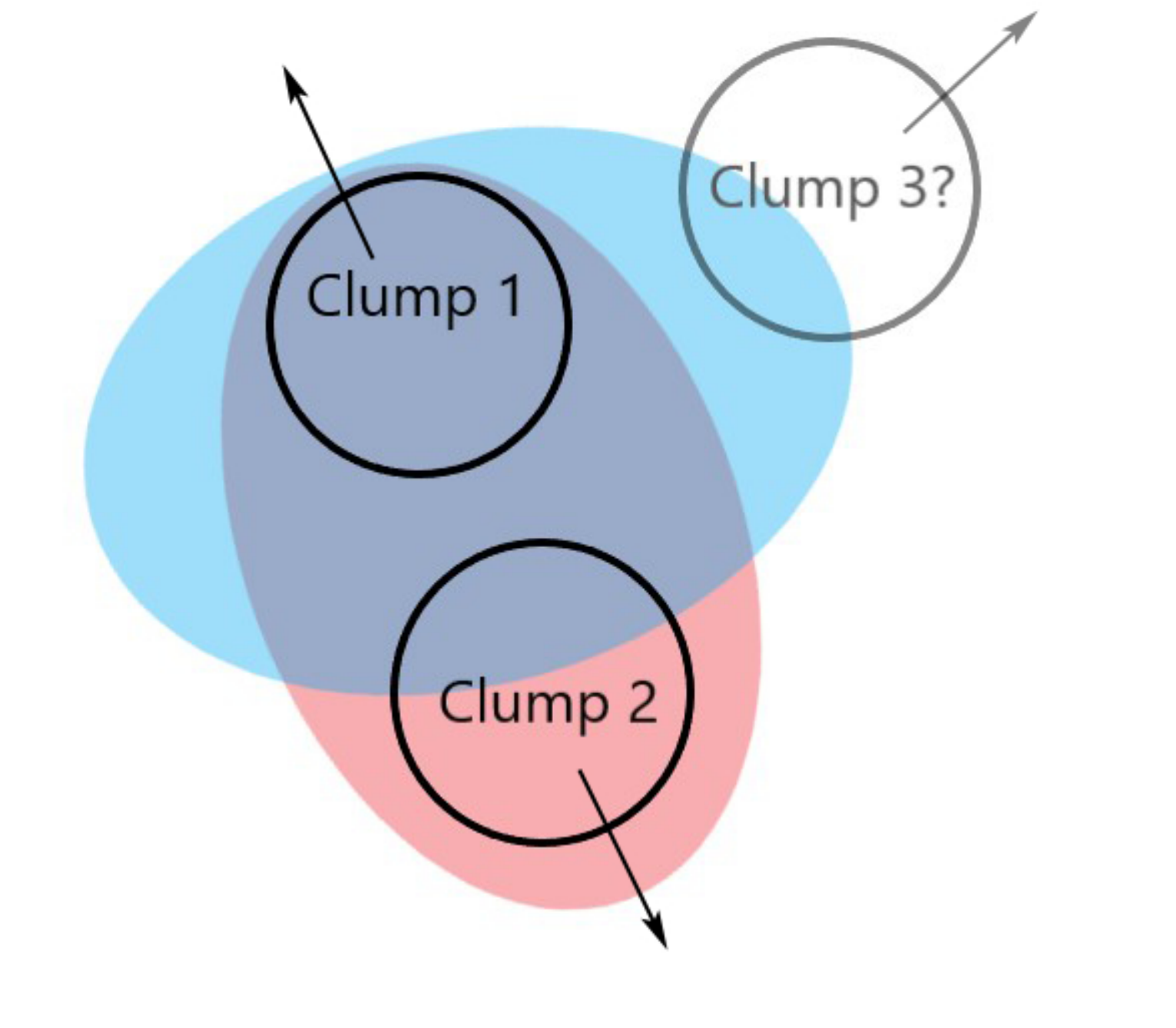} 
\caption{Dynamics of the merger in A523 as suggested by X-ray data. The radio halo is sketched in blue, while the thermal ICM emission in red.}
\label{fig:sketch_dyn}
\end{figure}

The pseudo-entropy map shows some lower entropy regions located both along the main merger axis and the main radio-halo axis. Low entropy regions at the center of the system should be highlighting some dense gas originally situated in the subclusters' cores. Regions in the perpendicular direction could instead be tracing some gas stripped from the core of a third subcluster, which is merging in a direction perpendicular to the one of the main merger (as sketched in Figure \ref{fig:sketch_dyn}), producing some turbulence responsible for the peculiar cluster radio emission. This scenario is further supported by the fact that the outliers from the best-fit relation found for the X-ray/radio surface brightness comparison are located in these lower-entropy regions (see Figure \ref{fig:lts_linefit}) and by the elongation with two tongues-like features by the radio halo, as depicted by both our radio image and the radio image presented by \cite{Golovich18}. A suggestion for this scenario was also put forward in G16 given the tentative evidence of a NW group and therefore the possibility that A523 is
actually forming at the crossing of two filaments along the SSW-NNE and ESE-WNW directions.
A better validation of this hypothesis would need more extended optical data.

\subsection{IC upper limits and B lower limits}
When fitting the \textit{NuSTAR} global spectrum (extracted from a central circular region with a radius of 5') with a simple 1T model with the addition of a power-law model, we find a $3\sigma$ upper limit on the 20-80 keV IC flux of $9.4 \times 10^{-13} \, \mathrm{erg} \, \mathrm{s}^{-1} \, \mathrm{cm}^{-2}$. If we assume that this flux is truly of non-thermal origin, and is produced by the same population of relativistic electrons responsible for the synchrotron emission, we can derive a lower limit on the cluster mean magnetic field, averaged over the whole radio emitting region. Assuming a radio spectral index of $\alpha = 1.0$, a total radio flux density at 1.4 GHz of $72 \pm 3$ mJy and making use of Eq. (34) of \cite{GF04}, this upper limit on the IC emission translates to a lower limit on the magnetic field strength of $B \gtrsim 0.15 \, \mu G$. However, the selected region is not isothermal, as is evident from both the \textit{XMM-Newton} and \textit{NuSTAR} temperature maps. Thus, the non-thermal flux found with the simple 1T+IC model, could just be mimicking the additional thermal components. 

This hypothesis is indeed confirmed when we fit the global spectrum with a 2T model and, more in detail, with a multi-temperature model derived from the \textit{NuSTAR} temperature map. These models successfully reproduce the observed spectrum by themselves, and the inclusion of an additional power-law model does not result in an appreciable and statistically significant improvement of the fit. The $3\sigma$ upper limit to the IC non-thermal emission in the 20-80 keV band in this case is in the range $\left[2.2 - 4.0\right] \times 10^{-13} \, \mathrm{erg} \, \mathrm{s}^{-1} \, \mathrm{cm}^{-2}$, even accounting for a possible small non-thermal bias when fitting the regions for the temperature map with simple 1T models over the whole energy range. This upper limit range for the IC emission corresponds to a lower limit range for the magnetic field strength of $[0.23 - 0.31] \, \mu G$. We also mention that our results on the IC upper limits are further supported by the simulations presented in Appendix \ref{sec:appendixB}.

Finally, when we analyze the ``maximum halo'' region (Section \ref{sec:max_halo}), we find a much lower upper limit, which translates into an higher lower limit to the magnetic field strength of $B \gtrsim 0.81 \, \mu G$.

The search for the detection of IC emission by the radio halo electrons in a galaxy cluster has not yet been successful, despite trying to maximizing our chances by choosing a source with an extreme radio-to-X-ray flux ratio and a possible low volume-averaged magnetic field spread on large spatial scales.

\subsection{The nature of the diffuse radio emission}
The diffuse radio emission in A523 is clearly peculiar. The working hypothesis in this and previous works (G11, G16) has been that this diffuse radio emission is a radio halo. Clearly there are many features which makes this interpretation not entirely satisfying: i) the detected polarization of the radio emission which can be explained by a particular configuration of the magnetic field. However we are also aware of the energetic constraints set on the distribution of the energetic electrons by the magnetic field configuration; ii) the fact that the source is an outlier both in the global scaling relations followed by radio halos and also locally in terms of the radio-X-ray surface brightness as shown in this work.

The radio relic interpretation is even more problematic.
As already discussed the position of the radio emission within the merging subclusters is difficult to reconcile in a bimodal merger scenario. A configuration of a merger mainly along the line of sight where a relic emission could cover both subclusters in projection is challenged by the optical data presented in G16 showing evidence of a merger mainly in the plane of the sky. This is supported by the  the relatively small difference in velocity of the two clumps, 100-650 km/s, by the small velocity difference by the two BCGs, 150km/s and by the dispersion velocity of the global system, 949 km/s, in broad agreement with the global mass and temperature estimates.
The fact that the system might be at the turnaround point is disfavoured by the two-body toy model discussed in G16.
This for example has to be compared with the case of A2256 as an example of of interpretation of diffuse radio emission as a projected relic along the line of sight. In that case there is a clear indication from galaxy redshift measurements of a merger along the line of sight, in particular the very large
dispersion velocity of 1200 km/s and the very high velocity difference between
the two BCG exceeding 2000 km/s (and by the offset between both of them and the X-ray emission, see \citealp{Golovich18} and references therein).

A possible alternative interpretation is that the
diffuse radio emission is re-energized plasma of AGN origin (\citealp{vanWeeren19} and references therein).
The properties of A523 will be rather extreme also in this hypothesis however a vast phenomenology of this class is recently emerging. Future radio observations, in particular with the LOw Frequency ARray (LOFAR; \citealp{vanHaarlem13}), will help to clarify this issue.

\section{Summary}
\label{sec:summary}
We performed a detailed analysis of a joint \textit{XMM-Newton} and \textit{NuSTAR} observation of the merging galaxy cluster Abell 523. Our findings are summarized as follows:
\begin{itemize}
    \item the radial profiles of temperature and abundance and the two dimensional maps of thermodynamical quantities confirm the merging state of the cluster. The entropy map suggests the presence of a secondary merging event;
    \item the complex merging state may explain the peculiarity of the radio halo emission of this system, in particular the fact that it is an outlier in the radio-X-ray brightness relation;
    \item tight upper limits on the presence of IC emission constrain the volume-averaged magnetic field to be greater than 0.2-0.3 $\mu G$ ($0.8 \mu G$ in the region with the brightest radio emission);
    \item given some elements of conflicting evidence against the radio halo interpretation of the diffuse radio emission, we leave open the possibility that it is due to re-energized plasma of AGN origin. Future radio observations, in particular with LOFAR, will clarify this issue.
\end{itemize}

\begin{acknowledgements}
We thank the referee, Scott Randall, for useful comments which improved the presentation of the paper.
Based on observations obtained with XMM-Newton, an ESA science mission with instruments and contributions directly funded by ESA Member States and NASA.
    
This research made use of data from the \textit{NuSTAR} mission, a
project led by the California Institute of Technology, managed
by the Jet Propulsion Laboratory, and funded by NASA. We
thank the \textit{NuSTAR} Operations, Software, and Calibration teams
for support with the execution and analysis of these observations.
This research has made use of the \textit{NuSTAR} Data Analysis
Software (NuSTARDAS) jointly developed by the ASI Science
Data Center (ASDC, Italy) and the California Institute of
Technology (USA).

We acknowledge financial contribution from the contracts NARO15 and NARO16 ASI-INAF I/037/12/0 and Fabrizio Fiore for his support to Italian researchers exploiting \textit{NuSTAR} data. 

M. Gaspari is supported by the Lyman Spitzer Jr. Fellowship (Princeton University) and by NASA Chandra GO7-18121X.
\end{acknowledgements}

\bibliographystyle{aa}
\bibliography{biblio}

\begin{appendix}
\section{Background models}
\label{sec:appendixA}
\subsection{XMM-Newton}
\begin{figure} [t]
\centering
\includegraphics[angle=270,width=\linewidth]{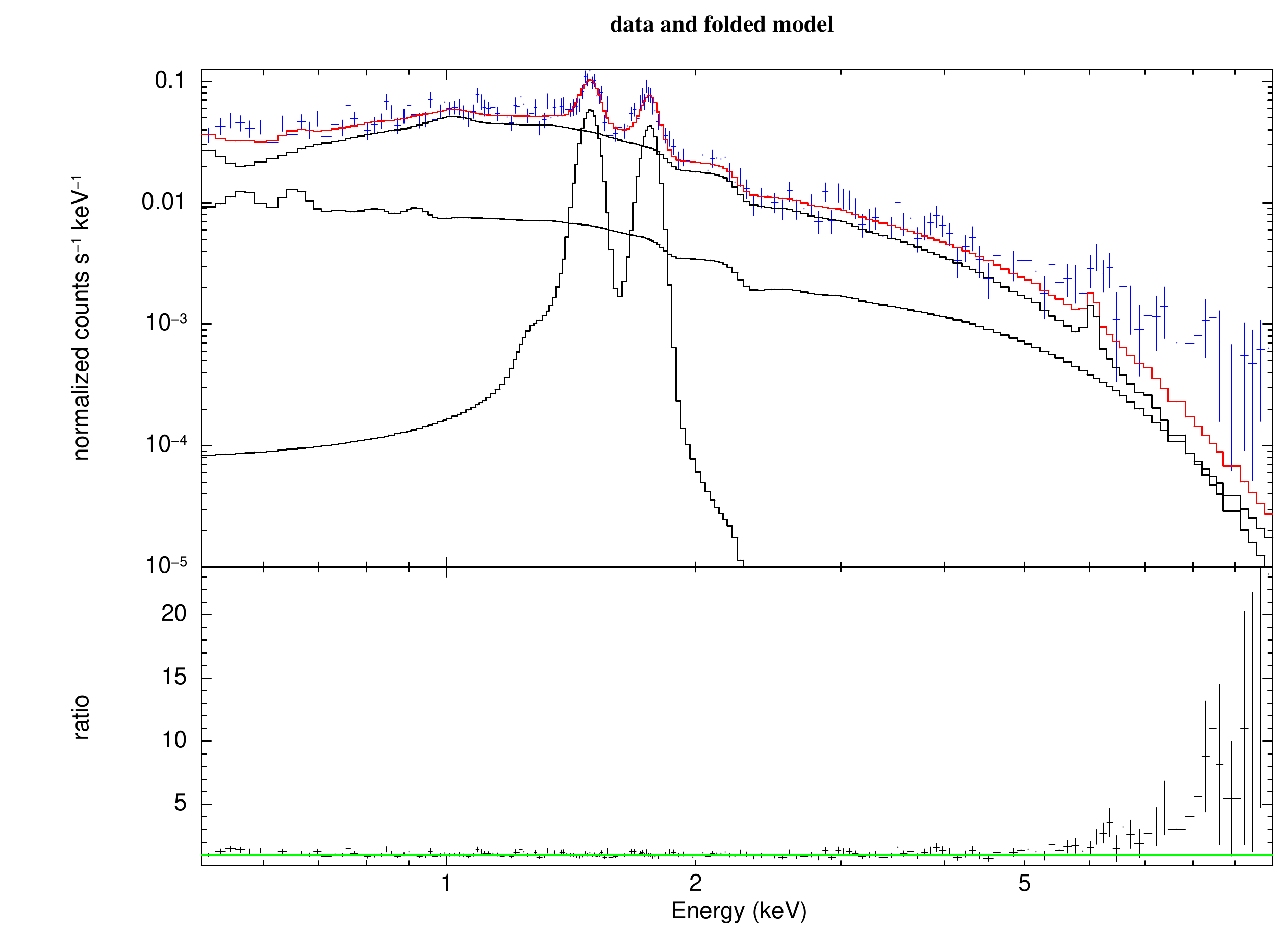}
\includegraphics[angle=270,width=\linewidth]{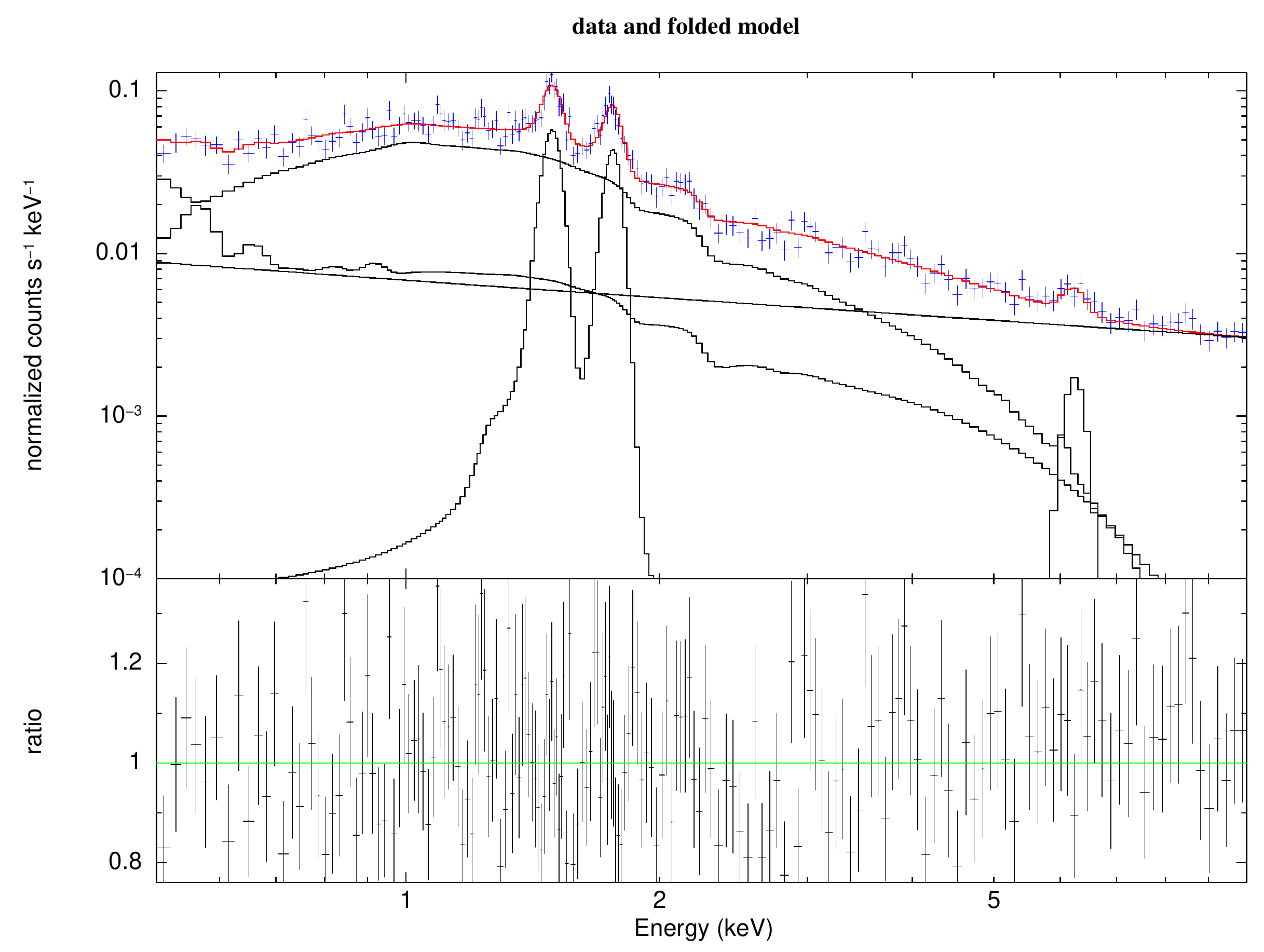}
\caption{Spectral fit for the 6th annulus and for only the MOS1 instrument, performed with two different methods. Top panel: background subtraction of the ESAS-produced background spectrum. Bottom panel: using the QPB model obtained from FWC observations, obtained as described in this section. In red we show the total best-fit model. Lower panels show the data to total model ratio.}
\label{fig:ESAS_nxb_fit}
\end{figure}

\textit{XMM-Newton}'s instrumental background is composed of fluorescence lines and non X-ray background (NXB), which is made of three components: the quiescent and cosmic-ray induced particle background (QPB), soft protons (SP) and a stable quiescent component of still unknown origin \citep{Salvetti17}. The QPB background can be produced with the ESAS routines \textit{mos-back} and \textit{pn-back}, with the exclusion of the fluorescence lines; the standard procedure \citep{Snowden08} requires that this spectrum is subtracted from the source spectrum for the spectral fit. However, for this particular analysis this method presents some issues, in particular for almost all the extracted spectra a clear counts excess in the higher energy range is visible (e.g. Figure \ref{fig:ESAS_nxb_fit}, left panel), even after providing a reasonable power-law model representing a possible soft proton flare residual. This was probably due to an incorrect modeling of the particle background in the ESAS procedure, giving the fact that at higher energies a number of florescence lines is expected and they are not accounted for in the produced spectrum. To overcome this problem, we used filter-wheel-closed observations close to A523's observations (REV 2966 and 2967) available in terms of revolution and exposure time, to model the spatial distribution of the QPB and of the florescence lines. The FWC spectra were modeled with Gaussian lines, taken from \citet{LeccMol08}, and a broken power-law with parameters left free to fit. These models were then used for the cluster spectral analysis, providing a sensible improvement in the fit quality (Figure \ref{fig:ESAS_nxb_fit}, right panel).

\begin{figure*} [ht]
\centering
\begin{multicols}{2}
\includegraphics[width=\linewidth]{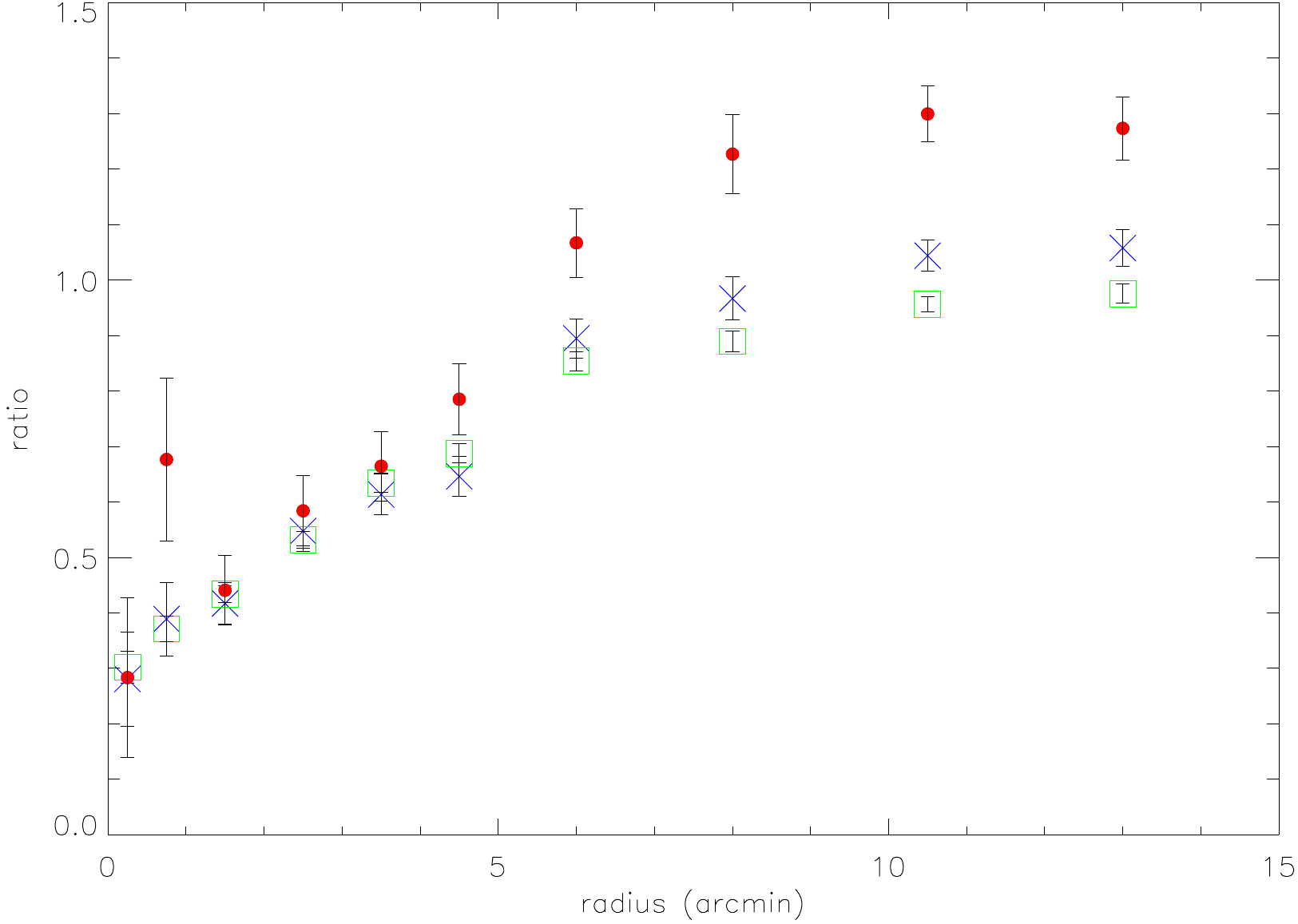}\par
\includegraphics[width=\linewidth]{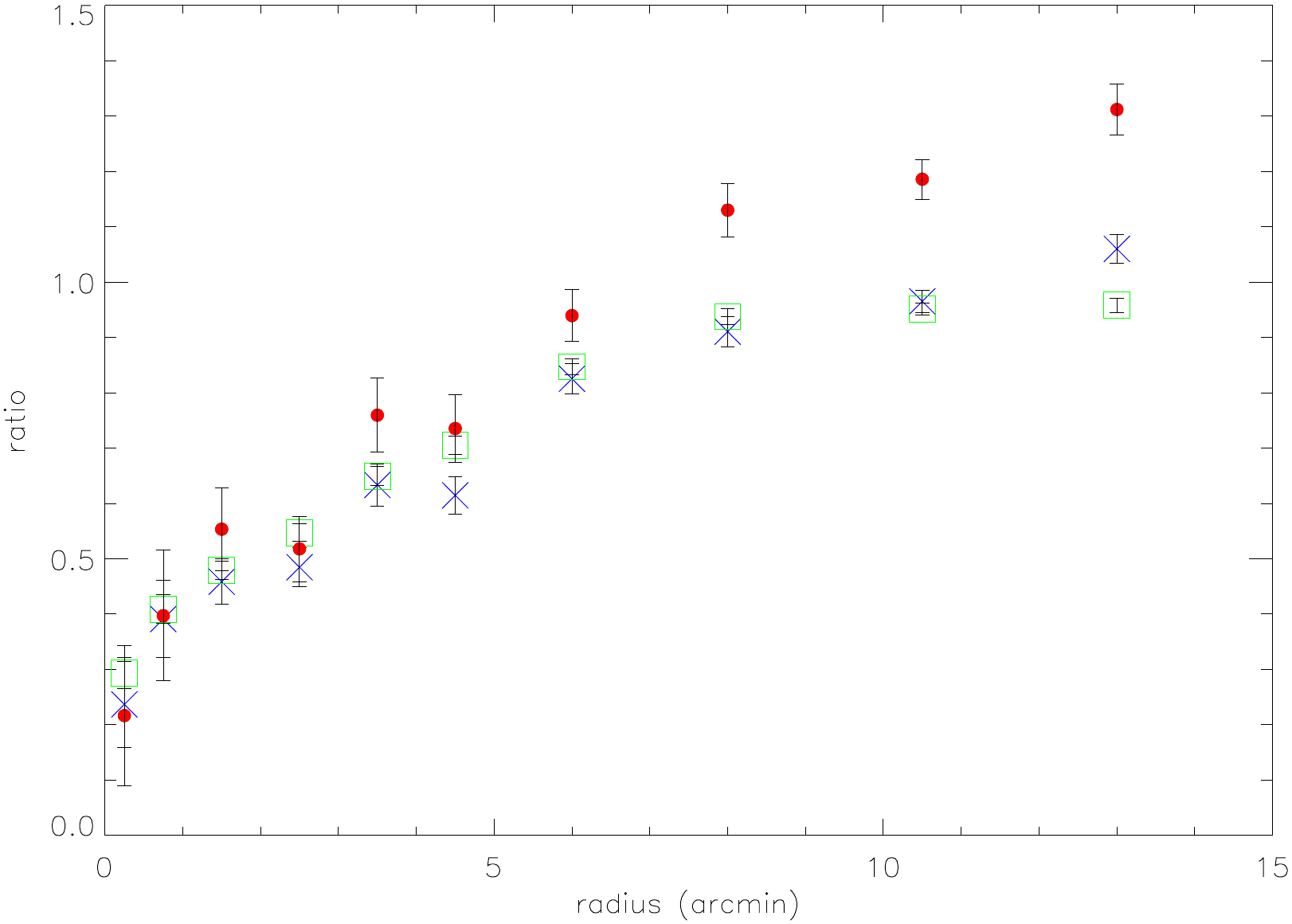}\par 
\end{multicols}
\vspace{0.25cm}
\includegraphics[width=0.5\linewidth]{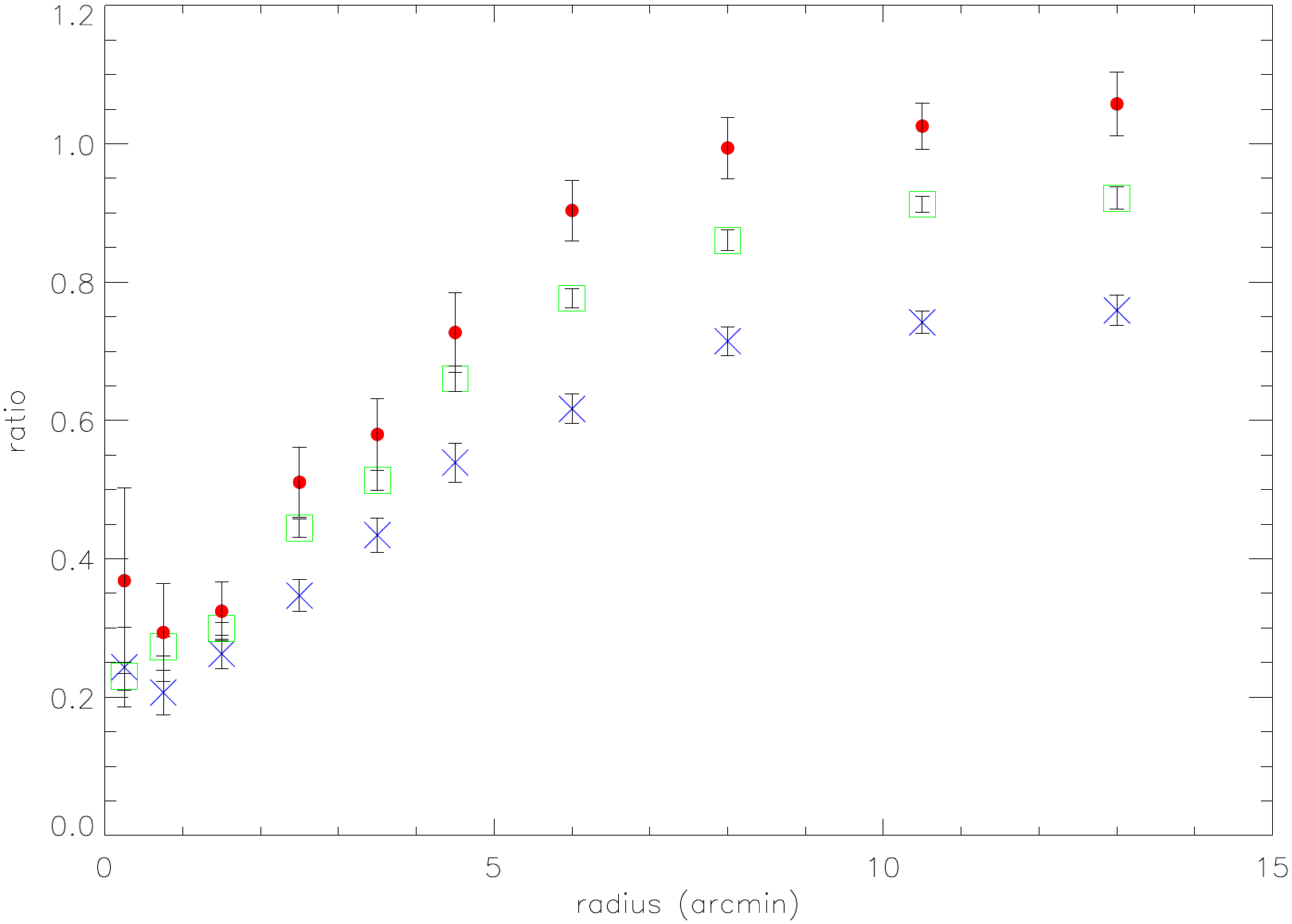}
\caption{Radial profiles extracted from the selected FWC observations, compared with the ESAS-produced background profile (green squares): REV2969 (red points); REV2877 (blue crosses, MOS only); REV2830 (blue crosses, pn only). Upper left panel: MOS1; upper right panel: MOS2; lower panel: pn. The data represent the  count rate ratios between the FWC observation and A523's observation, calculated in the same energy band (see text).}
\label{fig:FWC_radial}
\end{figure*}

In Figure \ref{fig:FWC_radial}, we show the radial profiles taken from the considered FWC observations, compared with the radial profiles of the ESAS-produced background. The observations considered correspond to the following revolutions: REV 2969 ($t_{exp} \sim 10$ ks), REV 2877 ($t_{exp} \sim 30$ ks, MOS only), REV 2830 ($t_{exp} \sim 38$ ks, pn only). The plots show the count-rate (cts/s) ratios of the FWC and ESAS-background spectra to A523's spectra extracted from the same annular regions, calculated in a line-free energy range (5-10 keV for the two MOS; 4-7 kev for the pn). As clear from the profiles, the REV 2877 observation shows a better agreement with the ESAS-produced background level for the MOS instruments, while REV 2969 looks more appropriate for the pn instrument.

\begin{figure}[th]
\includegraphics[width=\linewidth]{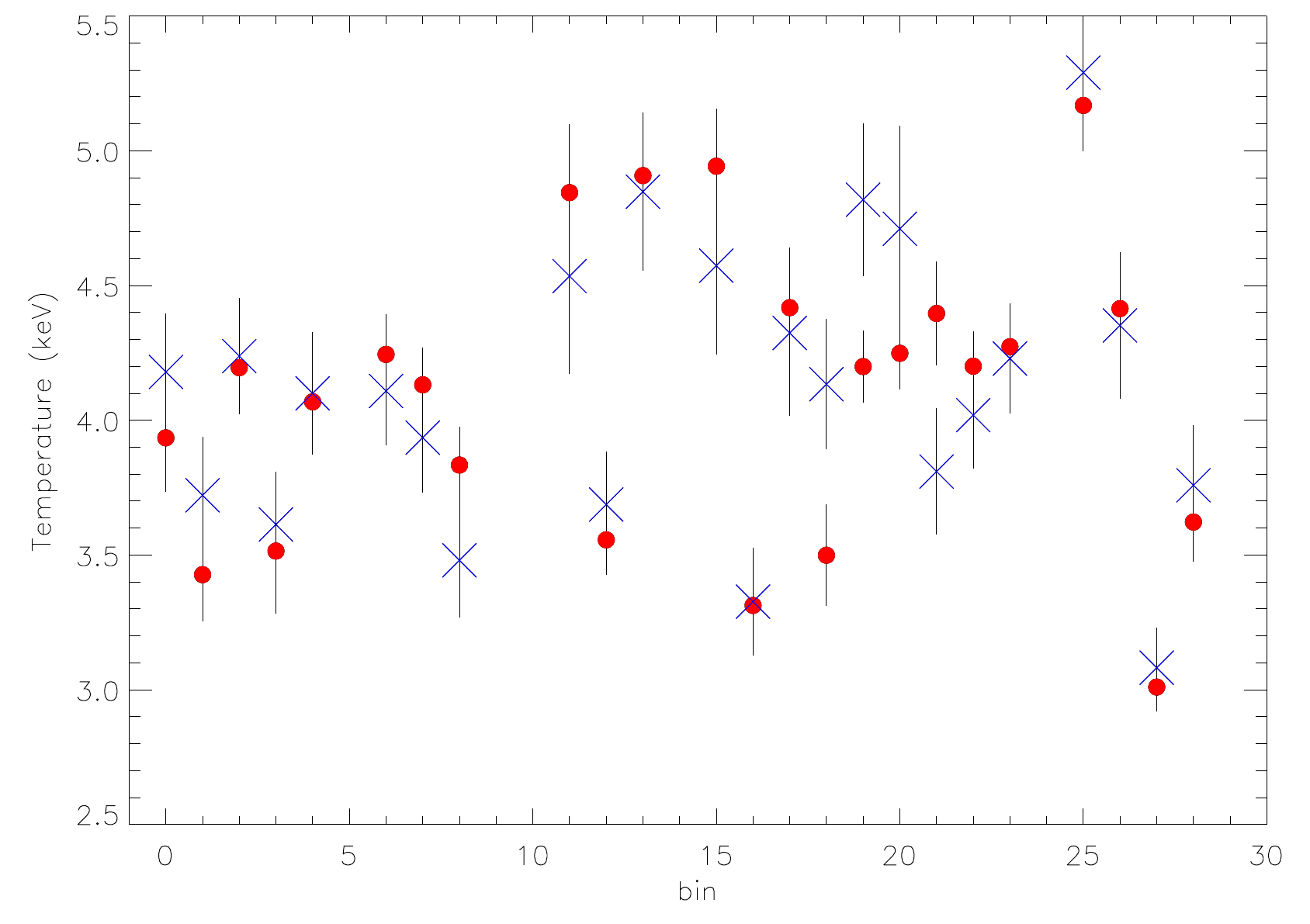}
\caption{Comparison between the best-fit temperature parameter extracted from the spectral analysis performed on the regions used for the \textit{XMM} thermodynamic maps, using both subtracting the ESAS-produced instrumental background (blue crosses) and the background modeling procedure described in this paper (red points).}
\label{fig:temp_bins_nxb-ESAS}
\end{figure}

Finally, we tested how our procedure affects the relevant spectral parameters. To do so, we compared the best-fit temperature parameter extracted from the very same regions used for the \textit{XMM} thermodynamic maps showed in Section \ref{sec:xmm_therm_maps}, using the two different procedures: the standard ESAS routine, with the subtraction of the instrumental background, and our procedure. In Figure \ref{fig:temp_bins_nxb-ESAS} we show the result of this comparison, where regions are numbered as in Figure \ref{fig:map_over_xmm}. From the plot, we can conclude that overall the effect of choosing one procedure over the other is typically lower or comparable to the parameter statistical uncertainty, with only the outer regions (numbered 18,19,20,21) showing a greater discrepancy. This latter result reflects the fact that the cluster outer regions, having a lower thermal emission, are more sensible to the instrumental background.

\subsection{NuSTAR}
In order to identify the better procedure to estimate the correct background parameters for \textit{NuSTAR}'s observation, we decided to adopt two different methods and compare their performances and results. 
\begin{figure*} [t]
\centering
\begin{multicols}{2}
\includegraphics[width=\linewidth]{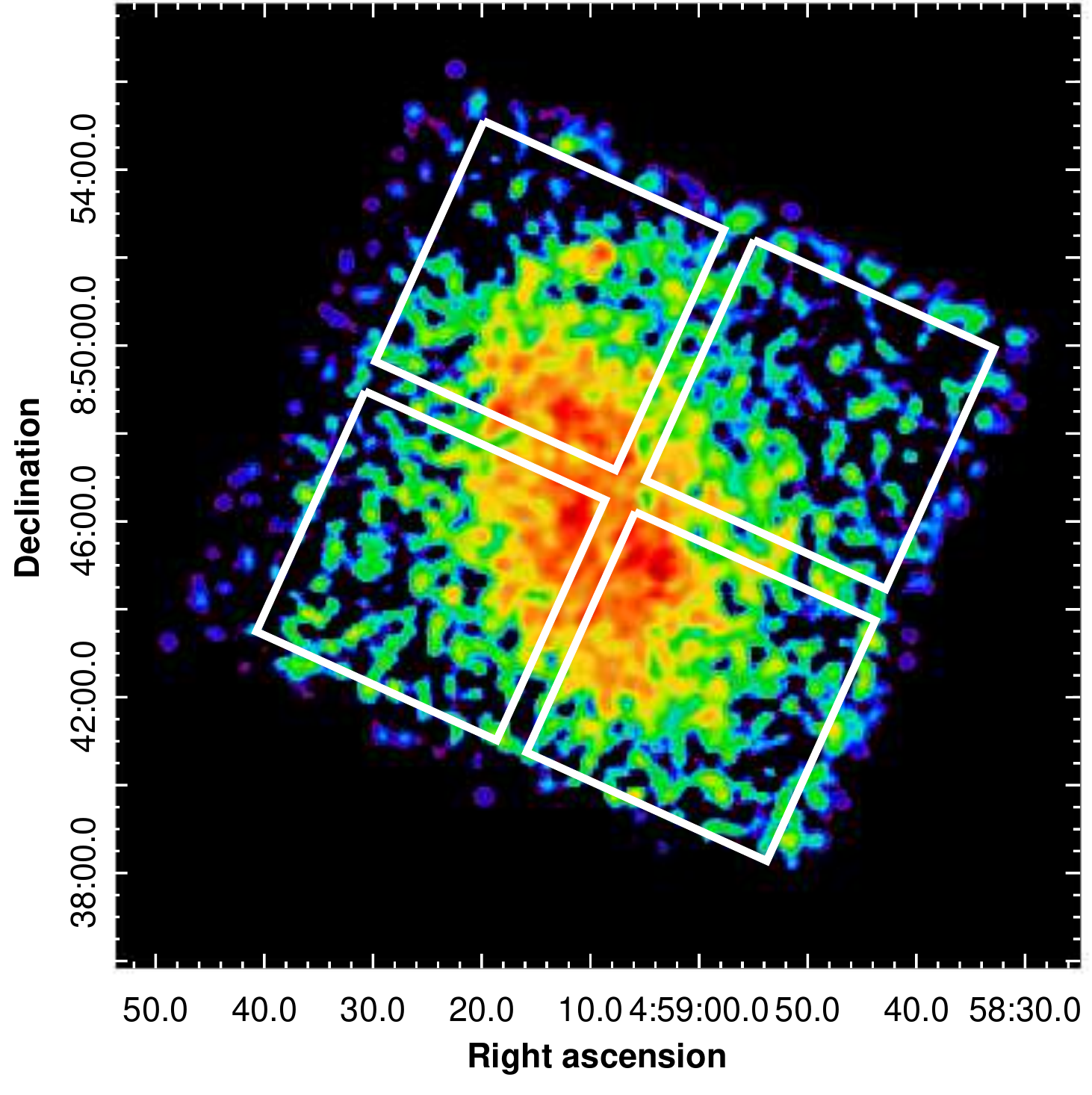}\par 
\includegraphics[width=\linewidth]{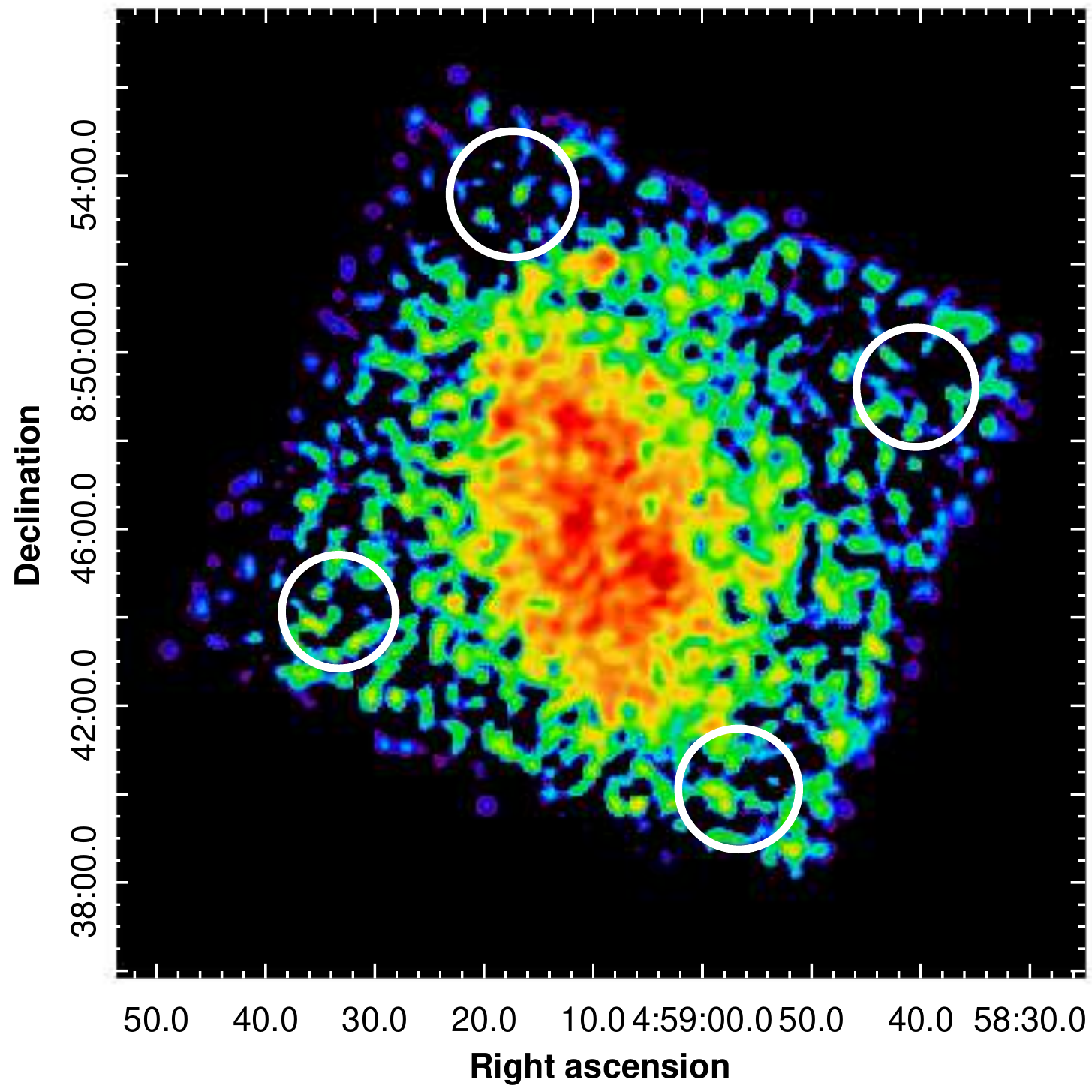}\par 
\end{multicols}
\caption{Regions considered for the background estimation for the two methods described in the text: \textit{fixed} (left panel) and \textit{corner} (right panel).}
\label{fig:bgd_reg}
\end{figure*}

The first method (\textit{fixed} background method) consists of using fixed empirical nominal models (for the Aperture and the fCXB components) based on blank field observations in fitting the spectra of four square regions that cover almost entirely the areas of the four detectors, as done by \citet{Gasta15} (see Figure \ref{fig:bgd_reg}, left panel). We accounted for the cluster emission by including a thermal \url{APEC} model \citep{Smith01}, fixing its parameters to the values extracted from the previous global analysis of a deep \textit{XMM-Newton} observation.

The second method (\textit{corner} background method) consists of adopting the standard \url{nuskybgd} procedure, starting from four small circular regions localized at the corners of each detector for both telescopes and observations, as shown in the right panel of Figure (\ref{fig:bgd_reg}), and leaving all the background components free to fit. To test if these circular regions were actually source-free, we performed another fit of their spectra freezing the Aperture and fCXB normalizations to the same nominal values adopted for the \textit{fixed} background estimation method, but without including the additional thermal model. In this case, the spectral fit does not show significant discrepancies from the fit performed with free parameters, and no clear excess of cluster emission at lower energies is visible. 

Based only on the results here presented, the two methods seem to perform equally well in fitting the spectra extracted from different regions across the FOV, provided that in the \textit{fixed} background method a proper thermal model accounting for the cluster emission is included. The \textit{corner} method could then in principle be more solid in estimating the background components relative to this particular observation, as it does not seem to need an additional thermal model and the background parameters are left free to fit. However, it's important to note that the small circular regions considered have a quite low statistic, compared to that of the bigger square regions of the \textit{fixed} method, giving poor constraints on the background parameters. In addition to that, even though a clear residual cluster emission is not immediately visible in their spectra, it's still not safe to affirm that these corner regions are really source-free.
\begin{figure*}
\centering
   \resizebox{0.7\linewidth}{!}{
            \includegraphics[angle=270,width=\linewidth]{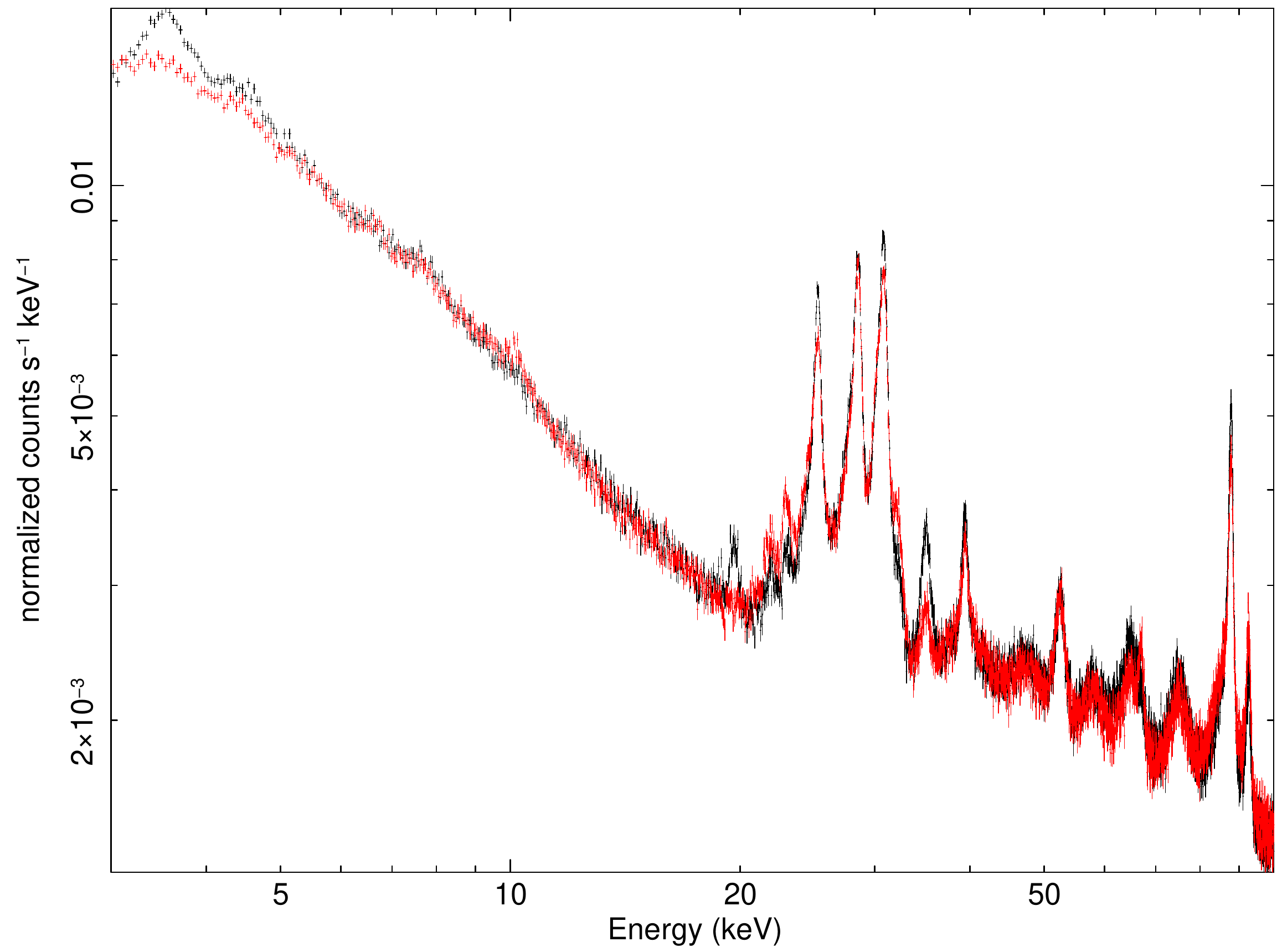}
      			
      			}
      			\caption{Comparison between the background spectra produced using the models from the two different methods of background estimation: $fixed$ method (red) and $corner$ method (black). The spectra in this image refer to a central circular region of 5 arcmin, covering the majority of the cluster emission, and for telescope A, ObsID 7012001004.}
         \label{fig:bgcomp_fixed_corn}
   \end{figure*}

In order to better investigate this possibility, we proceeded in producing the total background spectra for a centered circular region of 5 arcmin (covering almost all of the cluster emission), starting from the best-fit models of both the methods presented above, using \url{fakeit} in XSPEC. For each telescope and for both observations we then compared the simulated spectra from the two methods and calculated their net count rate (cts/s) ratios in two bands: 3.0-100.0 keV and 3.0-5.0 keV, the latter being motivated by the fact that in that band the spectra seem to differ the most in almost all cases (see an example in Figure \ref{fig:bgcomp_fixed_corn}). The calculated ratios are listed in Table \ref{tab:rates_ratios}. The count rate ratios seem to indicate that the two produced spectra are globally very similar, being in all the cases considered well within 2\%; the discrepancies arise in the narrow 3.0-5.0 keV band, where the ratios are less than 10\% or 15\% in the worst case. This greater difference is most probably due to the fact that the background spectra produced starting from the \textit{corner} estimation method (in black in Figure \ref{fig:bgcomp_fixed_corn}) are often over-estimated, because of some residual cluster emission which was not accounted for and incorrectly contributed to the Aperture background. Given this result, and the motivations outlined above, we decided to rely on the \textit{fixed} estimation method to simulate background spectra needed for A523's spectral analysis and for the production of the background images.
\begin{table}[th]
\medskip
\centering
\begin{tabular}{c c c c c} 
\toprule
 & \multicolumn{2}{c}{7012001002} & \multicolumn{2}{c}{7012001004} \\
Band & A& B & A & B \\
\midrule
3.0-100.0 keV & 1.019 & 1.007 & 0.984 & 0.982 \\
3.0-10.0 keV & 0.991 & 1.005 & 0.967 & 0.939 \\
\bottomrule
\end{tabular}
\caption{\label{tab:rates_ratios} Ratios of the model predicted rates from the backgrounds total models derived by the two methods described in this Section (\textit{corner} over \textit{fixed}).}
\end{table}

\section{Simulations}
\label{sec:appendixB}

In order to investigate the robustness of our upper limits on the non-thermal IC flux, we performed a series of ad hoc simulations of our \textit{NuSTAR} observation, using \textit{fakeit} in \url{Xspec}. In particular, we simulated the spectra for our inner 5' circular region, giving the same nominal background model that we used in our analysis and varying only the input thermal model. The purpose of this exercise is to test whether our upper limits are a reliable guess of the non-thermal IC flux in our cluster, and at what level we are able to effectively reproduce a multi-thermal spectrum. In this section we will present the various simulations and the cumulative results obtained when fitting the spectra with different thermal and non-thermal models.
\subsection{1T model}
We started simulating a simple single temperature model with a temperature of $kT = 5.2$ keV (i.e. the best-fit temperature found for our global analysis with a single temperature model). We then fitted to the simulated spectra a single temperature model, in order to test if we were able to reproduce the given input temperature, and a 1T plus power-law model with photon spectral index $\Gamma = 2$, to test if we could get a fictitious non-thermal IC flux in the 20-80 keV band even when not present and in a perfectly isothermal situation. 
\begin{figure*} [th]
\begin{multicols}{2}
\includegraphics[width=0.49\textwidth]{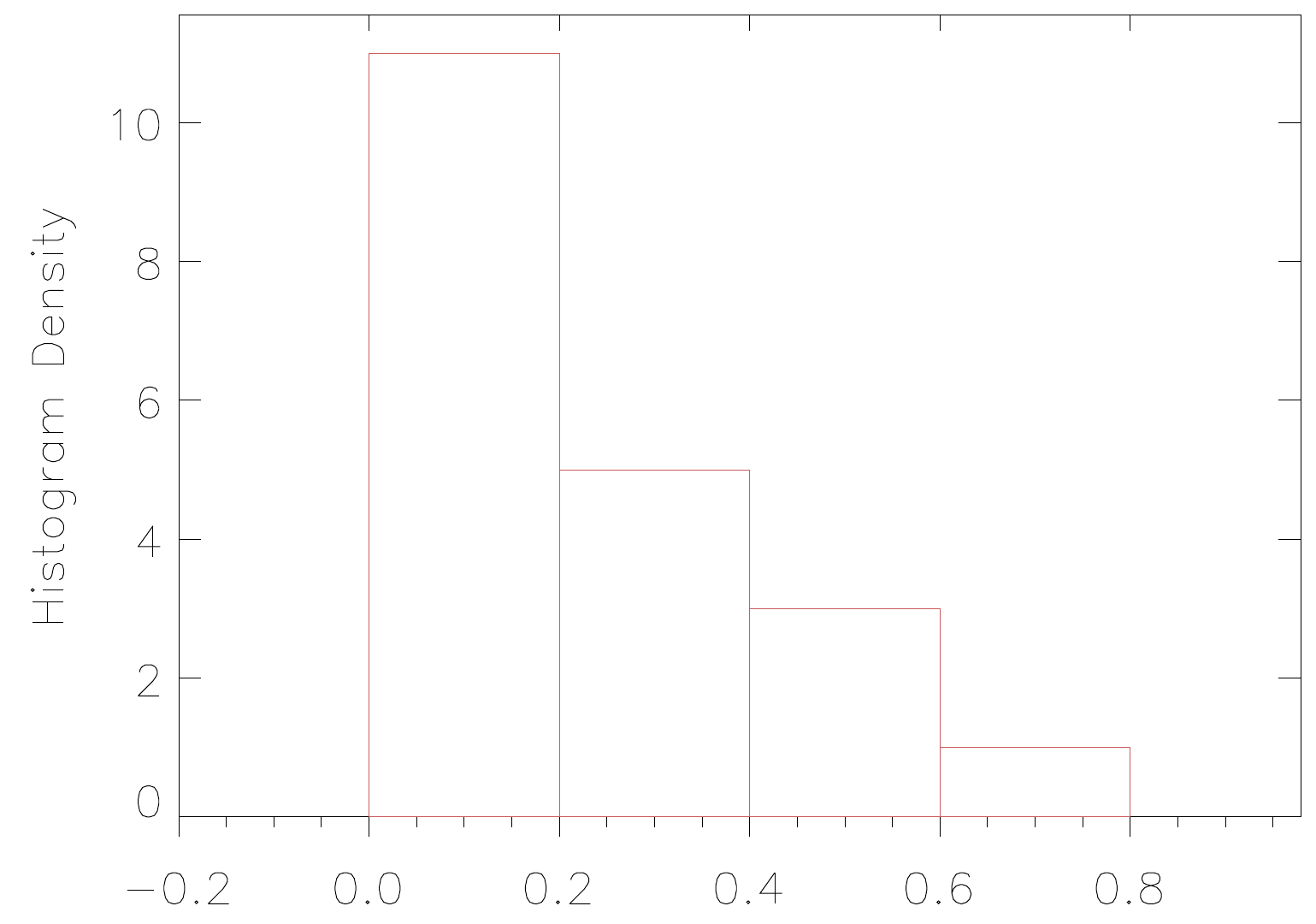}\par
\includegraphics[width=0.49\textwidth]{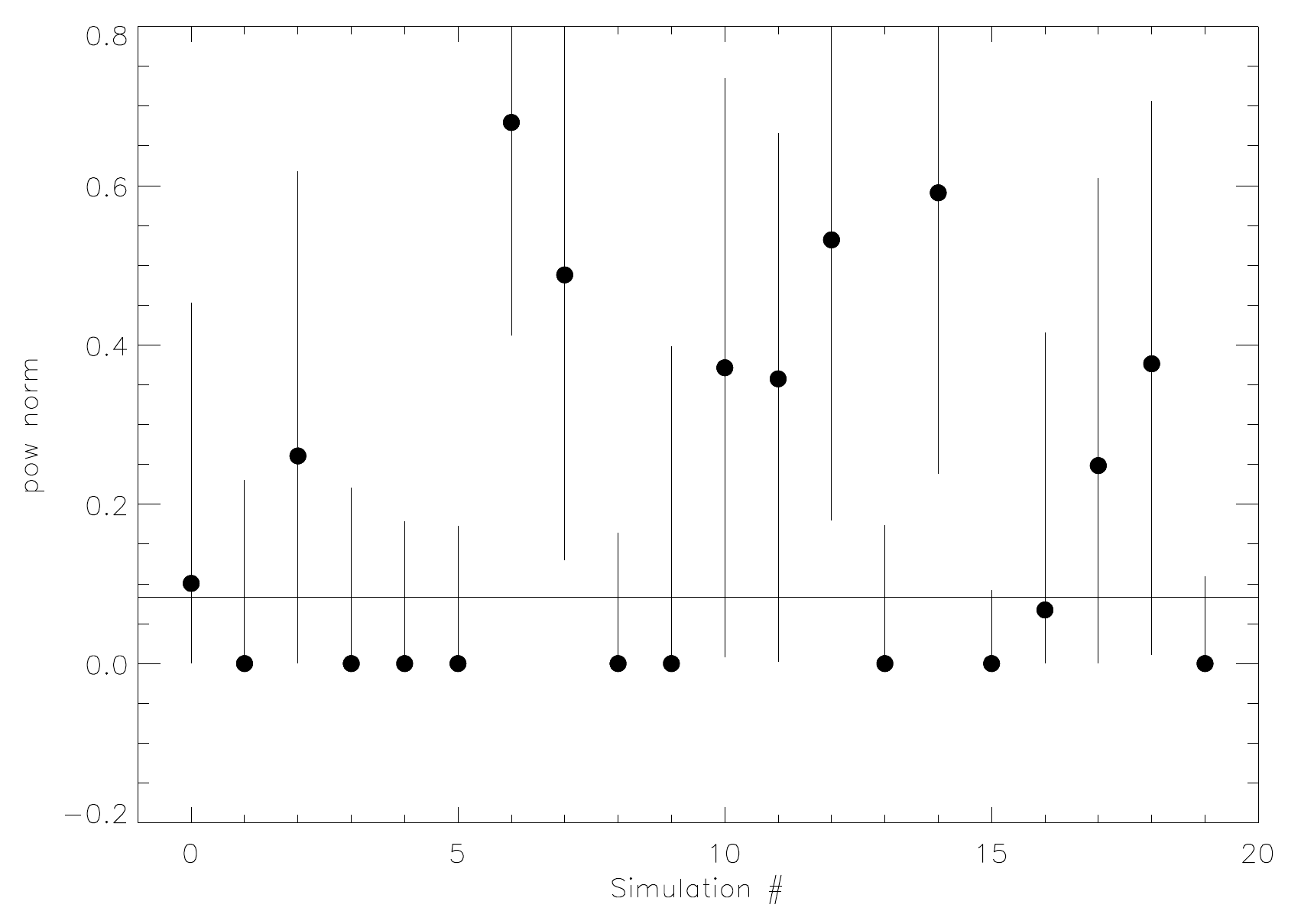}\par
\end{multicols}
\caption{Left panel: histogram density of the IC flux in the 20 - 80 keV band obtained fitting a 1T+IC model for the single temperature simulations. Right panel: IC fluxes with error bars for each simulation, the solid black line represents the mean value, averaged over all the simulations.}
\label{fig:simul_1T}
\end{figure*}

The first model resulted in a mean temperature value of $kT = 5.18 \pm 0.08$, averaged over all the simulations performed, which is perfectly consistent with the input parameter. The results of the 1T+IC model fitting for each simulation are presented in Figure \ref{fig:simul_1T}. While the majority of our simulations did not produce any detection of an IC flux, a small tail of non-zero flux values is present in the histogram density plot. However, these values have very large error bars and the weighted average is lower than our upper limits. We can thus conclude that in a perfectly isothermal cluster, our method does not detect any spurious non-thermal flux due to statistic.
\subsection{1T+IC model}
We also wanted to test if our spectrum allowed us to reproduce a non-thermal IC flux when this is actually present. To do that, we simulated a single temperature cluster ($kT = 4.7 $ keV) which also has a non-thermal component, represented by a power-law model which has the same photon spectral index of the IC component that we looked for in our global analysis ($\Gamma = 2.0$) and the same flux in the 20 - 80 keV band that we measured with a 1T+IC model in our analysis ($\approx 0.38 \times 10^{-12} \, \mathrm{erg} \, \mathrm{s}^{-1} \, \mathrm{cm}^{-2}$). 
\begin{figure*} [th]
\begin{multicols}{2}
\includegraphics[width=0.49\textwidth]{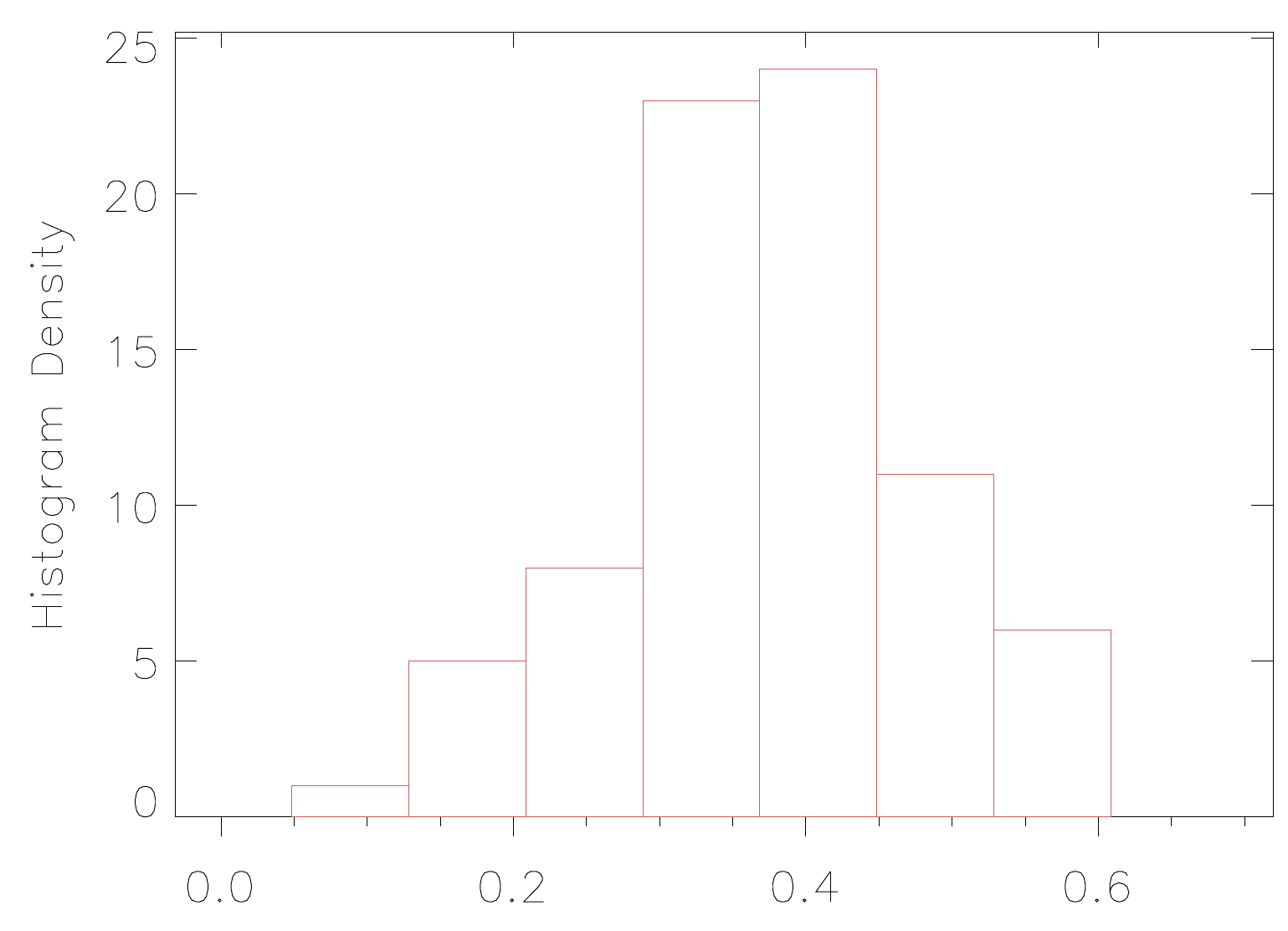}\par
\includegraphics[width=0.49\textwidth]{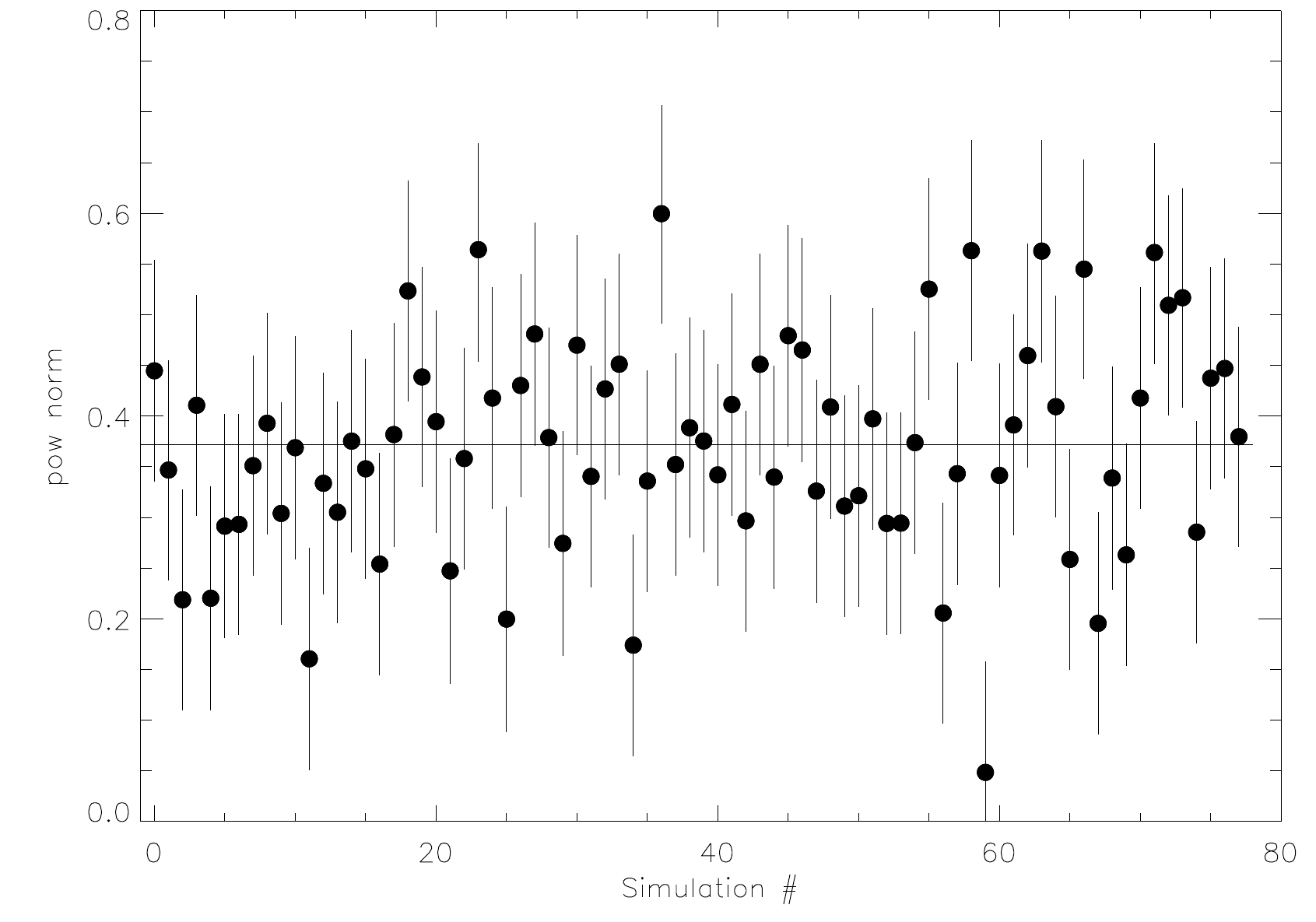}\par
\end{multicols}
\caption{Left panel: histogram density of the IC flux in the 20 - 80 keV band obtained fitting a 1T+IC model for the 1T+IC simulations. Right panel: IC fluxes with error bars for each simulation, the solid black line represents the mean value, averaged over all the simulations.}
\label{fig:simul_1T+IC}
\end{figure*}

The results are showed in Figure \ref{fig:simul_1T+IC}. We can clearly see that the weighted average IC flux value found from all our simulations is perfectly consistent with the input value, meaning that when a non-thermal component is actually present in our spectrum, we are able to detect it properly with our fitting procedure. We then tried to fit a simple single temperature model to this simulated spectrum. The mean temperature found with this procedure, averaging all the simulations, was of $\approx 5.2$ keV, which is consistent with the best-fit temperature found when fitting our global spectrum with a 1T model. This confirms that when a non-thermal component is actually present, fitting the whole spectrum with a single thermal model results in a best-fit temperature which is on average higher than the cluster's real temperature.

\subsection{2T model}
We proceeded further in simulating a two-temperature cluster, without any non-thermal component. We chose the two input temperature values to be of $kT_{1} = 4.0$ keV and $kT_{2} = 7.0$ keV, in accordance with our temperature map (see Fig. \ref{fig:temp_map_nu}). We then fitted these simulated spectra with three different models: 1T+IC (single temperature plus power-law), 2T, 1T.
\begin{figure*} [th]
\begin{multicols}{2}
\includegraphics[width=0.49\textwidth]{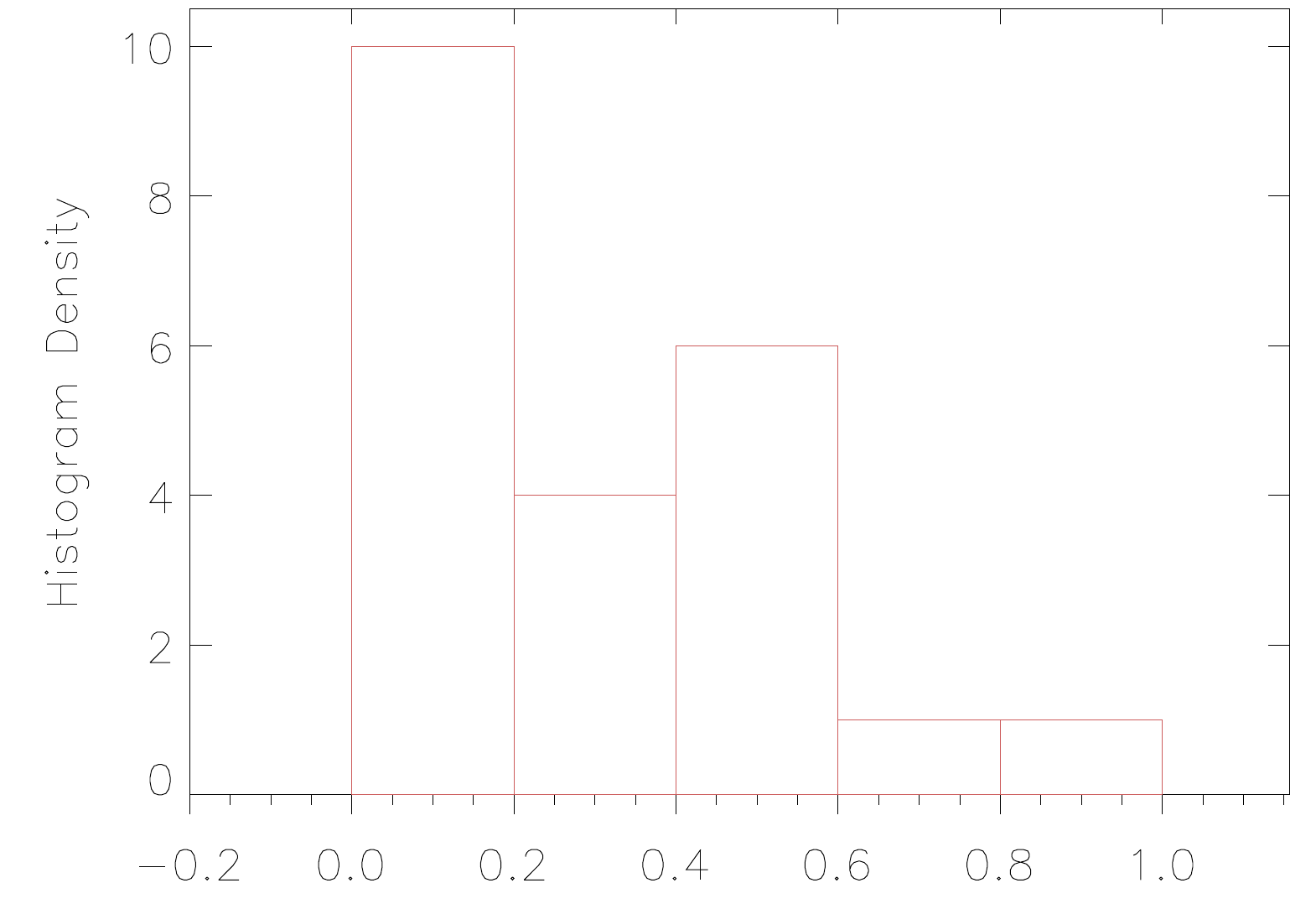}\par
\includegraphics[width=0.49\textwidth]{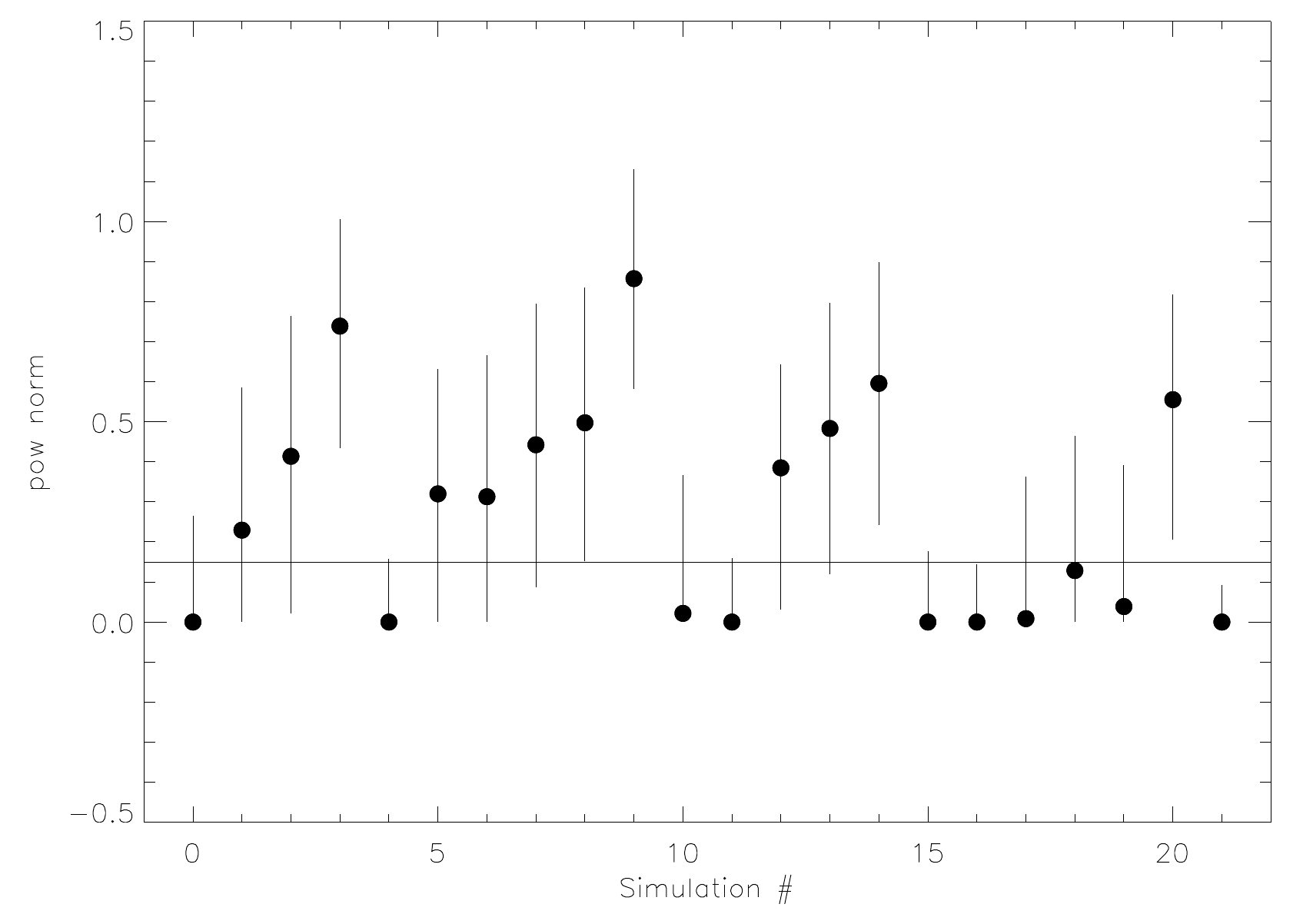}\par
\end{multicols}
\caption{Left panel: histogram density of the IC flux in the 20 - 80 keV band obtained fitting a 1T+IC model for the two-temperature simulations. Right panel: IC fluxes with error bars for each simulation, the solid black line represents the mean value, averaged over all the simulations.}
\label{fig:simul_2T}
\end{figure*}

The results for the first model are shown in Figure \ref{fig:simul_2T}. The tail of non-zero values looks slightly longer and higher than the one found using the same model in the simulated 1T spectra, which is expected. Again, the error bars on these higher values are quite large, and this is consistent with what found in our best-fit 1T+IC model for A523's global spectrum. We can thus conclude that, when fitting a single temperature model with a non-thermal component to a multi-temperature spectrum, there is some probability that the fitting procedure may produce a fictional detection of a non-thermal flux, trying to accommodate the fit for the missing thermal components, as we hypothesized in the text. The upper limit of our 1T+IC fit to the global spectrum cited in the text, could then include a significant bias of this nature, and should not be accounted as an effective limit on the purely non-thermal flux. We then tried to fit to our simulated spectra a two-temperature model, in order to test if we were able to reproduced the initial given temperatures. This was not the case, as the mean temperatures averaged for all our simulations are, $kT_{1} \approx 1.7$ keV and $kT_{2} \approx 5.5$ keV. Our spectra may then not be accurate enough for us to discriminate between two temperatures, and the two thermal parameters settle to a non-physical value and to an average temperature value for the whole cluster, respectively. This is indeed confirmed when we fit to this simulated spectra a simple 1T model, which results in a mean temperature of $\approx 5.2$ keV, which again is consistent with the average cluster value found in our global fit and with the second best-fit temperature found in this section.
\end{appendix}

\end{document}